\documentstyle[psfig]{mn}

\newcommand{\gsim}{\raisebox{-3.8pt}{$\;\stackrel{\textstyle >}{\sim}\;$}}
\newcommand{\lsim}{\raisebox{-3.8pt}{$\;\stackrel{\textstyle <}{\sim}\;$}}

\newcommand{\Msol}{$M_{\odot}$}
\newcommand{\Zsol}{$Z_{\odot}$}
\newcommand{\Lsol}{$L_{\odot}$}

\newcommand{\etal}{\mbox{{\rm et~al.\ }}}

\title{On the Mass--to--Light ratio and the Initial Mass Function \\
in disc galaxies}

\author[Portinari et al.]{ L.~Portinari$^1$,  
J.~Sommer--Larsen$^1$, R.~Tantalo$^2$ \\ 
       $^1$ Theoretical Astrophysics Center, Juliane Maries Vej 30,
 DK-2100 Copenhagen \O \\
       $^2$ Dipartimento di Astronomia, Universit\`a di Padova,
 Vicolo dell'Osservatorio 2, I-35122 Padova, Italy \\
E-mail: {\tt lportina,jslarsen@tac.dk; tantalo@pd.astro.it}
}

\date{\tt Submitted: July 2003. Accepted: September 2003}

\pubyear{2003}
\begin{document}
\maketitle
\title{On the M/L ratio and the IMF in disc galaxies}
%
%
\begin{abstract}
A low mass--to--light ratio for the stellar component of spiral galaxies
(M/L$\lsim$1 in the I--band) is advocated by various dynamical arguments 
and by recent cosmological simulations of the formation of these systems.
We discuss this possibility by means of chemo-photometric models for 
galactic discs, adopting different Initial Mass Functions. We show that 
a number of ``bottom--light'' IMFs (namely, with less mass locked in low--mass
stars than the standard Salpeter IMF), suggested independently in recent 
literature, do imply M/L ratios as low as mentioned above, at least
for late type spirals (Sbc/Sc). This conclusion still holds when the bulge 
contribution to mass and light is included.
We also predict the typical stellar M/L ratio, and correspondingly the 
zero--point of the Tully--Fisher relation, to vary considerably with Hubble 
type (about 0.5--0.7~mag in the red bands, from Sa to Sc type).

For some of the bottom--light IMFs considered, the efficiency of metal 
production tends to exceed what is 
typically estimated for spiral galaxies. Suitable tuning of the IMF mass 
limits, post--supernova fallback of metals onto black holes or metal outflows 
must then be invoked, 
to reproduce the observed chemical properties of disc galaxies.

In the appendix we provide M/L---colour relations to estimate the 
stellar M/L ratio of a galaxy on the base of its colours, for several IMFs.
\end{abstract}


\section{Introduction}
\label{sect:introduction}
\begin{figure}
\centerline{\psfig{file=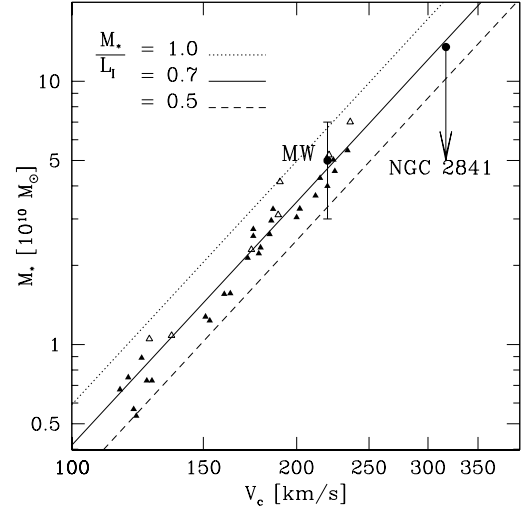,width=7.1truecm}}
\caption{Straight lines: location of the observed Tully--Fisher relation 
for Sbc--Sc disc galaxies by Giovanelli \etal (1997a,b; $h$=0.65 adopted here) 
in the stellar mass vs.\ circular velocity plane, for different assumptions 
about the stellar M/L$_I$ ratio. We show
also the location of simulated disc galaxies by Sommer--Larsen \etal (2003,
triangles), and of two observed galaxies: the Milky Way and NGC 2841 
(the latter rescaled to $h=0.65$).}
\label{fig:TF}
\end{figure}

Present cosmological simulations of the formation 
of disc galaxies within the Cold Dark Matter hierarchical scenario, 
suffer from the
angular momentum problem  (Navarro, Frenk \& White 1995).
Suggested ways out include early energy injection
from stellar feed-back (Sommer--Larsen, Gelato \& Vedel 1999; Sommer--Larsen,
G\"otz \& Portinari 2003; Thacker \& Couchman 2001) or Warm Dark Matter 
cosmology (Sommer--Larsen \& Dolgov 2001). 
Such N--body+SPH simulations not only improve significantly on the angular
momentum problem, but also 
compare successfully to the observed Tully--Fisher (TF) relation, 
{\it provided} the mass-to-light (M/L) ratio of the stellar component 
is low (Sommer-Larsen \& Dolgov 2001; Sommer-Larsen \etal 2003):
in the I--band, the inferred {\mbox{M/L$_I \sim$ 0.8}} (Fig.~\ref{fig:TF}). 
However, in cosmological simulations the need of a low baryonic M/L ratio 
may be partly due to the too dense cores of dark matter haloes:
in simulated galaxies, dark matter is
more concentrated and dynamically 
dominant in the central regions, than 
estimated from 
rotation curves (Navarro \& Steinmetz 2000a,b; Eke, Navarro \& Steinmetz 2001; 
Abadi \etal 2003; Salucci 2001; Salucci \& Borriello 2001; Bottema 2002; 
de Blok \& Bosma 2002; de Blok \etal 2001a,b; Marchesini \etal 2002).

The suggestion of a low stellar M/L ratio is nevertheless reinforced 
by a simple 
estimate of the stellar mass of our Galaxy, based on the local stellar 
density and on a plausible density profile for the Disc
(Sommer-Larsen \& Dolgov 2001).
The stellar mass obtained in this way for the Milky Way is of the order of 
$5 \times 10^{10}$~\Msol,
and if our Galaxy is to lie on the 
observed TF relation together with other spirals, its M/L$_I$ ratio
must be around 0.75 (see Fig.~\ref{fig:TF}). Considering all uncertainties,
Sommer-Larsen \& Dolgov (2001) estimate for the {\it total} baryonic matter 
in the Milky Way (inclusive of stars, gas and stellar remnants)
{\mbox{M/L$_I = 0.87 \pm 0.57$}}. 

Similar conclusions can be drawn from the Sb galaxy NGC 2841:
estimates of its disc mass range between 
{\mbox{$5 \, h_{50}^{-1} \times 10^{10}$~\Msol}} and
{\mbox{$11.7 \, h_{75}^{-1} \times 10^{10}$~\Msol}}\footnote{We indicate 
with $h=h_{100}$, with $h_{50}$ and with $h_{75}$ the Hubble constant 
in units of 100, 50 and {\mbox{75 km sec$^{-1}$ Mpc$^{-1}$}} }
and its circular velocity is 317~km~sec$^{-1}$
(Kent 1987; Benson \etal 2000 and references therein), 
which also points to a {\mbox{M/L$_I \lsim$ 0.7}} (Fig.~\ref{fig:TF}).
Regarding the zero--point of the Tully--Fisher relation, however, some
uncertainties exist and are discussed in Appendix~A.

An independent indication of a low M/L ratio for the baryonic
component of spiral discs relies on stability arguments.
Bar instability of isolated discs was one of the reasons 
to introduce massive dark halos historically (Ostriker \& Peebles 1973); 
the argument still holds,
in the sense that if too much mass is stored in the luminous disc
component, strong bars should form.
Based on this, Efstathiou, Lake \& Negroponte (1982) put an upper limit 
of {\mbox{M/L$_B \leq 1.5 h$}} to the disc M/L ratio, 
that is {\mbox{M/L$_B \leq 1$}} for $h=0.65$.

The M/L ratio of the stellar component is also relevant to
the issue as to whether discs are maximal or {\mbox{sub-maximal}},
or to what extent the luminous, baryonic mass dominates
the dynamics and the rotation curve in the inner regions of the disc.
Even under the hypothesis of maximal discs, Bell \& de Jong (2001)
point out that lower M/L ratios for the stellar component are implied,
than those predicted with a Salpeter IMF.
And discs may well be sub-maximal (Bottema 1993, 1997; Courteau \& Rix 1999), 
although this is still much debated 
(Ratnam \& Salucci 2000; Fuchs 2002, 2003a,b; Masset \& Bureau 2003; 
Kranz, Slyz \& Rix 2003).
For his favoured sub-maximal disc case, with the luminous component
contributing 63\% of the rotation curve at maximum, Bottema (2002) 
suggests a stellar {\mbox{M/L$_I \sim 0.82$}} for spirals in the Ursa Major
cluster.
For the Milky Way itself, a variety of mass models is compatible
with available kinematic data (Binney \& Merrifield 1998; Dehnen \& Binney 
1998), so that on this base no clear discrimination is possible between 
the maximal or sub-maximal case; though recent combined results on luminosity
and mass distribution, gas flows and microlensing seem to favour a maximal 
disc (Gerhard 2002).

Two recent, detailed studies of individual spirals also point
toward low M/L ratios. A dynamical analysis of the isolated 
Sc galaxy NGC 4414 indicates an intrinsic stellar M/L$\sim$1 
in the B, V and I band (Vallejo, Braine \& Baudry 2002). 
For the Sab spiral 2237+0305,
Huchra's lens, Trott \& Webster (2002) combine dynamical and gravitational
lensing constraints to derive {\mbox{M/L$_I \sim$1.1}} for the disc.

In the light of these results, the obvious question arises whether M/L ratios 
so low are plausible from the point of view of the stellar content and
of the photometric evolution of galaxies; this relies heavily on the stellar
Initial Mass Function (IMF) of stars in disc galaxies. A standard Salpeter 
IMF, extended over the typical stellar mass range [0.1--100]~\Msol\
certainly yields much higher M/L ratios than those mentioned above
(see \S\ref{sect:b-models}).
On the other hand, there is by now a general consensus
that a single--slope power law IMF over the whole 
stellar mass range is not adequate for the local nor for other environments. 
The slope of the IMF appears in fact to become shallower
below 1~\Msol, 
in the local environment (Miller \& Scalo 1979; Scalo 1986; Kroupa, Tout
\& Gilmore 1993; Reid \& Gizis 1997; Gould \etal 1997, 1998; Chabrier 2001) 
as well as in globular clusters and in the Galactic Bulge
(Piotto \& Zoccali 1999; Zoccali \etal 2000). 
Besides, a turn-over in the regime of brown dwarfs is favoured by recent 
results (Chabrier 2002). 
The observed IMF, that is, appears to be ``bottom-light'' with respect 
to a single power-law, Salpeter-like IMF; for recent reviews, see Kroupa 
(2002); Chabrier (2003). From the theoretical point of view, the flattening 
and turn--over of the IMF
at low masses may be related to the Jeans scale (thermal support)
and to the scale of magnetic support against gravitational collapse 
(Larson 1998; Padoan \& Nordlund 2002).

In this paper we investigate
what kind of IMF is required
to obtain a low M/L ratio (say, $\sim$0.8 in the I--band) in a typical 
Sbc--Sc galaxy.
We consider 
different IMFs among those suggested
in literature from a variety of independent studies. For each of these IMFs, 
we calculate chemical evolution models with infall, metallicity
gradients, and star formation histories representative of 
late--type spiral discs; then we compute the corresponding broad--band 
luminosities by convolving suitable Single Stellar Population (SSP) models
with the star formation and metal enrichment histories.

The paper is organized as follows. In Section~2 we discuss which Hubble types 
and photometric passbands are most relevant for our present analysis
of the M/L ratio in galactic discs. In Section~3 we present the various IMFs
considered in our chemo--photometric models and in Section~4 the corresponding
sets of SSPs. In Section~5 we present simple exponential models
with fixed metallicity, to give a first overview of the problem and to discuss
the solidity of the predicted M/L ratios with respect to the
adopted stellar models and spectral library. In Section~6 we introduce our
full chemo--photometric models for disc galaxies, and the corresponding 
calibration on observational constraints. In Section~7 we present our results
on M/L ratios obtained from chemo--photometric models with different IMFs.
In Section~8 we add the contribution of the bulge to the global M/L ratio
of galaxies. Finally, in Section~9 we outline summary and conclusions.
In Appendix~A we discuss observational uncertainties in the zero--point of the
TF relation. In Appendix~B we provide M/L---colour relations to estimate the 
stellar M/L ratio of a galaxy on the base of its colours.

\section{Focusing on late Hubble spiral types}
\label{sect:Hubble}
The M/L ratio of the stellar component of a galaxy depends on two basic
quantities: the IMF and the star formation history (SFH); the latter sets the
relative weight of young, bright and blue stellar populations with respect
to old, dim and red ones. 
The SFH depends on Hubble type, as clearly indicated
by the variation of typical colours along the Hubble sequence (Roberts
\& Haynes 1994). Hence, the M/L ratio of the stellar mixture is also expected
to depend on Hubble type. This effect is indeed observed in the empirical
TF relation, where a systematic, negative offset 
in luminosity is observed when going from Sbc--Sc spirals to Sb to earlier 
types (Rubin \etal 1985; Giovanelli \etal 1997a); and the offset correlates 
with the B--R colour as expected from different SFHs in different Hubble types
(Kannappan, Fabricant \& Franx 2002).
Giovanelli \etal (1997a,b) correct for this morphological dependence,
and provide a TF relation (the one shown in Fig.~\ref{fig:TF}) 
typical for late spiral types. This allows us to focus on the M/L ratio of
Sbc--Sc spirals.

Having selected our fiducial Hubble type, we need to characterize it
for the sake of the chemo--photometric models to be developed. 
The historical, B--band based picture of the Hubble sequence 
as a sequence of bulge--to--disc ({$\cal B/D$}) ratios is not supported 
by recent surveys in redder bands; in the I or K bands, which are better 
mass tracers, $\cal B/D$ correlates fairly weakly with spiral type, 
and with a huge scatter, thus proving to be a poor morphological
indicator (de Jong 1996a; Sommer--Larsen \etal 2003
and references therein; see also \S\ref{sect:bulges}). 
Kennicutt, Tamblyn \& Congdon (1994, hereinafter KTC94) 
demonstrated that the sequence 
of spiral types is rather a sequence of different SFHs in the discs 
themselves, as traced by the so--called {\it birthrate parameter}
\begin{equation}
\label{eq:b-SFR}
b = \frac{SFR}{<SFR>}
\end{equation}
defined as the ratio between the present and past average star formation
rate (SFR). KTC94 estimated the {\mbox{$b$--parameter}} both on the base 
of the equivalent width of the H$_{\alpha}$ line, and from the ratio between 
the H$_{\alpha}$ luminosity (tracer of the present SFR) and the V--band
luminosity (tracer of the global underlying stellar mass in the disc). 
The two methods are not completely independent, for EW(H$_{\alpha}$) is also
a flux ratio between the H$_{\alpha}$ line and the nearby red continuum
(corresponding to the R--band, roughly), but they give consistent results.
From KTC94, the $b$--parameter increases from $b < 0.2$ for Sa--Sab discs, 
to $b \sim$0.3--0.4 for Sb discs, to $b \sim$0.8--1 for Sbc--Sc discs,
while the trend breaks down toward later Hubble types, which display more
irregular and episodic star formation histories; see also Sommer--Larsen
\etal (2003).

Henceforth, in gross terms our chemo--photometric models will be taken 
to represent typical Sbc--Sc spirals when their SFHs correspond 
to a $b$--parameter {\mbox{$\sim$0.8--1}}.

\subsection{Focusing on the I--band}
\label{sect:Iband}
\begin{figure}
\centerline{\psfig{file=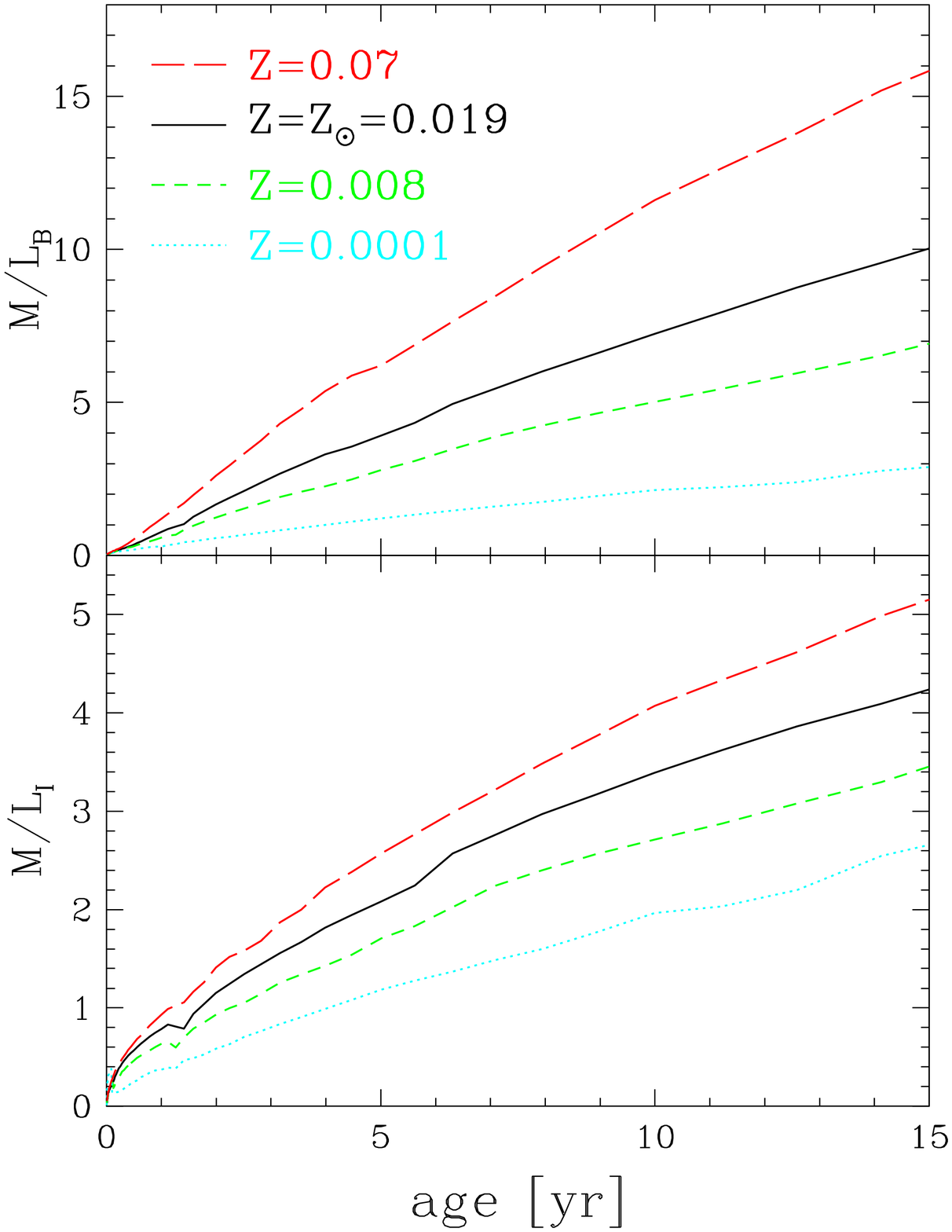,width=6.8truecm}}
\caption{M/L ratio as a function of age for SSPs of different metallicity.
Within this wide metallicity range (from $\sim 10^{-2}$ to 3.5 times solar), 
the M/L ratio in the I--band varies only by a factor of 2 ({\it lower panel}) 
as opposed to a factor of 5 in the B--band
{\mbox{({\it upper panel}).~~~~~~~~~~~~~~~~~~~~~~~~~~~~~~~~~~~~~~~~~~~~~~~}}
{\mbox{These M/L ratios are computed}} from the mass of living stars 
on the SSP, not including the remnants of more massive, dead stars.
The mass of remnants is however included in the exponential and 
chemo--photometric models in the remainder of the paper.}
\label{fig:ML_IB_Z}
\end{figure}

In this paper we will mainly discuss the M/L ratio in the {\mbox{I--band,}}
firstly because the observational evidence quoted in 
\S\ref{sect:introduction}, including the fiducial TF, refers to the I--band. 
Secondly, red bands
like the I--band are good mass tracers because they are less sensitive than
bluer bands to recent SF episodes, as well as to metallicity
effects (Fig.~\ref{fig:ML_IB_Z}). Thirdly, compared to redder, near infrared
bands,
the integrated I--band luminosity at intermediate and old ages is not
so sensitive to the specific treatment of the  Asymptotic Giant Branch
(AGB) phase (see \S\ref{sect:AGB}). 
Finally, red bands like the I--band are less affected by dust extinction
and corresponding corrections. The 
empirical TF relation by Giovanelli et~al., as well as other observational
evidence quoted in the introduction, are already corrected for extinction.

\section{The Initial Mass Function(s)}
\label{sect:IMFlist}
\begin{figure}
\psfig{file=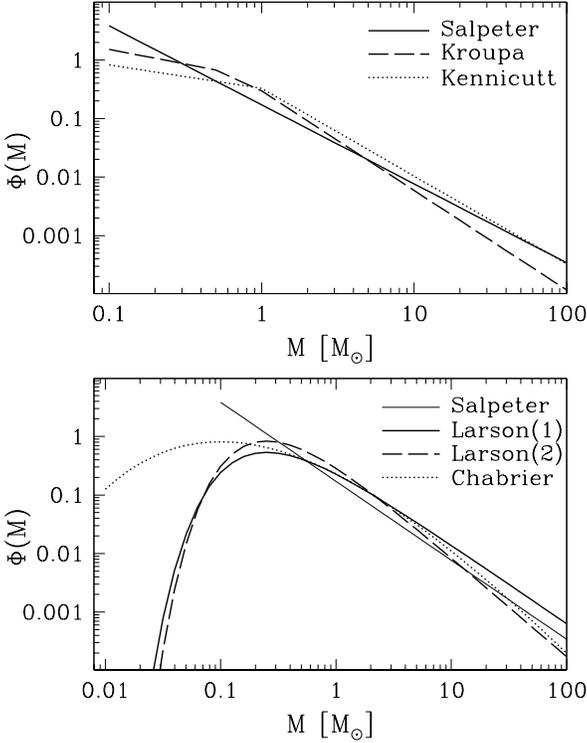,width=7.9truecm}
\caption{Comparison between the 6 different IMFs considered in this paper,
all normalized to and integrated mass of 1~\Msol. {\it Top panel}: the three 
``power--law'' IMFs {\bf (1--3)}, defined in the range 
{\mbox{[0.1--100]~\Msol}}.
{\it Bottom panel}: the three ``exponential'' IMFs {\mbox{\bf (4--6)}}, 
defined in the range [0.01--100]~\Msol; Larson~(1) and {\mbox{Larson~(2)}} 
refer to the original and ``modified'' Larson IMF, respectively; 
the Salpeter IMF is also repeated for comparison.}
\label{fig:IMF}
\end{figure}

In this section we discuss the IMFs we have considered in turn in our
chemo-photometric models, together with their respective 
theoretical/observational background. We will present sets of models with three
``power-law IMFs'' and three ``exponential IMFs'', displayed in 
Fig.~\ref{fig:IMF}. In the following, $M$ will indicate the stellar mass
in solar units.
%
%
\begin{description}
\item[{\bf (1)}]
A Salpeter (1955) IMF, extended over the typical stellar mass--range 
[0.1--100]~\Msol:
\[ \Phi(M) = C_s \, M^{-1.35}~~~~~~~~~~~~C_s=0.1716 \] 
where $C_s$ is a normalization coefficient fixed so that the IMF is 
normalized to unit mass when integrated between the low and high stellar
mass ends.\\
Although the original Salpeter IMF was derived only for stars between
0.4 and 10~\Msol\ and, as mentioned in the introduction 
(\S\ref{sect:introduction}), there is substantial
evidence that a single power--law over the entire mass range 
is not adequate for the observed IMF, this functional form
is still widely adopted in literature and it is useful to include it in 
the present study, for the sake of comparison to other IMFs.
\item[{\bf (2)}]
The Kroupa (1998) IMF, derived for field stars in the Solar Neighbourhood
and often adopted in chemical evolution models for disc galaxies (e.g.\
Boissier \& Prantzos 1999):
\[ \begin{array}{l l}
\Phi(M) = &  \left\{
\begin{array}{l l}
C_{kr1} \, M^{-0.5}  \, &  \, 0.1 < M < 0.5 \\
C_{kr} \, M^{-1.2}  \, &  \, 0.5 < M < 1 \\
C_{kr} \, M^{-1.7}  \, &  \, 1 < M < 100 
\end{array} \right.
\end{array} \]
\[ C_{kr1} = 0.480~~~~~~~~~~~~~~~~~~~C_{kr} = 0.295 \]
where $C_{kr1}$ and $C_{kr}$ are fixed so as to guarantee normalization and 
continuity of the IMF over the range [0.1--100]~\Msol.\\
This IMF is steeper than the Salpeter one in the high--mass
range {\mbox{($M > 1$~\Msol)}} but it flattens out progressively at low masses 
{\mbox{($M < 1$~\Msol).}} 
The steep slope in the high--mass range was actually taken after Scalo (1986).
In a more recent determination Kroupa (2001) finds instead
a shallower slope of $1.3$, close to the Salpeter value, but 
the steeper Scalo slope is recovered if unresolved binary 
systems are taken into account (see also Kroupa 2002).
\item[{\bf (3)}]
The Kennicutt (1983) IMF, often advocated as adequate to reproduce the global
properties of spiral galaxies (e.g.\ KTC94, Sommer-Larsen 1996), again with 
the typical mass range [0.1--100]~\Msol.
\[ \begin{array}{l l}
\Phi(M) = &  \left\{
\begin{array}{l l}
C_{k83} \, M^{-0.4}  \, &  \, 0.1 < M < 1 \\
C_{k83} \, M^{-1.5}  \, &  \, 1 < M < 100 \\ 
\end{array} \right.
\end{array} \]
\[ C_{k83} = 0.328 \]
This IMF is somewhat shallower in the high--mass range than the local one, as
seems to be required to account for the observed H$_{\alpha}$ luminosities
and equivalent widths of external disc galaxies. The flattening below 
1~\Msol\ is introduced after the observational evidence for the local IMF 
(see KTC94 for details). 
\item[{\bf (4)}]
The Larson (1998) IMF, over the mass range {\mbox{[0.01--100]~\Msol,}} that is
down to the sub-stellar regime (which in this case has, however, 
a negligible contribution to the mass budget):
\[ \Phi(M) = C_{L1} \, M^{-1.35} \, exp \left( - \frac{M_{L1}}{M} \right) \]
\[ C_{L1}=0.317~~~~~~~~~~~~~~~~~~~M_{L1} = 0.3375 \] 
This IMF recovers a Salpeter power--law at high masses, while at low masses
the exponential cut-off produces a characteristic peak mass, or mass scale,
{\mbox{$M_p = M_{L1}/1.35 = 0.25$~\Msol}} in agreement with some observational
evidence for the Solar Neighbourhood.
An exponential cut-off in the low--mass range is
favoured also by recent determinations of the local IMF down to the brown dwarf
regime (Chabrier 2001, 2002; see below).\\
This functional form of the IMF was originally introduced by Larson (1986)
with the suggestion that the characteristic mass scale $M_{L1}$ may 
be larger for higher temperatures of the parent gas, 
as expected for the Jeans mass scale. This would imply
a top--heavy IMF in the early galactic phases, at high redshifts and/or 
in low--metallicity environments. A suitable variation of the typical 
mass scale $M_{L1}$
could reproduce the systematic variations of M/L ratios in elliptical
galaxies (Chiosi et al.\ 1998) and explain the chemical enrichment of the 
intra--cluster medium (Chiosi 2000; Moretti, Portinari \& Chiosi 2003; 
Finoguenov, Burkert \& B{\"o}hringer 2003), or be responsible for the possible 
presence of dark massive objects in the Galactic halo 
(Chabrier 1999)\footnote{Notice that the Macho--inspired halo IMF suggested
by Chabrier (1999) and mentioned here, peaking at 1.7~\Msol, differs from the 
IMF derived by Chabrier (2001, 2002) for the Solar Vicinity, peaking 
around 0.1~\Msol\ and used below in this paper.}.\\
In the present study however, we do not consider variations in the IMF with
time or metallicity, and within each chemo-photometric model we assume the
IMF to be constant.
\item[{\bf (5)}]
A modified Larson IMF:
\[ \Phi(M) = C_{L2} \, M^{-1.7} \, exp \left( - \frac{M_{L2}}{M} \right) \]
\[ C_{L2}=0.4337~~~~~~~~~~~~~~~~~~~M_{L2} = 0.425 \] 
This IMF maintains the same functional form as the original Larson (1998) IMF,
but with a steeper slope at high masses in better agreement with
local determinations (Scalo 1986; Kroupa 2001). The mass scale $M_{L2}$
is chosen so as to maintain the same peak mass $M_p=M_{L2}/1.7 = 0.25$~\Msol\
as in the previous case.
\item[{\bf (6)}]
The IMF suggested by Chabrier (2001, his case IMF3):
\[ \Phi(M) = C_C\, M^{-2.3} \, 
exp \left[ - \left( \frac{M_C}{M} \right)^{0.25} \right] \]
\[ C_C=40.33~~~~~~~~~~~~~~~M_C=716.4 \] 
over the mass range [0.01--100]~\Msol.
Chabrier (2001) adopts this functional form, a power--law with an
exponential cut-off, after Larson (1986) and tunes its parameters  
to fit  the local field IMF in the low--mass range ($M<1$~\Msol), showing
also that this form of the IMF is valid down to the brown dwarf regime
(Chabrier 2002). In the range of massive stars 
the steep Scalo/Kroupa slope is recovered. 
\end{description}

\section{The Single Stellar Populations}
\label{sect:SSP}
A Single Stellar Population (SSP) is a coeval, chemically homogeneous assembly
of stars with masses distributed according to the IMF; the integrated spectrum
and luminosity $L_{SSP}(\tau,Z)$ of a SSP of age $\tau$ and metallicity $Z$
can be computed on the base of stellar isochrones, combined with a spectral 
library. SSPs are the ``building blocks'' of spectro--photometric models 
of more complex systems, like galaxies:
by convolving SSPs with the SFH $\Psi(t)$ and the metal enrichment history 
$Z(t)$ provided by chemical models, the corresponding global 
luminosity in any given pass-band can be calculated as:
\begin{equation}
\label{eq:Ltot}
L = \int_0^T \Psi(t) \, L_{SSP}[T-t,Z(T-t)] \, dt
\end{equation}
where T is the present age of the galaxy.

For the purpose of this paper, we computed 
SSPs for the IMFs discussed in \S\ref{sect:IMFlist}, and
for metallicities  $Z$=10$^{-4}$, $4 \times 10^{-4}$, 0.001, 0.004, 0.008, 
0.019(=$Z_{\odot}$), 0.04, 0.07, 0.1.
Our SSPs are based on the latest Padua isochrones 
(Girardi \etal 2000, 2002; {\mbox{Salasnich}} \etal 2000; 
{\sf http://pleiadi.pd.astro.it}), recently extended to 
$Z$=0.1 by L.~Girardi (private communication). These isochrones are constructed
from stellar tracks with updated opacities, equation of state and neutrino 
cooling, and with a moderate amount of convective overshoot; see 
the above mentioned papers for all details. In this work we consider only 
isochrones with solar--scaled metallicity; $\alpha$--enhanced compositions 
may be found for stars at low metallicities and/or in the inner disc regions, 
where the SFH is more rapid (see \S\ref{sect:chemo-photo}). Recently, stellar
tracks
and isochrones with $\alpha$--enhanced composition have become available
for $Z \geq$0.008 (Salasnich \etal 2000); for lower metallicities,
suitable conversion formul\ae\ from the $\alpha$--enhanced to the 
solar--scaled case can be applied (Salaris, Chieffi \& Straniero 1993).
We will consider the role of $\alpha$--enhanced compositions in future models.

For the sake of the present study, most notable in these isochrones is that 
the AGB phase includes
up-to-date prescriptions for the Thermally Pulsing (TP) regime 
(Girardi \& Bertelli 1998); the contribution of AGB stars is very important
for the evolution of the NIR flux and of visual--NIR colours 
(Girardi \& Bertelli 1998; Schulz \etal 2002; Mouhcine \& Lan\c{c}on 2002),
and is potentially interesting also for the red bands considered in  
the present paper. The effect of including full semi--analytical models
of the TP-AGB phase in the isochrones and SSPs 
will be discussed in \S\ref{sect:AGB}, by considering the isochrones 
by Marigo \& Girardi (2001) as an alternative set.
We stress that a consistent inclusion of all post--main 
sequence phases of stellar evolution in the computation of isochrones 
and SSPs is crucial for a correct estimate of the integrated luminosity 
and of M/L ratios (Buzzoni 1999, 2002).

The transformation
from theoretical quantities ($L_{bol}$, $T_{eff}$, $g$) into 
magnitudes and colours is performed through the spectral library
compiled by Girardi \etal (2002), a suitable extension of
Kurucz' ATLAS9 library (Castelli, Gratton \& Kurucz 1997). From the same 
library 
we derive for the solar model of the Padua tracks (log~$T_{eff}$=3.762, 
log~$g$=4.432, L.~Girardi, priv.\ comm.) the solar absolute magnitudes listed 
in Table~\ref{tab:magsol}, first column. To express integrated luminosities 
of photometric models in solar units, we consider in fact more consistent
to refer to the theoretical solar magnitudes derived from the same spectral 
library, rather than to the observed ones or to the magnitudes given by more 
detailed solar spectrum models; differences are anyways well below 0.1~mag
in the bands of interest. 
Uncertainties related to
the use of different libraries are briefly discussed in \S\ref{sect:Lejeune},
by means of SSPs calculated with the Lejeune library.

For further details on the calculation of SSPs, the reader is referred to 
Bressan, Chiosi \& Fagotto (1994) and Tantalo \etal (1996).
For consistency
with previous work, in the SSPs the adopted solar bolometric magnitude
and the definition of the bolometric corrections 
are slightly different than those in Girardi \etal (2002). However,
when allowance is made for the different definition of the zero points,
the integrated luminosities are identical in the two cases
and the predicted M/L ratios are unaffected, as expected.

\begin{table}
\begin{center}
\begin{tabular}{ | c c c | }
\hline
 & Girardi/Castelli & Lejeune \\
\hline
bol & 4.720 & 4.720 \\
U   & 5.666 & 5.576 \\
B   & 5.448 & 5.446 \\
V   & 4.788 & 4.807 \\
R   & 4.413 & 4.433 \\
I   & 4.065 & 4.080 \\
J   & 3.640 & 3.653 \\
H   & 3.295 & 3.307 \\
K   & 3.263 & 3.252 \\
\hline
\end{tabular}
\end{center}
\caption{Solar magnitudes in the main Johnson-Cousins-Glass bands as predicted
by the stellar libraries in use}
\label{tab:magsol}
\end{table}

\begin{figure*}
\centerline{\psfig{file=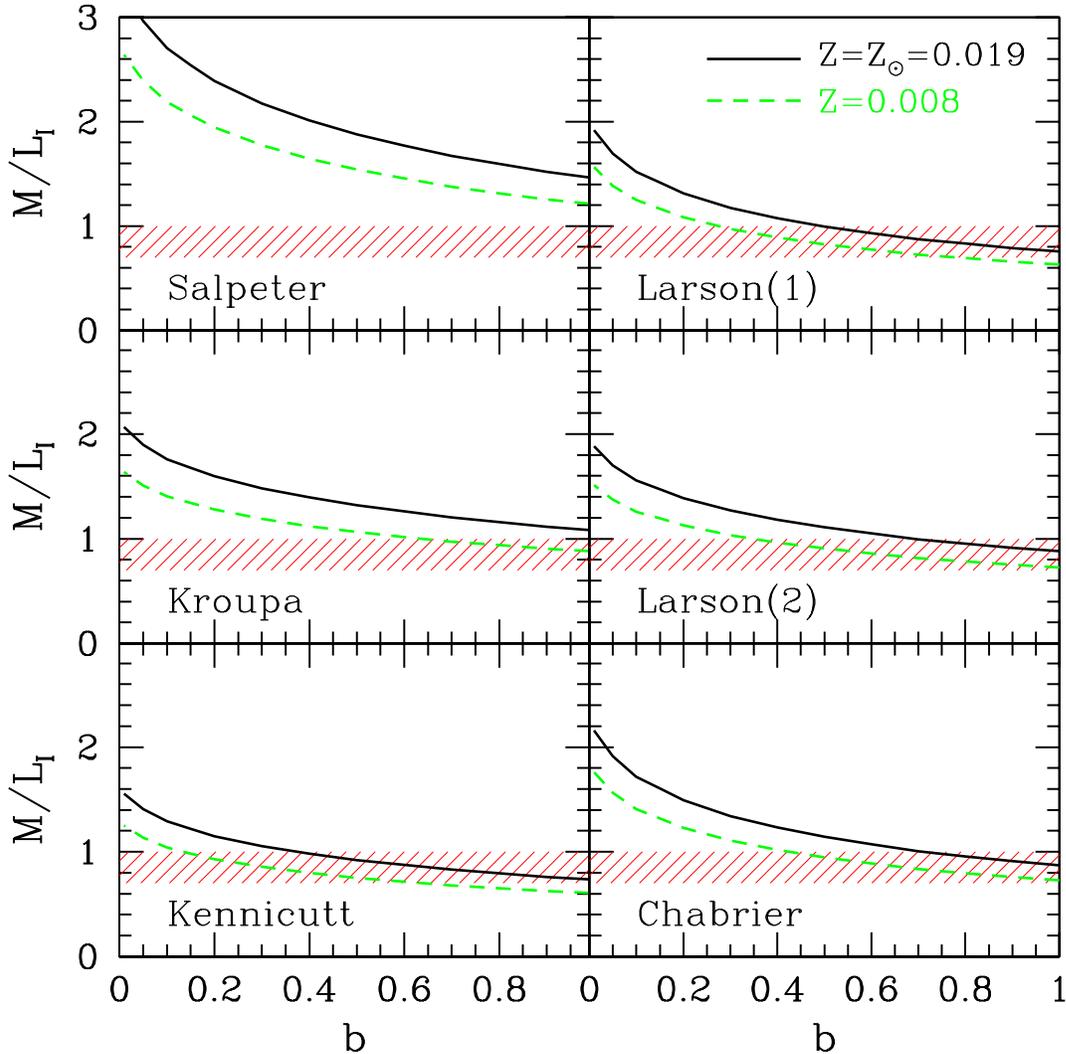,width=14.9truecm}}
\caption{M/L ratio in the I--band for simple exponential models with varying
$b$--parameter; different panels correspond to models with different IMFs,
as indicated.
The shaded area marks the range M/L$_I$=0.7--1 favoured by observations
of Sbc--Sc discs ($b = 0.8-1$).}
\label{fig:b-models}
\end{figure*}

\section{A discussion of simple models}
\label{sect:b-models}
For a preliminary study of the effect of different IMFs
on the M/L ratio, we first resort to very simple 
models, adopting 
an exponentially decaying star formation rate
\[ \Psi(t) \propto e^{-\frac{t}{\tau_{SF}}}\]
where different timescales $\tau_{SF}$ correspond to different
values of the $b$--parameter (Eq.~\ref{eq:b-SFR}). The age of the model
discs is assumed to be $T$=10~Gyr, as estimated for the bulk of the stellar
population in the disc of the Milky Way (Carraro 2000).
Exponentially decaying SFRs
with varying timescales have often been used as effective prescriptions
to reproduce the Hubble sequence in spectro--photometric models 
(e.g.\ Larson \& Tinsley 1978; Bruzual 1983; 
Guiderdoni \& Rocca--Volmerange 1987; Bruzual \& Charlot 1993; KTC94;
M\"oller, Fritze--v.Alvensleben \& Fricke 1997; Bell \& de Jong 2000). 
Our preliminary ``exponential'' models 
do not include chemical evolution, but we can learn about the role 
of metallicity by adopting SSPs of different metallicities in 
Eq.~\ref{eq:Ltot}.
These simple models are also useful to assess uncertainties related
to the detailed prescriptions for the TP-AGB phase 
and to the adopted stellar library (\S\ref{sect:AGB} and \S\ref{sect:Lejeune}).

Fig.~\ref{fig:b-models} shows the stellar (i.e. stars+remnants) M/L ratio 
in the I--band for our exponential models as a function of the corresponding 
$b$--parameter; different panels correspond to models with different IMFs.
Metallicities between $\sim$half solar and solar are considered representative
of the bulk of stars in galactic discs. The shaded area indicates the range
of observed M/L$_I$=0.7--1 we want to reach with our
Sbc--Sc disc models {\mbox{($b = 0.8-1$,}} see \S\ref{sect:Hubble}).

It is immediately apparent how a Salpeter IMF yields a M/L ratio much higher 
than desired. The local Kroupa IMF has still a slightly high 
{\mbox{$M/L_I \gsim 1$}} while all the other ``bottom--light'' IMFs 
generally reach
M/L$_I \leq$1 for the models corresponding to Sbc/Sc spirals.
This overview promotes the development of more refined models 
for a detailed prediction of M/L ratios with bottom--light IMFs
(\S\ref{sect:chemo-photo}).

\begin{figure}
\centerline{\psfig{file=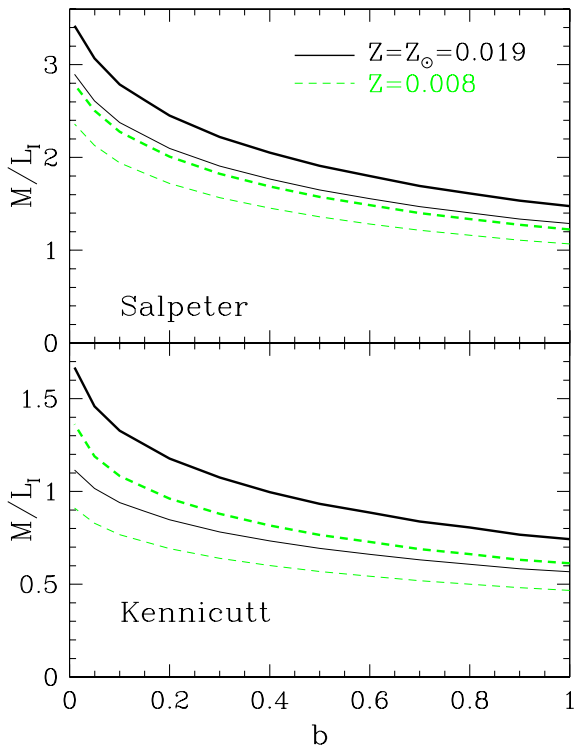,width=7.9truecm}}
\caption{I--band M/L ratio for the global stellar population (stars+remnants,
thick lines) compared to that of the sole living stars (thin lines). The mass
of remnants typically contributes for a 10\%, but it can reach a 30\% for
bottom--light IMFs, such as the Kennicutt IMF, and SFHs skewed to old
populations (low $b$--parameters).}
\label{fig:MLstars}
\end{figure}

Fig.~\ref{fig:MLstars} shows for two examples, the Salpeter 
and the Kennicutt
IMFs, the predicted M/L ratios of the global stellar population 
(i.e.\ stars and remnants, thick lines) compared to the M/L ratio of the sole 
population of living
stars (thin lines). Neglecting the mass of stellar remnants
(white dwarfs, neutron stars and black holes) induces an underestimate of the
M/L ratios typically of 10\%, which can increase up to 30\%
depending on the IMF and SFH (cf.\ models with low $b$--parameter and Kennicutt
IMF). This illustrate the importance of calculating self--consistently the 
contribution of stellar remnants when discussing the M/L ratio of the 
stellar/baryonic component of galaxies.

\subsection{Dependence of the M/L ratio on Hubble type}
\label{sect:Hubbledependence}
\begin{figure}
\centerline{\psfig{file=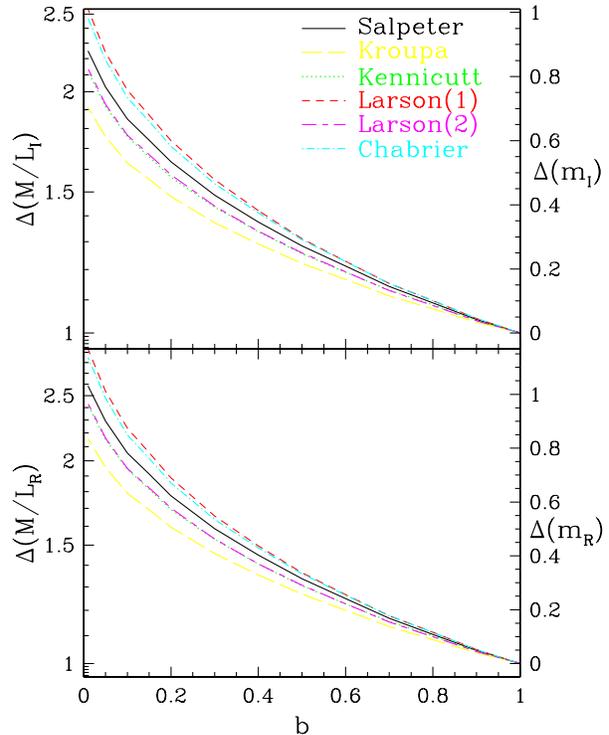,width=7.9truecm}}
\caption{Relative M/L ratio normalized to the value corresponding to
$b=1$ models. {\it Top panel:} I--band; {\it bottom panel:} R--band.
Results are plotted for different IMFs as indicated, for solar metallicity.}
\label{fig:relativeML}
\end{figure}

It is interesting to consider the variation of stellar M/L ratios
with Hubble type, here parameterized by the $b$--parameter indicative
of the different star formation 
histories. Fig.~\ref{fig:relativeML} shows the M/L ratio as a function of $b$,
{\it normalized} to the value corresponding
to $b$=1. The scale on the right axis indicates the corresponding shift
in magnitude. The plot is for different IMFs and for solar metallicity;
results for {\it relative} M/L ratios do not change considering the case 
$Z=0.008$ instead. In the I--band 
(top panel in Fig.~\ref{fig:relativeML}) with respect to Sbc--Sc spirals
we predict the M/L ratio to increase by a factor of 1.3--1.5 for Sb spirals 
($b \sim$0.35) and of 1.7--2.2 for Sa-Sab spirals ($b \sim$0.1).
This corresponds to a systematic magnitude offset of 0.3--0.4 mag and 
of 0.6--0.8 mag for Sb and for Sa-Sab spirals, respectively, 
as a result of the difference in characteristic SFH.
These offsets are somewhat reduced when bulges are included in the computation
of the total M/L ratios of galaxies (\S\ref{sect:bulge-Hubble}), but they
remain significant.
The offsets we predict are larger than the empirical ones found by Giovanelli 
\etal (1997a): 0.1~mag for Sb spirals and 0.32~mag for earlier types.
However, the extent of systematic offsets in the TF relation is still a matter 
of debate: for instance, larger offsets in the nearby R--band are found by
Kannappan \etal (2002). Their offset of 0.76 mag for Sa galaxies 
is in good agreement with our predictions in the R--band at 
$b \sim 0.1$ (Fig.~\ref{fig:relativeML}, lower panel).

\subsection{The treatment of the AGB phase}
\label{sect:AGB}
For SSPs of intermediate age, the luminosity (in particular the bolometric 
and NIR)
is dominated by the contribution of stars in the AGB phase. Hence careful
modelling of these phases of stellar evolution is very important when 
determining M/L ratios, especially in the NIR.

As underlined in \S\ref{sect:SSP}, our SSPs are based on isochrones
(Girardi \etal 2002)
which explicitly include the TP-AGB phase with a synthetic algorithm
based on the core growth, the core mass--luminosity relation and the
initial--final core mass relation (Girardi \& Bertelli 1998).
Recently, Marigo \& Girardi (2001) discussed self-consistency between 
chemical and photometric evolution in galactic models.
They refined the calculation of the ``stellar fuel'' (amount of hydrogen 
or helium consumed by nuclear burning) adding the 
contribution of the chemical yields dredged up and lost by the stars,
to the classic fuel estimate based only on the
initial--final core mass relation.
According to the Fuel Consumption Theorem (Renzini \& Buzzoni 1986),
this additional fuel corresponds to additional emitted luminosity (mostly
in the NIR for AGB stars).
Marigo \& Girardi also provide isochrones including detailed 
semi--analytical models for the TP-AGB phase,
improving upon the simple synthetic algorithm adopted by Girardi \etal

\begin{figure}
\psfig{file=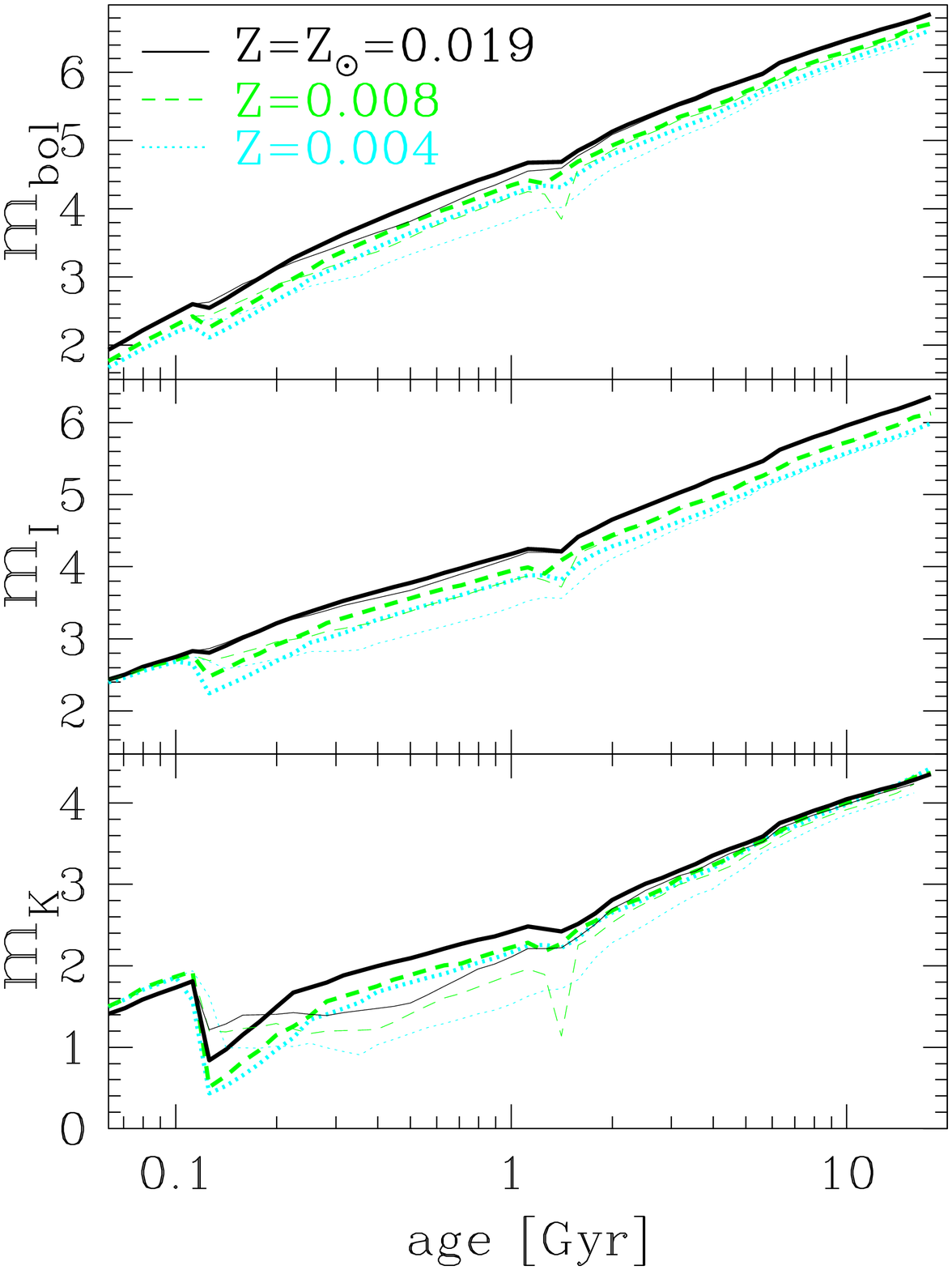,width=8.5truecm}
\caption{Comparison between the G-SSPs and the M-SSPs (thick and thin lines,
respectively) in bolometric, I and K
magnitude. Detailed modelling of the TP-AGB phase (M-SSPs) leads to a larger
overall luminosity contributed by AGB stars, mostly 
{\mbox{in the K and NIR bands.~~~~~~~~~~~~~~~~~~~~~~~~~~~~~~~~~~~~~~~~~~~~}}
{\mbox{For the M-SSPs with $Z=0.008$,}} notice the occurrence of a ``red 
spike'' around 1.5~Gyr, a feature discussed by Girardi \& Bertelli (1998).
This feature is very short lived and does not influence the bulk of emitted
light integrated over time.}
\label{fig:marigoSSP}
\end{figure}

In this Section, we check the effect of a different modelling of the
TP-AGB phase on the predicted M/L ratios in galactic discs. We calculated
a set of SSPs based on the isochrones by Marigo \& Girardi (M-SSPs) to be 
compared to our reference SSPs based on the isochrones by Girardi \etal
(G-SSPs).
Fig.~\ref{fig:marigoSSP} compares the luminosity evolution of the G-SSPs
(thick lines) to that of the M-SSPs (thin lines) for the three metallicities
available for the M-SSPs; we adopt a total SSP mass of 1~\Msol\ and the 
Salpeter IMF for the sake of 
illustration. With the M-SSPs, although the onset of the AGB contribution 
is more gradual (with a smoother phase transition due to the overluminosity
effect, see Girardi \& Bertelli 1998), 
the overall luminosity produced in the course of evolution is larger,
as expected from the larger ``fuel'' actually consumed.
The effect is generally negligible in the I--band (and in all other optical 
bands) with the exception of very low metallicities, 
but it is highly significant in the NIR; in the K--band, the M-SSPs can be 
up to 0.6~mag brighter.

\begin{figure}
\psfig{file=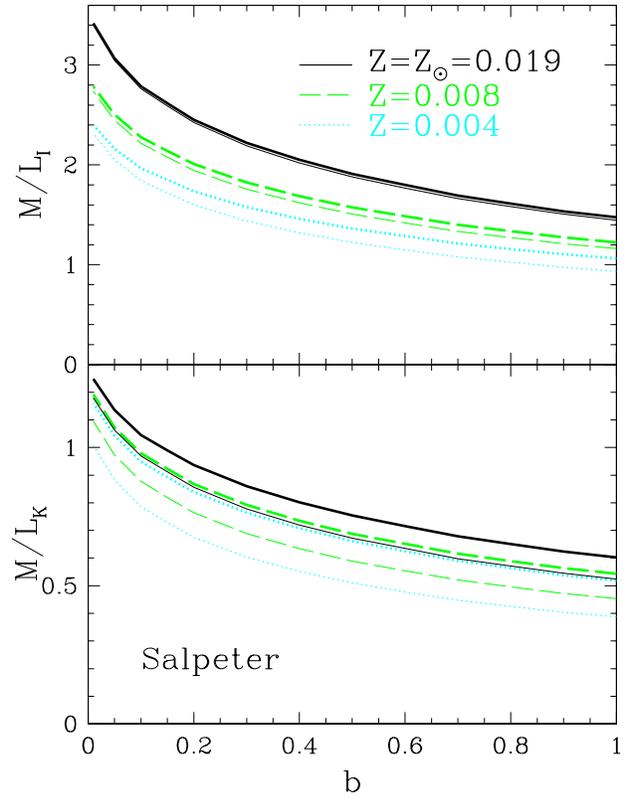,width=8.5truecm}
\caption{M/L ratios for our exponential disc models, as obtained with the
G-SSPs and the M-SSPs (thick and thin lines, respectively).}
\label{fig:marigoML}
\end{figure}

Fig.~\ref{fig:marigoML} compares the M/L ratios of our simple exponential
models for discs, as predicted with the G-SSPs and with the M-SSPs. 
In the I--band the difference is minor, and negligible for the relevant 
metallicities $Z \geq 0.008$, while in the K--band it becomes significant
(up to 20--30\%). The Salpeter IMF is adopted for the sake of 
illustration, but the situation is similar for all the other IMFs.

In summary, the effect of detailed TP-AGB modelling is negligible 
in the I--band (and in optical bands in general), making from this point of
view the I--band luminosity superior to NIR luminosities for mass estimates.

Stellar and photometric evolution in the TP--AGB phase can be further altered
by dusty circumstellar envelopes (Bressan, Granato \& Silva 1998;
Mouhcine 2002; Piovan, Tantalo \& Chiosi 2003) or different opacities and mass
loss rates in the carbon star phase (Marigo 2002; Marigo, Girardi \& Chiosi 
2003). 
Again, the effects are seen mostly in the NIR colours
but they should be negligible for the integrated I--band luminosity, 
where the contribution of AGB stars is not so dominant.
We conclude that estimates of 
M/L ratios in the I--band are robust with respect to uncertainties in the
modelling of the TP-AGB phase.

\subsection{An alternative spectral library}
\label{sect:Lejeune}
An alternative spectral library has been recently provided by Lejeune,
Cuisinier \& Buser (1997, 1998) and has been widely used in population 
synthesis studies 
(Boissier \& Prantzos 1999; Liu, Charlot \& Graham 2000; Schulz \etal 2002).
This is also an extension of the synthetic Kurucz' ATLAS9 library, but 
with additional empirical corrections to reproduce the observed 
color--temperature relations of individual stars. However, 
Girardi \etal (2002) argue against {\it a posteriori} calibrations
of theoretical libraries, while Westera et al.\ (2002) find that 
it is not possible to attain a semi-empirical library that could 
reproduce both empirical colour--temperature relations and, 
when combined with current isochrones, the observed colour--magnitude 
diagrams of star clusters.

\begin{figure}
\psfig{file=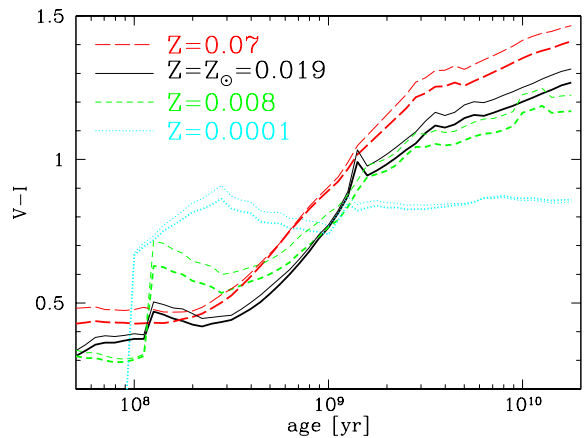,angle=270,width=8truecm}
\caption{Comparison between the G-SSPs and the L-SSPs (thick and thin lines,
respectively) in V--I. The Lejeune library predicts redder V--I colours, 
hence more light in the I--band. The Kroupa IMF is adopted here for the sake 
of example, however the age evolution of V--I is insensitive to the IMF.}
\label{fig:lejeuneV-I}
\end{figure}

\begin{figure}
\centerline{\psfig{file=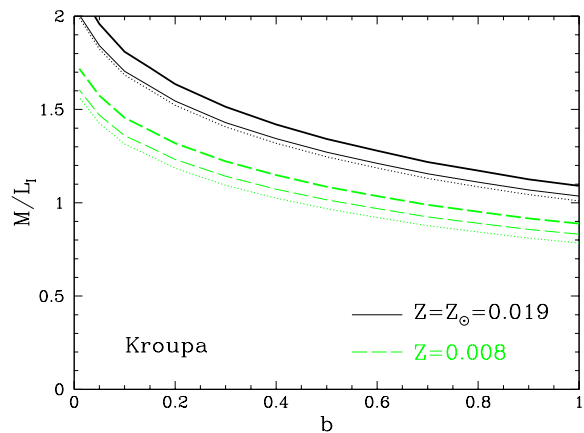,angle=270,width=8truecm}}
\caption{M/L ratios for our exponential disc models obtained with 
the L-SSPs (thin lines) as compared to the ``standard'' {\mbox{G-SSPs}} 
(thick lines). Dotted lines
for the LM-SSPs. The Kroupa IMF is adopted here for the sake of example.}
\label{fig:lejeuneML}
\end{figure}

Thus, the present status of semi--empirical libraries is not yet fully
satisfactory; it is still useful, however, to discuss these alternative 
libraries in connection with
uncertainties in integrated magnitudes due to the spectral input. 
To this purpose we calculated an alternative set of SSPs based on
the same isochrones as the G-SSPs (see \S\ref{sect:SSP}) but adopting the
semi--empirical library by Lejeune \etal (1997, 1998). We indicate this
alternative set as L-SSPs; the corresponding solar magnitudes, derived
for the solar model linked to the Lejeune library, are listed in 
Table~\ref{tab:magsol}. The difference between the G-SSPs and the L-SSPs
is prominent in the V--I colour (Fig.~\ref{fig:lejeuneV-I}) making the L-SSPs
brighter in the I--band by up to 0.1~mag at intermediate and old ages
(since the absolute V magnitude is no different between the two sets). This
is purely an effect of the spectral library, and makes our model M/L ratios
in the I band lower by a 5--10\%, when the L-SSPs are adopted 
(Fig.~\ref{fig:lejeuneML}).

Finally, we checked whether a more refined treatment of the AGB phase
would have a sizeable effect on the predicted M/L ratios with
the Lejeune library, where red giants are more luminous 
in the I--band. The effect of a detailed inclusion of the TP-AGB
phase, from the isochrones by Marigo \& Girardi (2001), was found to be small
in the I--band when the Girardi/Castelli library was adopted (\S\ref{sect:AGB}
and Fig.~\ref{fig:marigoML}). If the Marigo \& Girardi isochrones 
are combined with the Lejeune library (set LM-SSPs), again the effects on the
I--band M/L ratio is small with respect to the results from the L-SSPs
(Fig.~\ref{fig:lejeuneML}). In conclusion, with present models the I--band
luminosity is more affected by the adopted spectral library than by the
detailed modelling of the AGB phase.

Since the situation with semi--empirical libraries is still somewhat
dichotomic (Westera \etal 2002), in the remainder of the paper 
we will adopt as our standard set the G-SSPs based on the theoretical 
spectral library
by Girardi/Castelli. However, from the discussion above and from 
Figs.~\ref{fig:marigoML} and~\ref{fig:lejeuneML}, it is apparent that current
alternative spectral libraries (and, marginally, a more refined
modelling of the AGB phase) tend to increase the predicted I--luminosity. 
Therefore, our predictions for M/L ratios in the I band will be 
conservatively high, though not by large (at most 10\%).

\section{Chemo--photometric models}
\label{sect:chemo-photo}
\begin{figure}
\centerline{\psfig{file=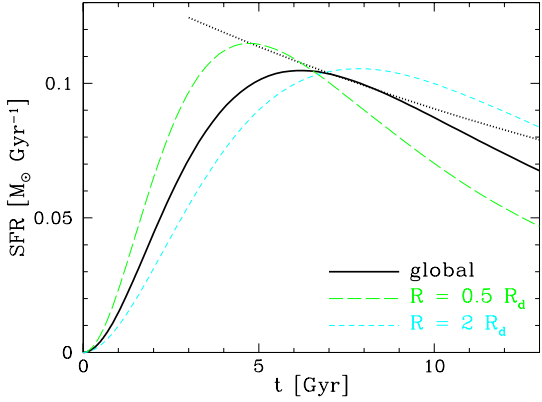,angle=270,width=7.5truecm}}
\caption{Star Formation History for the disc of model {\sf SalpA}
(between $R=0.5-3~H_B$, see \S\protect{\ref{sect:calibration}}). 
For comparison, the dotted line shows the SFH of a simple exponential model 
with the same $b$=0.79 and age 10~Gyr.
{\mbox{The long-dashed and short-dashed lines}} illustrate the different pace
of the SFH at different galactocentric distances, characteristic of inside--out
models. For the sake of comparing the respective shape, all the SFR curves 
are normalized to the same integrated SFH of 1~\Msol.}
\label{fig:sfhSalpA}
\end{figure}

More realistic photometric models for galactic discs should include star
formation histories and corresponding chemical evolution consistently. 
From chemical evolution studies of the Solar 
Neighbourhood, especially from the G--dwarf problem (Lynden--Bell 1975;
Tinsley 1980; Pagel 1997 and references therein), as well as from dynamical
models of the formation of galactic discs (Larson 1976; Sommer--Larsen 1991;
Burkert, Truran \& Hensler 1992;
Sommer--Larsen \etal 2003) it is well--known that galactic discs form
gradually by slow accretion (``infall'') of primordial or low--metallicity
gas. Correspondingly, the SFR is expected first to increase,
following the increase of available gas, reach a maximum and later 
decrease due to gas consumption (Fig.~\ref{fig:sfhSalpA}). 

Moreover, the metallicity and colour gradients observed in the Galactic disc
as well as external disc galaxies, indicate that
the star formation and chemical enrichment history typically proceeds 
at a different (slower) pace at increasing galactocentric radius
(inside--out scenario, Matteucci \& Fran\c{c}ois 1989; Chiappini,
Matteucci \& Gratton 1997; Fig.~\ref{fig:sfhSalpA}; but see also 
Sommer-Larsen \etal 2003).
A realistic SFH for a disc galaxy is therefore more complex 
than the monotonic, exponential form adopted in the simple models of 
\S\ref{sect:b-models}.

As a base for our photometric models, we develop a series of
chemical evolution models with infall, and with a multi--zone radial
structure so that gradients of metallicity and SFH can be reproduced.
The reader not keen on the details of model construction and calibration 
onto the observable constraints, nor on the detailed discussion of 
the sets of models with each specific IMF, can directly find an overview
of the results in \S\ref{sect:overview}.

We make use of the chemical evolution code by Portinari, Chiosi \& Bressan 
(1998); Portinari \& Chiosi (1999). 
The disc is assumed to have an exponential surface density profile:
\begin{equation}
\label{eq:discprofile}
\sigma(R) = \sigma_d \, e^{- \frac{R-R_d}{R_d}}
\end{equation}
where $\sigma(R)$ is the present--day total surface density distribution, 
$R_d$ is the scale-length of the mass distribution and $\sigma_d$ is the  
surface density at a distance $R=R_d$ from the center.
The disc is typically divided into 26 zones, $0.2 \, R_d$ wide, 
between $R=0$ and $R=5 \, R_d$.

In each zone infall is assumed to occur, as customary, at an exponentially 
declining rate:
\[ \frac{d \sigma}{dt} \propto exp \left( -\frac{t}{\tau_{inf}} \right) \]
where the infall timescale may vary with radius, 
{\mbox{$\tau_{inf}=\tau_{inf}(R)$}}, typically
increasing outward in the inside--out scenario. A similar behaviour of the
infall process, exponentially declining rates with timescales increasing
outward, is also found in some recent cosmological/hydro--dynamical 
simulations of the formation of galactic discs (Sommer--Larsen \etal 2003).

We assume $T=13$~Gyr for the age of the models, as suggested by the 
$\Lambda$CDM simulations
of Sommer--Larsen et~al., where the SF activity associated to the main
galaxy typically starts at a redshift of 4--5. This is larger than
the age of 10~Gyr assumed in our exponential models in \S\ref{sect:b-models},
but one should keep in mind that the latter represents the age of the 
{\it bulk} of local disc stars.
In an infall scheme, the formation of the disc is described since 
the very early stages of gas accumulation; SF develops gradually, 
reaches its maximum after a few Gyrs
and then declines, so that the the bulk of stars form in fact during the last
$\sim$10~Gyr (Fig.\ref{fig:sfhSalpA}).

As to the SF law, we adopt the form suggested by 
Dopita \& Ryder (1994):
\[ \frac{d \sigma_*}{dt} \propto \, \sigma^{1/3} \sigma_g^{5/3} \]
depending both on the gas surface density $\sigma_g$ 
and on the total surface density $\sigma$; $\sigma_*$ is the surface density
of stars. This is a Schmidt--like law (Schmidt 1959)
with an efficiency decreasing at increasing radius through the dependence on 
$\sigma(R)$, as
supported by both theoretical and empirical arguments (Dopita \& Ryder 1994).
The resulting SFHs are necessarily faster in the inner regions and slower 
in the outer ones where the surface density is lower; this SF law results in
metallicity gradients comparable to the observed ones 
(Portinari \& Chiosi 1999). With a convenient normalization 
(see Portinari \& Chiosi 1999 for details) we express this law:
\[ \frac{d}{dt} \sigma_*(R,t) = \nu \, 
\frac{\sigma^{1/3}(R,t) \, \sigma_g^{5/3}(R,t)}{\sigma(R_d,T)} \]
where $\nu$ is the star formation efficiency in units of $[t^{-1}]$.
The code performs detailed calculations of the delayed gas and metal 
restitution from stars of various lifetimes, including supernov\ae\ of type Ia,
and of the evolution of the abundances of the main chemical elements; 
we refer to Portinari \etal (1998) for all details on the chemical model 
and to Portinari \& Chiosi (1999) for the implementation of the 
SF law and of the multi--zone scheme. With respect to the above mentioned 
papers, 
the present models are updated in the input yields of low and 
intermediate mass stars (Marigo 2001), which are consistent with the
TP-AGB calculations included in the M-SSPs (\S\ref{sect:AGB}).

Our models are a two--parameter family, with parameters:
\begin{center}
\begin{tabular}{l p{7truecm}}
$\tau_{inf}$ & infall timescale (in Gyr; not to be confused with the SF 
               timescale
	       $\tau_{SF}$ of the simple models in \S\ref{sect:b-models}) \\
$\nu$	     & SF efficiency (in Gyr$^{-1}$) \\
\end{tabular}
\end{center}
This pair of parameters will be tuned, for each selected IMF, so as to
reproduce two main constraints discussed below, the first selecting 
the relevant SFHs, the second setting the level of metal enrichment
and the gas fraction.

\subsection{Constraints on the SFH}
\label{sect:SFH-constraint}
From the SFH of each model we can define its {\mbox{$b$--parameter}}: 
\begin{equation}
\label{eq:b-psi}
b = \frac{\psi(T)}{<\psi>}~~~~~~~~~~~~~~~~~~~~~~<\psi> \,=\,
\frac{1}{T} \int_0^T \psi(t) \,dt
\end{equation}
where $\Psi(t)$ is the instantaneous global SFR. 
We select the models with $b$=0.8--1 as representative of Sbc--Sc discs,
as discussed in \S\ref{sect:Hubble}; the corresponding typical intrinsic
B--V colour is (B--V)$_0 \sim$0.55 (Roberts \& Haynes 1994).

\subsection{Constraints on chemical evolution}
\label{sect:chem-constraint}
A second constraint is the ``typical chemical properties'' of an Sbc--Sc disc. 
Non--barred spirals exhibit an oxygen gradient in the gas component
that seems to be
roughly independent of Hubble type, once radii are expressed in units of 
scalelength (in the B--band) rather than in physical distance units like 
kiloparsecs. The typical average value is:
\[ \frac{d}{dR} \left[ \frac{O}{H} \right] \,=\, 
- 0.2~{\rm dex} \, h_B^{-1}\]
with values ranging between --0.1 and --0.3~dex~$h_B^{-1}$
(Garnett \etal 1997; van Zee \etal 1998; Prantzos \& Boissier 2000).
Oxygen can be taken as tracer of the global metal content, as it is the
most abundant metal and it generally  contributes roughly half of the overall 
metallicity. In the following the term ``metallicity'' will be often used
in place of ``oxygen abundance'', for the sake of simplicity.

The typical disc gas metallicity, usually defined as the central value
extrapolated
from the observed gradient or, more directly, the metallicity at a 
galactocentric distance $R=h_B$, shows a marked 
dependence on mass. The samples by Garnett \etal and van Zee et~al., 
as well as those by KTC94 and Roberts \& Haynes (1994) considered above
in the paper,
mostly consist of galaxies with magnitude around $M_B = -21$ 
(when all rescaled to the same $H_0$=50). Therefore we will discuss
mainly bright galaxies with sizes comparable to the Milky Way, while we do 
not investigate in this paper
mass--dependent behaviours. The dependence of the typical metallicity on Hubble
type seems to be a secondary relation, mostly a consequence of 
the mass--metallicity relation through the Hubble type--mass relation; 
there is in fact a tendency
for earlier Hubble types to be more massive, though with a large overlap 
in mass between types.

For Milky--Way sized galaxies, from the samples of Garnett \etal and 
van Zee \etal we derive a typical metallicity of Sbc--Sc discs of:
\[ 12 + log \left( \frac{O}{H} \right) = 9.1~{\rm dex~~~~~~~at}~R=h_B \]
We will use this typical metallicity and metallicity gradient to constrain
our chemical evolution models. 

Another significant constraint for chemical models is the gas fraction,
that is the final ratio between the gas mass and the total baryonic mass
(Tinsley 1980; Pagel 1997)
\[ f_{gas} = \frac{M_{gas}}{M_{gas}+M_*} \]
in obvious notation.
Hence we need an estimate of the typical gas fraction in Sbc--Sc 
discs. This in turn requires an estimate of the stellar mass $M_*$,
which can only be obtained from the observed luminosity
and an assumed stellar M/L ratio (one cannot use dynamical mass estimates
because of the unknown contribution of dark matter). 
However, since the M/L ratio is supposed to be
a prediction of our models we should make no {\it a priori} assumptions
about it, and 
in place of the usual gas mass fraction we consider as a constraint
the gas--to--luminosity fraction
\[ f_{L, gas} = \frac{M_{gas}}{L_B} \]
This quantity, besides being independent 
of any assumed M/L ratio, has the additional advantage of being independent 
of distance and hence of $H_0$.
As the mass of gas (HI+H$_2$) is derived
from the observed flux in the 21 cm line and in the CO lines, which have the
same dependence on the assumed distance as $L_B$, their ratio is distance 
independent.

Roberts \& Haynes (1994) give for Sbc--Sc galaxies a typical value of 
\[ \frac{M(HI)}{L_B} = 0.30^{+0.18}_{-0.12} \, \frac{M_{\odot}}{L_{\odot}} \]
Comparable values can be derived from the samples of de Mello \etal (2002a,b)
and just slightly lower ones from that of KTC94. 
The masses of neutral hydrogen quoted here are the global masses 
for the whole HI disc. For the sake of chemical evolution, however, 
the significant constraint is the gas fraction within the area 
of active star formation, i.e.\ limited to within the optical radius.
Some corrections to the above quantities should be applied to account for the
fact that the HI disc generally extends farther than the optical disc.
Considering the sample of spirals of KTC94, 
it is possible to correct from the total gas mass to the 
gas mass within the optical radius, using the ``Roberts times''
listed in their Table~5.
We obtain typical corrections of 0.6 for field spirals and 0.8 for cluster
spirals, whose outer gaseous discs are presumably stripped because of the
dense environment, and hence less extended beyond 
the optical disc. Since the large sample of Roberts \& Haynes, from the
RC3 catalogue, contains both field and cluster spirals, we apply an average
correction of 0.7 to derive the gas masses within $R_{25}$. This agrees
also with the corrections of 0.66--0.72 derived by Sommer--Larsen (1996).

As to the molecular component, Young \& Knezek (1989) give for Sbc--Sc galaxies
\[ \frac{M(H_2)}{L_B} = 0.16^{+0.02}_{-0.02} \, \frac{M_{\odot}}{L_{\odot}}\]
We find similar values in the sample of KTC94 and somewhat lower values 
($\sim$0.1) in the sample by of de Mello et~al.\footnote{Recently Boselli, 
Lequeux \& Gavazzi (2002), applying a calibrated conversion factor $\chi$ 
between CO emission and $H_2$ mass depending on metallicity, UV field and 
galaxy luminosity, derived a much lower typical $M(H_2)/M(HI) \sim 0.15$.
They suggest that previous studies, adopting a constant conversion factor,
overestimated the molecular mass in bright galaxies, while underestimating it
in faint galaxies, by a factor of 2--3. If so,
our estimates of the gas--to--luminosity fractions above are upper limits, 
and the bulk of the gas would lie in the HI component.}
Adding the two components, we derive:
\[ \frac{M(HI+H2)}{L_B} = 0.3-0.37 \]
and including the standard factor of 1.4 for the contribution of other
elements (mostly helium) to the total gas mass, we finally derive
\[ f_{L,gas} \sim 0.42-0.52 \]
as a typical gas--to--luminosity fraction for Sbc--Sc galaxies.

This value is broadly compatible with the analogous quantity in our Milky Way
Disc. The masses of neutral and molecular hydrogen in our galaxy, within the
optical radius of the disc, are {\mbox{$M(HI)=4.3 \times 10^9$ \Msol}} and
{\mbox{$M(H_2)=1.3-1.7 \times 10^9$ \Msol}} (Binney \& Merrifield 1998; 
Dame 1993).
We can estimate the total B--band luminosity of the Milky Way as follows:
the K--band luminosities of the Disc and Bulge are, 
respectively, {\mbox{$L_{K,{\cal D}}=4.9 \times 10^{10}$ \Lsol}} and 
{\mbox{$L_{K,{\cal B}}=1.1 \times 10^{10}$ \Lsol}} (Binney \& Merrifield 1998; 
Kent, Dame \& Fazio 1991); considering (B--K)=3.51 as typical for a Sbc--Sc 
spiral (de Jong 1996b), we get for our Milky Way a global 
{\mbox{$L_B \sim 1.8 \times 10^{10}$ \Lsol}. Including the standard
1.4 factor for the helium correction to the gas mass, we derive 
for the Milky Way $f_{L,gas} \sim 0.45$.

Finally, as we model the SFH and the chemical evolution of galactic discs, 
we should adopt for the gas--to--luminosity fraction the luminosity of the
disc only. So,
we must correct for the contribution of bulges to the
total B--luminosity. For Sbc--Sc galaxies, the average bulge contribution
is estimated to be {\mbox{$L_{B,{\cal B}}/L_B \sim$10--20\% }}(Simien \&
de Vaucouleurs 1986; de Jong 1996a), which implies 
\[ f_{L,gas,{\cal D}} \sim 0.45-0.55 \]
We therefore adopt $0.5 \pm 0.05$ as our constraint for the chemical
enrichment level. However, this ``gas fraction'' constraint is much less firm
than the observed metallicities, for it has been derived by assembling a number
of independent observational studies and applying various corrections.

\subsection{Calibration of the models}
\label{sect:calibration}
For each of the six considered IMFs, we develop a grid of models with
different timescales $\tau_{inf}$. For each model, the SF efficiency
$\nu$ is fixed so that the final gas metallicity is $12+log(O/H) \sim$9.1~dex
at $T$=13~Gyr and at a radius $R=h_B$ (see the discussion in
\S\ref{sect:chem-constraint} and Fig.~\ref{fig:gradOHSalpA}). 
Among these models with specific ($\nu, \tau_{inf}$) combinations, 
we select those whose global SFHs correspond
to $b$=0.8--1, that is to Sbc--Sc discs. This corresponds to selecting
tendentially models with long infall timescales, as larger $\tau_{inf}$'s 
induce longer SFHs and hence larger $b$ values. 
The parameters and characteristics 
of the selected models are listed in Table~\ref{tab:models}.

The adopted SF law often generates metallicity gradients in broad agreement 
with the observed ones, for it results in an inside--out SFH even when
$\tau_{inf}$ is constant with radius.
Only when necessary, we include a radial increase 
of the infall timescale $\tau_{inf}(R)$ reinforcing the inside--out
behaviour (see model {\sf SalpC} in Fig.~\ref{fig:gradOHSalpC}).

To illustrate our procedure, here we describe in detail the calibration of
the models with the Salpeter IMF.
The comparison to the results for the other
``bottom--light'' IMFs will be discussed in the next Section.

Model {\sf SalpA} adopts $\tau_{inf}=5$~Gyr, and the SF efficiency 
is fixed to be $\nu=0.47$~Gyr$^{-1}$ to get an oxygen abundance of 9.1~dex at 
the reference radius $R=h_B$
(Fig.~\ref{fig:gradOHSalpA}). 
We calculate the disc photometry by applying Eq.~\ref{eq:Ltot} to each 
individual annulus with its specific SFH $\Psi(R,t)$ and metal enrichment 
history $Z(R,t)$, to obtain the surface brightness profiles in 
Fig.~\ref{fig:profilSalpA}. With an underlying exponential mass distribution, 
the light profile resulting from chemo--photometric models is not perfectly 
exponential, but close enough that a luminosity scalelength can be defined 
at least within $\sim 2-3~R_d$ (see also Boissier \& Prantzos 1999 for similar 
results). In model {\sf SalpA}, we derive a B--band scalelength 
{\mbox{$h_B \sim 0.95 R_d$}} within the inner $\sim 3 \, R_d$.
Fig.~\ref{fig:profilSalpA} also shows that the redder the band, 
the shorter the corresponding scalelength,
as expected and as observed also in the Milky Way (Boissier \& Prantzos 
1999 and references therein). 

\begin{figure}
\psfig{file=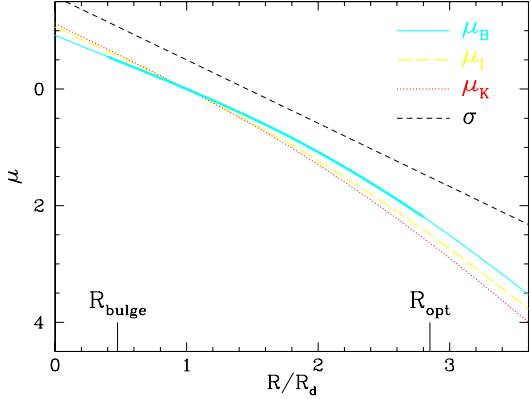,angle=270,width=7.5truecm}
\caption{Surface brightness in B, I and K--band for model {\sf SalpA}, all 
normalized to $\mu=0$ at $R=R_d$. The thick line marks the B--band profile
in the significant ``disc region'' between the bulge and the optical radius
$R_{opt}$ (see text). Also shown is the underlying exponential mass profile
$\sigma(R)$, vertically displaced by --0.5 for clarity.}
\label{fig:profilSalpA}
\end{figure}

\begin{figure}
\psfig{file=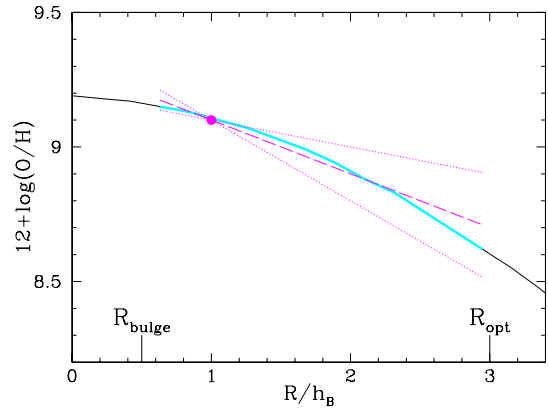,angle=270,width=7.5truecm}
\caption{Metallicity gradient in the present--day gas phase for model 
{\sf SalpA}, with radius in units
of B--band scalelength $h_B$. The significant ``disc region'' is marked 
as a thick line. The solid dot marks the observational calibration point 9.1 
for metallicity at $R=h_B$. The dashed line is the observed average gradient
{\mbox{--0.2~dex/$h_B$}}, the dotted lines mark the limits of the broad range 
of observed gradients, between --0.1 and --0.3~dex/$h_B$.}
\label{fig:gradOHSalpA}
\end{figure}

\begin{figure}
\psfig{file=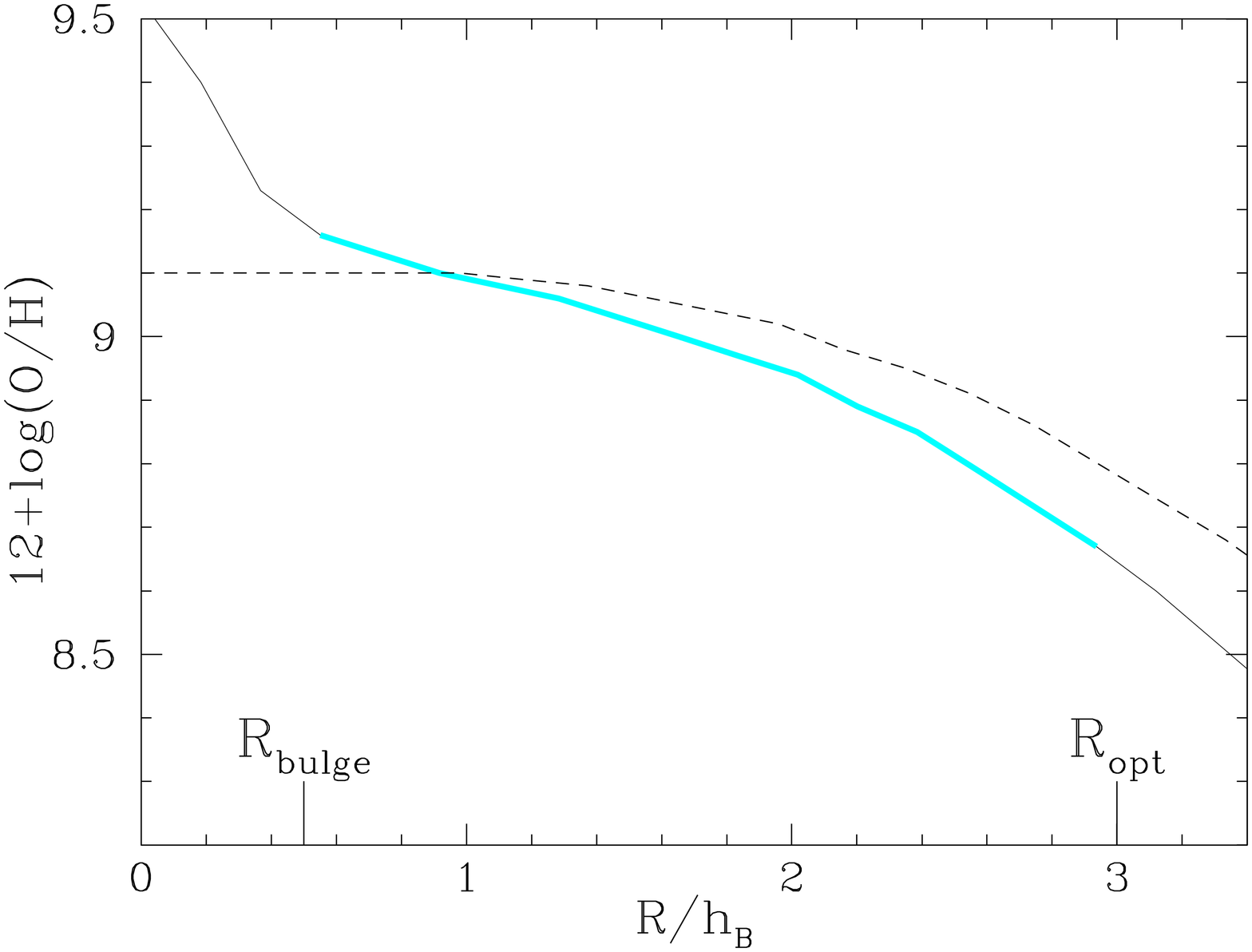,angle=270,width=7.5truecm}
\caption{Metallicity gradient for model {\sf SalpC}, with radially increasing
infall timescale (Table~\protect{\ref{tab:models}}), compared to a model with 
a constant infall timescale $\tau_{inf}=10$~Gyr (dashed line).}
\label{fig:gradOHSalpC}
\end{figure}

Computing $h_B$ is necessary to define the limit of the optical disc, 
which is the region to be compared to actual observations. A typical central 
(extrapolated) value of the surface brightness of discs is 
$\mu_{B0}$=21.7~mag~arcsec$^{-2}$ (Freeman 1970), although with a large 
scatter between 20 and 22~mag~arcsec$^{-2}$ (for Hubble types up to Sc,
de Jong 1996a); the optical radius $R_{25}$ corresponds to 
$\mu_B$=25, or $R_{25} \sim 3-4 h_B$ from 
{\mbox{$\mu_B(R)=\mu_{B0}+1.086~R/h_B$}}. For the following,
we define operationally the optical disc as the region within $R_{opt}=3~h_B$;
we verified that taking 3.5 or 4~$h_B$ as the limit would not influence 
the global M/L ratios or gas fractions, since the outskirts of the disc
are characterized by a low surface density and brightness, and do not
contribute significantly to the global values.

We also exclude from our analysis the innermost 
$0.5 \, h_B$, where the bulge typically lies in real disc galaxies.

Also the metallicity constraints derived from observations are expressed
in terms of the {\mbox{B--band}} scalelength $h_B$, rather than the 
scalelength 
$R_d$ of the underlying mass distribution (Eq.~\ref{eq:discprofile}). 
In Fig.~\ref{fig:gradOHSalpA} the 
oxygen abundance profile is plot as a function of $R/h_B$, like
actual observations; the model gradient is well within the typical
observed range.

Longer infall timescales tend to produce shallower metallicity gradients
at the present time. With {\mbox{$\tau_{inf}=10$~Gyr}}, the gradient is very 
shallow if $\tau_{inf}$ is constant over the disc; we obtain a gradient of 
about {\mbox{--0.2~dex/$h_B$}} only by adopting
a radially increasing infall timescale
(model {\sf SalpC}, Fig.~\ref{fig:gradOHSalpC}; Table~\ref{tab:models}). 

\begin{table*}
\begin{center}
\begin{tabular}{ |ll|rcc|ccccc| }
\hline
\multicolumn{1}{c}{$^{(1)}$} & \multicolumn{1}{c}{$^{(2)}$~~~~~} & 
\multicolumn{1}{c}{$^{(3)}$} & $^{(4)}$ & $^{(5)}$ & $^{(6)}$ & $^{(7)}$ & 
$^{(8)}$ & $^{(9)}$ & $^{(10)}$ \\
\multicolumn{1}{c}{Model} & \multicolumn{1}{c}{IMF~~~~~} & 
\multicolumn{1}{c}{$\tau_{inf}$} & $\nu$ & $\tau_{inf}(R)$ & $\frac{h_B}{R_d}$
& $b$ & $\frac{M_*}{L_I}$ & (B--V)$_0$ & $\frac{M_{gas}}{L_B}$ \\
\hline
{\sf SalpA} & Salpeter [0.1--100] &  5 & 0.47 &       constant        & 0.95 & 0.79 & 1.54 & 0.55 & 1.14 \\ 
{\sf SalpB} & Salpeter [0.1--100] &  7 & 0.69 &       constant        & 0.97 & 0.86 & 1.52 & 0.54 & 0.90 \\ 
\medskip
{\sf SalpC} & Salpeter [0.1--100] & 10 & 1.3  & $\propto \frac{R}{h_B}$ & 1.09 & 0.90 & 1.51 & 0.53 & 0.64 \\ 

\medskip
{\sf SalpD} & Salpeter [0.1--70]  & 10 & 1.8  & 
$\left\{ 
\begin{array}{l l}
\propto \frac{R}{h_B} & (R<h_B) \\
\propto \left( \frac{R}{h_B} \right)^2 & (R>h_B)
\end{array}
\right. $
& 1.19 & 0.78 & 1.58 & 0.53 & 0.51 \\ 

{\sf SalpE} & Salpeter [0.09--100]  & 10 & 1.7  & 
$\left\{ \begin{array}{l l}
\propto \frac{R}{h_B} & (R<h_B) \\
\propto \left( \frac{R}{h_B} \right)^2 & (R>h_B)
\end{array}
\right.$
& 1.18 & 0.79 & 1.67 & 0.53 & 0.56 \\ 
\hline
{\sf KrouA} & Kroupa  [0.1--100] &  2 & 0.75 &       constant        & 1.07 & 0.32 & 1.37 & 0.69 & 0.56 \\ 

{\sf KrouB} & Kroupa  [0.2--100] &  3 & 0.78 &        constant        & 1.05 & 0.45 & 1.06 & 0.66 & 0.50 \\

{\sf KrouC} & Kroupa  [0.3--100] &  4 & 0.75 &        constant        & 0.99 & 0.61 & 0.84 & 0.63 & 0.47 \\

\medskip
{\sf KrouD}  & Kroupa [0.35--100] &  5 & 0.82 & $\propto \surd \overline{\frac{R}{h_B}}$ & 1.02 & 0.73 & 0.73 & 0.60 & 0.43 \\ 

{\sf KrouE}  & Kroupa  [0.4--100] &  6 & 0.8 & $\propto \surd \overline{\frac{R}{h_B}}$ & 0.99 & 0.84 & 0.66 & 0.58 & 0.42 \\ 
\hline
{\sf KennA} & Kennicutt [0.1--100] &  3 & 0.21 &       constant        & 0.80 & 0.84 & 0.75 & 0.55 & 1.06 \\ 
{\sf KennB} & Kennicutt [0.1--100] &  4 & 0.25 &       constant        & 0.80 & 0.93 & 0.73 & 0.53 & 0.95 \\ 
{\sf KennC} & Kennicutt [0.1--100] &  5 & 0.29 &       constant        & 0.81 & 1.02 & 0.71 & 0.62 & 0.86 \\ 

{\sf KennD} & Kennicutt  [0--100]  &  4 & 0.32 &       constant        & 0.88 & 0.82 & 0.92 & 0.55 & 0.93 \\ 
{\sf KennE} & Kennicutt  [0--100]  &  5 & 0.38 &       constant        & 0.89 & 0.89 & 0.90 & 0.54 & 0.84 \\ 
{\sf KennF} & Kennicutt  [0--100]  &  7 & 0.49 &       constant        & 0.89 & 1.00 & 0.87 & 0.53 & 0.72 \\ 

{\sf KennG} & Kennicutt  [0.1--30] &  5 & 0.61 &       constant        & 0.96 & 0.78 & 0.74 & 0.56 & 0.50 \\ 
{\sf KennH} & Kennicutt  [0.1--35] &  7 & 0.61 &       constant        & 0.92 & 0.97 & 0.69 & 0.53 & 0.51 \\ 
{\sf KennI} & Kennicutt [0.05--35] &  5 & 0.54 &       constant        & 0.94 & 0.81 & 0.79 & 0.56 & 0.58 \\ 
{\sf KennJ} & Kennicutt [0.05--35] &  7 & 0.78 & $\propto \surd \overline{\frac{R}{h_B}}$ & 1.00 & 0.90 & 0.76 & 0.54 & 0.47 \\ 
\hline
{\sf LarsA} & Larson [0.01--100] &  1 & 0.085 &       constant        & 0.70 & 0.82 & 0.84 & 0.51 & 2.25 \\ 
{\sf LarsB} & Larson [0.01--100] &  2 & 0.10  &       constant        & 0.70 & 0.94 & 0.78 & 0.49 & 1.98 \\ 
\medskip
{\sf LarsC} & Larson [0.01--22]  &  5 & 0.60  & $\propto \surd \overline{\frac{R}{h_B}}$ & 0.98 & 0.81 & 0.78 & 0.51 & 0.56 \\ 

\medskip
{\sf LarsD} & Larson [0.01--22]  &  6 & 0.79  & $\propto \surd \overline{\frac{R}{h_B}}$ & 1.01 & 0.83 & 0.79 & 0.51 & 0.47 \\ 

{\sf LarsE} & Larson [0.01--23]  &  7 & 0.68  & $\propto \surd \overline{\frac{R}{h_B}}$ & 0.98 & 0.95 & 0.74 & 0.49 & 0.52 \\ 
\hline
\medskip
{\sf LmodA} & mod.\ Larson [0.01--100] & 4 & 0.54  &                 constant                 & 0.97 & 0.68 & 0.97 & 0.59 & 0.67 \\ 

\medskip
{\sf LmodB} & mod.\ Larson [0.01--100] & 5 & 0.70  & $\propto \surd \overline{\frac{R}{h_B}}$ & 1.01 & 0.74 & 0.93 & 0.57 & 0.58 \\ 

{\sf LmodC} & mod.\ Larson [0.01--100] & 7 & 1.4   &         $\propto \frac{R}{h_B}$          & 1.10 & 0.77 & 0.95 & 0.56 & 0.39 \\ 
\medskip
{\sf LmodD} & mod.\ Larson  [0.2--100] & 6 & 0.65  &                 constant                 & 0.96 & 0.86 & 0.82 & 0.56 & 0.56 \\ 

\medskip
{\sf LmodE} & mod.\ Larson  [0.2--100] & 7 & 0.78  & $\propto \surd \overline{\frac{R}{h_B}}$ & 0.99 & 0.90 & 0.80 & 0.55 & 0.51 \\ 

{\sf LmodF} & mod.\ Larson  [0.2--100] & 8 & 0.92  & $\propto \surd \overline{\frac{R}{h_B}}$ & 1.01 & 0.93 & 0.80 & 0.54 & 0.46 \\ 
\hline
{\sf ChabA} & Chabrier [0.01--100] & 3 & 0.24  &                 constant                 & 0.83 & 0.78 & 0.95 & 0.52 & 1.12 \\ 
{\sf ChabB} & Chabrier [0.01--100] & 4 & 0.28  &                 constant                 & 0.83 & 0.88 & 0.91 & 0.51 & 0.99 \\ 
{\sf ChabC} & Chabrier [0.01--100] & 5 & 0.33  &                 constant                 & 0.84 & 0.95 & 0.88 & 0.50 & 0.89 \\ 

{\sf ChabD} & Chabrier  [0.01--32] & 6 & 0.69  &                 constant                 & 0.97 & 0.83 & 0.94 & 0.52 & 0.57 \\ 

{\sf ChabE} & Chabrier  [0.01--32] & 7 & 0.84  & $\propto \surd \overline{\frac{R}{h_B}}$ & 1.01 & 0.87 & 0.91 & 0.51 & 0.51 \\ 
{\sf ChabF} & Chabrier  [0.01--33] & 8 & 0.92  & $\propto \surd \overline{\frac{R}{h_B}}$ & 1.01 & 0.92 & 0.90 & 0.50 & 0.48 \\ 
\hline
\end{tabular}
\end{center}
\caption{List of models with calibrated parameters (columns 3 to 5)
and results (columns 6 to 10). 
(1) Model name. 
(2) Adopted IMF with corresponding mass limits.
(3) Infall timescale $\tau_{inf}$ at $R=h_B$, in Gyr.
(4) SF efficiency $\nu$, in Gyr$^{-1}$.
(5) Radial dependence of the infall timescale.
(6) Scale--length of the B--band profile.
(7) $b$--parameter of the SFH (Eq.~\protect{\ref{eq:b-psi}}).
(8) M/L ratio in the I--band for the stellar component (stars+remnants).
(9) Intrinsic (B-V) colour of the disc (including the central regions, i.e.\ 
    between $R=0-3~h_B$).
(10) Gas--to--luminosity fraction.}
\label{tab:models}
\end{table*}

\section{Model results}
In this Section we present the resulting M/L ratios and chemical
properties of our full chemo--photometric models. First we discuss results 
for the Salpeter IMF, then for the other, ``bottom--light'' IMFs. All models
are calibrated to reproduce the observed metallicity distributions and SFHs,
as described in \S\ref{sect:calibration}; the calibrated parameters
and model results are all listed in Table~\ref{tab:models}.
\subsection{Salpeter models}
\label{sect:Salp-models}
The I--band M/L ratios for the disc models {\mbox{\sf SalpA-B-C}}, adopting 
the Salpeter IMF with the standard mass limits {\mbox{[0.1---100]~\Msol,}} 
are shown as solid circles in Fig.~\ref{fig:MLI_salp} as a function of the
corresponding $b$--parameter. Also shown as a
continuous thin line is the prediction
from the simple exponential models of \S\ref{sect:b-models}, with solar 
metallicity. The full chemo--photometric models confirm quite well
the high $M/L_I \sim$ 1.5 predicted by the simple models for the Salpeter case.

One should not infer, however, from the agreement with the simple models
that photometric computations at constant solar metallicity 
(like those in \S\ref{sect:b-models}) are sufficient. In fact, if
we calculate the luminosity of the full chemical models {\mbox{\sf SalpA-B-C}}
using only SSPs with $Z=$\Zsol, we obtain higher M/L ratios (open circles in 
Fig.~\ref{fig:MLI_salp}), mainly because of a slightly larger average age 
of stars in realistic models (with $T=13$~Gyr) with respect to the simple 
exponential ones (with $T=10$~Gyr). In full chemo--photometric models, 
this is compensated by the fact that an important fraction of
the long--lived stars contributing to the I--band luminosity forms at
metallicities below solar, lowering the M/L ratio with respect to
calculations at fixed $Z=$\Zsol\ (solid circles vs.\ open circles; 
see also the M/L ratios at $Z$=0.008 in Fig.~\ref{fig:b-models}). 
So, the agreement with
the simple exponential models with solar metallicity is the complex result 
of combining a more realistic, 
radially dependent SFH, with metallicity--dependent photometric calculations. 

The crosses in Fig.~\ref{fig:MLI_salp} represent the result of calculating
the photometry of models {\mbox{\sf SalpA-B-C}} including only SSPs 
of metallicity $Z \leq$\Zsol, adopting SSPs of $Z$=\Zsol\
for larger metallicities. The difference with respect to the reference models
is marginal, indicating that super--solar and solar metallicity stars 
contribute to the I--luminosity in a comparable way. However, neglecting 
super--solar
metallicities in the photometry has noticeable effects when considering, 
for instance, the B--luminosity 
(Fig.~\ref{fig:MLB_salp}). This is because the B--band is both more 
sensitive to metallicity than the I--band (Fig.~\ref{fig:ML_IB_Z}) and 
more influenced by young stars, which have a metallicity
that is higher on average and reaches super--solar values in the inner
disc regions. Therefore, we stress the importance of covering the full
metallicity range, from sub--solar to super--solar metallicities, when
computing the photometric properties of disc galaxies.

\begin{figure}
\centerline{\psfig{file=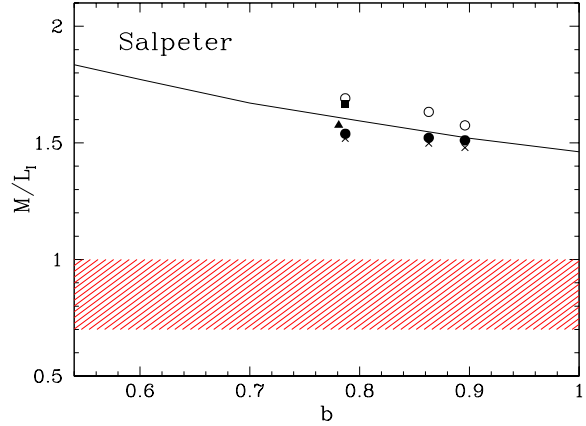,angle=-90,width=8truecm}}
\caption{M/L ratio in the I--band for chemo--photometric models with the
Salpeter IMF. The shaded area marks the range M/L$_I$=0.7--1 favoured by
observations of Sbc--Sc discs ($b=0.8-1$).
{\it Solid circles}: models {\sf SalpA-B-C}  adopting the standard IMF mass 
range [0.1--100]~\Msol. 
{\it Open circles}: models {\sf SalpA-B-C} with photometry calculated 
at constant metallicity $Z=Z_{\odot}=0.019$.
{\it Crosses}: models {\sf SalpA-B-C} with photometry calculated including
only SSPs with $Z\leq Z_{\odot}$.
{\it Triangle}: model {\sf SalpD} with a smaller upper mass limit
$M_s=70$~\Msol.
{\it Square}: model {\sf SalpE} with a smaller lower mass limit
$M_i=0.09$~\Msol.
{\it Thin line}: prediction from simple exponential models with $Z=$~\Zsol.}
\label{fig:MLI_salp}
\end{figure}

\begin{figure}
\centerline{\psfig{file=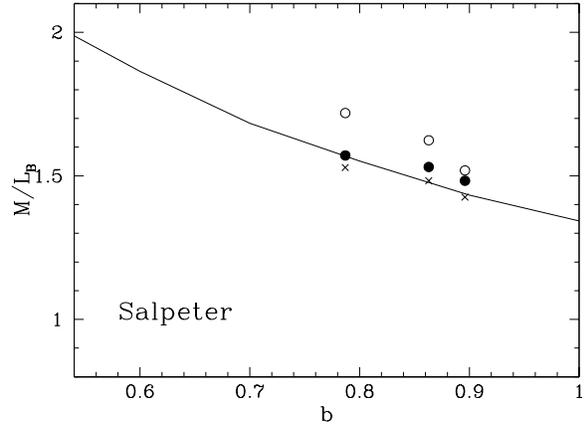,angle=-90,width=8truecm}}
\caption{M/L ratio in the B--band for chemo--photometric models with the
Salpeter IMF. Symbols as in Fig.~\protect{\ref{fig:MLI_salp}}.}
\label{fig:MLB_salp}
\end{figure}

\begin{figure}
\centerline{\psfig{file=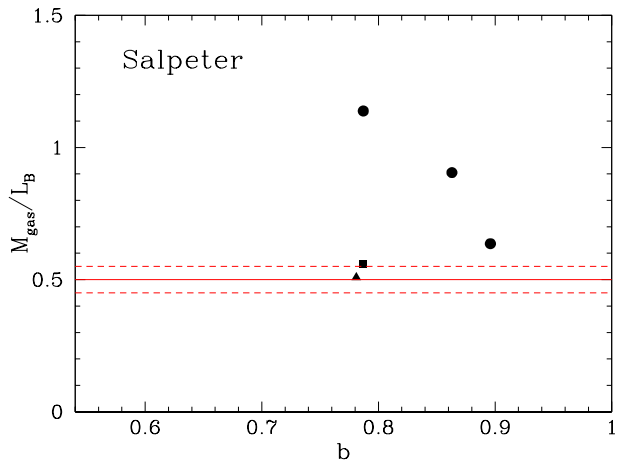,angle=-90,width=8.5truecm}}
\caption{Gas--to--luminosity fraction for chemo--photometric models with the
Salpeter IMF. Symbols as in Fig.~\protect{\ref{fig:MLI_salp}}.
The horizontal lines mark the observational range of 0.5$\pm$ 0.05 for
the gas fraction in Sbc-Sc discs ($b=0.8-1$).}
\label{fig:MgasLB_salp}
\end{figure}

Fig.~\ref{fig:MgasLB_salp} shows the resulting gas--to--luminosity fraction 
$M_{gas}/L_B$ for the chemo--photometric models, compared to the observational
estimates for Sbc-Sc discs (\S\ref{sect:chem-constraint}). 
Models with the Salpeter IMF between [0.1--100]~\Msol\ (solid circles)
tend to be too rich in gas with respect to observations. Considering that 
for a given set of stellar yields, the final metallicity increases with 
decreasing gas fraction (Tinsley 1980), our results indicate that 
with this IMF the global metal production (yield) is somewhat high:
if we had imposed the models to reach the observed gas fractions,
the final oxygen abundance in the gas would have been higher than observed. 
This finding 
is in line with known results for the Solar Neighbourhood: a Salpeter IMF
extended up to 100~\Msol\ produces too much oxygen to be compatible with
the local evidence, while better agreement is found with an upper mass limit
of 40--50~\Msol (Tsujimoto \etal 1997; Thomas, Greggio \& Bender 1998; Gratton 
\etal 2000).

The efficiency of metal enrichment from a stellar population
can be estimated by computing the so--called global (or net) yield 
(Tinsley 1980; Pagel 1997): 
\begin{equation}
\label{eq:yield}
y_Z = \frac{1}{\alpha} \, \int_{M(T)}^{M_s} p_Z(M) \Phi(M) \, dM
\end{equation}
where $M(T)$ is the stellar mass of lifetime corresponding to the age $T$
of the system (the galaxy), and $p_Z(M)$ is the mass fraction of new metals 
ejected by a star of mass $M$; hence the integral expresses the amount 
of metals globally produced by a stellar generation over the age of the galaxy.
The symbol $\alpha$ represents the fraction of mass locked up in
``ever--living'' low mass stars or remnants.
In the Instantaneous Recycling Approximation, the net yield can be directly
used to predict the metallicity evolution (see Tinsley 1980; Pagel 1997 for 
details); with more complex chemical networks taking into account finite
stellar lifetimes, delayed gas restitution and the different timescales 
of release of different elements, as in our models, things are not as 
straightforward but
the net yield per stellar generation still provides a useful insight
on the efficiency of metal enrichment. In particular it underlines that
the latter does not depend merely on the amount of metals produced per
mass involved in star formation, but on the ratio between this and
the mass that remains forever locked in stars.
The yield can thus be modified by altering the amount of metals produced
(for instance reducing/increasing the number of massive stars) 
and/or by changing the locked--up fraction.

For the Salpeter models, a lower yield seems to be necessary since oxygen
is somewhat over-produced.
We calculate model {\sf SalpD} decreasing the upper mass end
$M_s$ of the IMF to reduce the metal production, and obtain a corresponding 
gas fraction in agreement with observations (solid triangle in 
Fig.~\ref{fig:MgasLB_salp} and~\ref{fig:MLI_salp}). 
Alternatively, in model {\sf SalpE} we decrease the lower mass end $M_i$ 
and thereby increase the locked--up fraction, decrease the net metal yield 
and obtain again a gas fraction close to the observed values 
(solid square in Fig.~\ref{fig:MgasLB_salp}).

Notice however that, with the Salpeter IMF,
a very restricted range of parameters (mass limits and infall timescales)
can yield a Sbc/Sc--type SFH ($b \sim 0.8$) together with metallicities 
and gas fractions both close to the observational values. 
What matters for the metallicity--gas fraction relation is in fact 
the yield, and models {\sf SalpD} and {\sf SalpE} 
have a similar oxygen yield (Table~\ref{tab:IMFyields}). However,
when this condition is met by decreasing $M_i$ (model {\sf SalpE}), 
the M/L ratio increases for we are shifting the IMF toward star masses 
that contribute no light (model {\sf SalpE}, solid square in 
Fig.~\ref{fig:MLI_salp}).

\begin{figure}
\centerline{\psfig{file=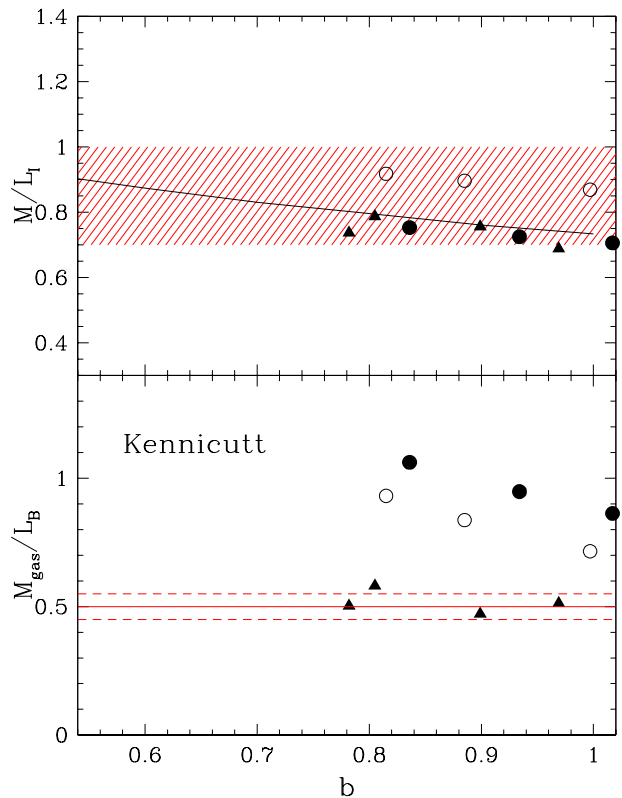,width=8.5truecm}}
\caption{{\it Upper panel}: M/L ratio in the I--band for chemo--photometric 
models with the Kennicutt IMF; {\it lower panel}: 
corresponding gas--to--luminosity fractions.
The shaded area in the top panel and the dashed lines in the lower panel mark 
the observational ranges for Sbc-Sc discs ($b$=0.8--1).
{\it Solid circles}: models {\sf KennA-B-C} adopting the standard IMF mass 
range [0.1--100]~\Msol. 
{\it Open circles}: models {\sf KennD-E-F} adopting the lowest possible
mass end, with range [0--100]~\Msol.
{\it Solid triangles}: models {\sf KennG-H-I-J} with mass limits tuned 
to reproduce the observed gas fractions ($M_i$=0.05--0.1~\Msol, 
$M_s$=30--35~\Msol).}
\label{fig:models_kenn}
\end{figure}

\begin{figure}
\centerline{\psfig{file=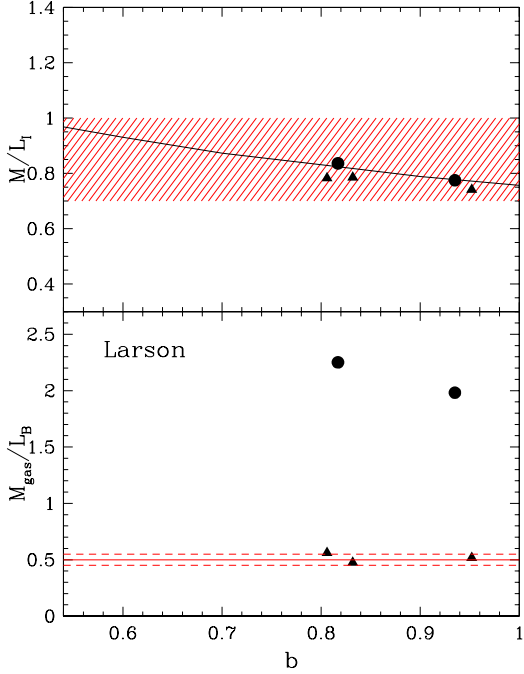,width=7.2truecm}}
\caption{{\it Upper panel}: M/L ratio in the I--band for chemo--photometric 
models with the Larson IMF; {\it lower panel}: corresponding 
gas--to--luminosity fractions. The shaded area in the top panel and the 
dashed lines in the lower panel mark the observational ranges for Sbc-Sc discs
($b$=0.8--1).
{\it Circles}: models {\sf LarsA-B} adopting the IMF mass
range [0.01--100]~\Msol. 
{\it Triangles}: models {\sf LarsC-D-E} with upper mass limit tuned 
to reproduce the observed gas fractions ($M_s$=22--23~\Msol).}
\label{fig:models_lars}
\end{figure}

\begin{figure}
\centerline{\psfig{file=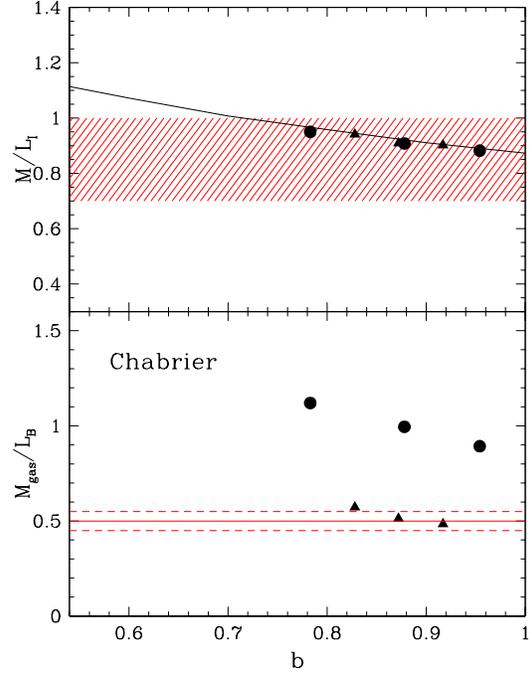,width=7.2truecm}}
\caption{Same as Fig.~\protect{\ref{fig:models_kenn}}
to~\protect{\ref{fig:models_lmod}}, but for the Chabrier IMF.
{\it Circles}: models {\sf ChabA-B-C} adopting the full IMF mass
range [0.01--100]~\Msol. 
{\it Triangles}: models {\sf ChabD-E-F} with upper mass limit 
$M_s$=32--33~\Msol, tuned to reproduce the observed metallicities 
and gas fractions with high $b$--parameters.}
\label{fig:models_chab}
\end{figure}

\subsection{Kennicutt models}
The I--band M/L ratios of models {\mbox{\sf KennA-B-C}}, adopting 
the Kennicutt IMF and standard limits {\mbox{[0.1---100]~\Msol,}} 
are shown as solid circles in Fig.~\ref{fig:models_kenn} (upper panel); 
they lie close to the predictions from the simple exponential 
models of \S\ref{sect:b-models} (thin line) and within the range favoured
by observations.

The lower panel shows that the corresponding gas--to--luminosity fractions
$M_{gas}/L_B$ are higher than observed, by a factor of two or so.
Hence, the net yield of this IMF is also too high (see the discussion
in the previous section for the Salpeter models).
Although the slope at the high--mass end (--1.5) is steeper than 
in the Salpeter IMF, so that one might expect less massive stars and less 
oxygen production, for the bottom--light Kennicutt IMF the smaller number of 
low--mass stars results in a low locked--up fraction and hence a high net 
yield overall (Table~\ref{tab:IMFyields}).
The slope at the low--mass end is so shallow that even extending the IMF down
to $M_i$=0 would increase the locked--up fraction only marginally, and would
not improve much the situation (models {\mbox{\sf KennD-E-F}}, open circles
in Fig.~\ref{fig:models_kenn}).
To reduce the global yield, the oxygen production per stellar generation
must also be reduced, by lowering the high--mass end.

In models {\mbox{\sf KennG-H-I-J}} (triangles in Fig.~\ref{fig:models_kenn})
we tune the mass limits of the IMF so that we obtain a SFH typical of 
Sbc/Sc discs ($b=0.8-1$) together with a
suitable gas--to--luminosity fraction $M_{gas}/L_B \sim 0.5$.
The corresponding M/L ratios are low, $M/L_I$=0.7--0.8, still
in agreement with the observed range.

\subsection{Larson and Chabrier models}
Models with the Larson or Chabrier IMF and mass limits [0.01--100]~\Msol\ 
({\sf LarsA-B} and {\sf ChabA-B-C} in Table~\ref{tab:models})
have a low M/L ratio, in very good agreement with the 
corresponding simple models of \S\ref{sect:b-models}; they are displayed
as solid circles in Fig.~\ref{fig:models_lars} and~\ref{fig:models_chab}, 
respectively. However, in both cases the corresponding gas--to--luminosity 
fraction is exceedingly high, larger than the observed one by a factor of 2 
for the Chabrier IMF and a factor of 4 for the Larson IMF.
This again indicates that the global net yield
is too high with these IMF (Table~\ref{tab:IMFyields}). 
This is no surprise for the Larson IMF which has a Salpeter slope 
at high masses --- and the Salpeter IMF already
had a somewhat high oxygen yield, \S\ref{sect:Salp-models} ---
and in addition, due to the exponential cut-off at low masses, 
its locked--up fraction is much smaller, leading to a very high net yield.
The Chabrier IMF is steeper at high masses, yet the low locked--up fraction
produces a high yield. 

Both IMFs have a sharp cut-off at low masses, so that even reducing 
the low mass end down to $M_i=0$ has no effect on the locked--up
fractions, and the only way to reduce the net yield is to lower the
upper mass end $M_s$. Models {\sf LarsC-D-E} with $M_s=22-23$~\Msol,
and models {\sf ChabD-E-F} with $M_s$=32--33~\Msol\
reach the correct gas fraction, while the corresponding I--band M/L ratios
are relatively unaffected by this tuning of the upper mass limits (triangles 
in Fig.~\ref{fig:models_lars} and Fig.~\ref{fig:models_chab}; see also
Tsujimoto \etal 1997).

\begin{figure}
\centerline{\psfig{file=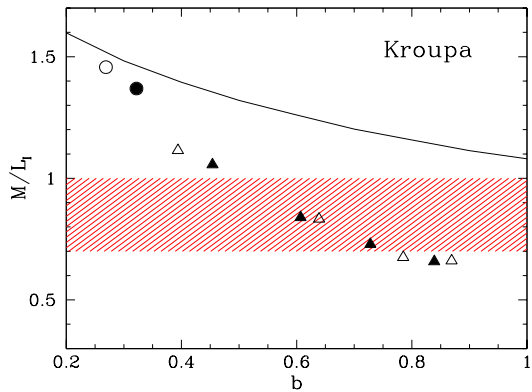,angle=-90,width=7.3truecm}}
\caption{M/L ratio in the I--band for chemo--photometric models with the
Kroupa IMF. The shaded area marks the range M/L$_I$=0.7--1 favoured by
observations of Sbc--Sc discs ($b=0.8-1$).
{\it Circles}: models adopting the standard IMF mass range [0.1--100]~\Msol;
they lie quite close to the predictions from the
exponential models from \S\protect{\ref{sect:b-models}}
({\it thin line}).
{\it Triangles}: models adopting a higher low mass end $M_i$=0.2--0.4
to reproduced the observed metallicities at higher $b$--values; the M/L ratio
correspondingly decreases because of the lower locked--up fraction.
{\it Solid symbols}: models {\sf KrouA-B-C-D-E} listed in 
Table~\protect{\ref{tab:models}}, with a gas--to--luminosity fraction 
$M_{gas}/L_B \sim 0.5$. 
{\it Open symbols}: other calculated models, with slightly different
gas fraction.}
\label{fig:MLI_krou}
\end{figure}

\subsection{Kroupa and modified--Larson models}
\label{sect:Kroupa_models}
The Kroupa IMF and the modified--Larson IMFs are characterized by a rather 
steep slope of --1.7 for $M > 1$~\Msol. Combined with the locked--up 
fractions corresponding to the standard mass limits of [0.1--100]~\Msol,
such steep slopes result in low net yields compared to the IMFs previously 
considered, especially for oxygen which is produced in massive stars
(Table~\ref{tab:IMFyields}).
 --- Actually, for the modified--Larson IMF the reference mass limits are
[0.01--100]~\Msol, however due to the sharp cut-off below the peak mass 
$\sim$0.25~\Msol, changing the low--mass end of this IMF to $M_i$=0.1 
(or 0)~\Msol\ has negligible effects on the results.

As a consequence, with these IMFs the typical metallicities observed 
in disc galaxies can be obtained only
for short infall timescales, and correspondingly low {\mbox{$b$--parameters}}
(models {\sf KrouA} and {\sf LmodA-B-C} in Table~\ref{tab:models}, displayed 
as solid circles in Fig.~\ref{fig:MLI_krou} and~\ref{fig:models_lmod}). 
Longer infall timescales tend
in fact to dilute the metals produced and lower the final metallicity, for 
a given net yield.

To reach higher $b$--values ($b=0.8-1$), longer infall timescales are 
required; at the same time, the net yield must be increased to reproduce 
the typical oxygen abundance 
of 9.1~dex at $R=h_B$. This we can achieve by increasing
the lower mass end $M_i$, which corresponds to decreasing the locked--up 
fraction and also the M/L ratio with respect to the standard case.

As to the Kroupa case, models {\sf KrouB-C-D-E} correspond to increasing
values of the $b$--parameter (solid triangles in Fig.~\ref{fig:MLI_krou}).
For simplicity, in Table~\ref{tab:models} we list only Kroupa models 
with a gas--to--luminosity fraction close to the observed value 
$M_{gas}/L_B \sim 0.5$ (solid symbols in Fig.~\ref{fig:MLI_krou}); 
other models with slightly different gas fractions
(open symbols) do not appreciably change the trends of the M/L ratios 
vs.\ $b$--parameter.
A good Sbc/Sc model ($b \sim 0.85$) is {\sf KrouE}, but such high values 
of $b$, combined with the constraints on oxygen production,
are reached only with an uncomfortably large $M_i=0.4$~\Msol. This limit,
however, should not be taken at face value: a sharp cut off at low 
masses is just a simplistic way to reduce the locked--up fraction so as 
to increase the yield. Stars below $\sim$0.8~\Msol\ do not contribute
metals nor substantial luminosity, and for chemical and photometric evolution
their mass distribution is irrelevant, what matters is just
their global mass. Any other distribution of masses at the low mass end, 
with the same locked--up fraction, would leave our results unchanged
(see also Tsujimoto \etal 1997).
Not surprisingly, with such a low locked--up fraction, the resulting M/L ratio 
is very low ($\lsim$0.7, Fig.~\ref{fig:MLI_krou}). 

\begin{figure}
\centerline{\psfig{file=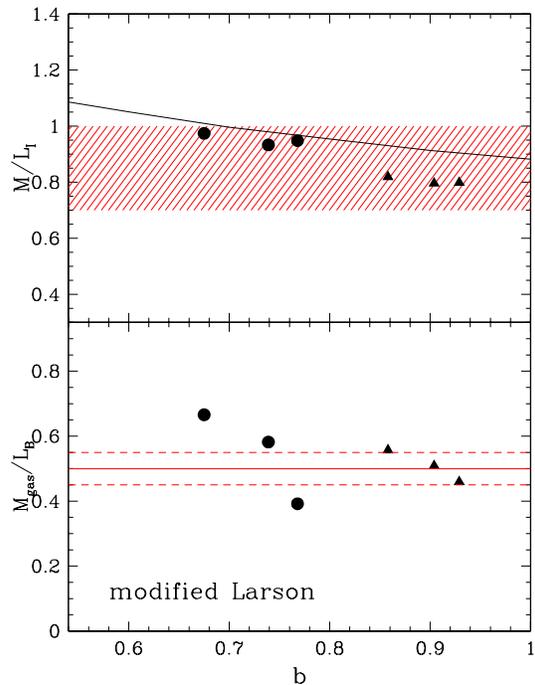,width=7.3truecm}}
\caption{Same as Figs.~\protect{\ref{fig:models_kenn}}, 
\protect{\ref{fig:models_lars}}
and~\protect{\ref{fig:models_chab}}, but for the modified Larson IMF.
{\it Circles}: models {\sf LmodA-B-C} adopting the full IMF mass
range [0.01--100]~\Msol. 
{\it Triangles}: models {\sf LmodD-E-F} with a lower mass limit of
$M_i$=0.2~\Msol, tuned to reproduce the observed metallicities 
and gas fractions with high $b$--parameters.}
\label{fig:models_lmod}
\end{figure}

For the modified--Larson IMF, the net yield is larger than in the Kroupa case, 
thanks to the exponential cut--off at low masses and the correspondingly
lower locked--up fraction (for the same mass limits). In this case, a less 
extreme value of the
low mass end $M_i$=0.2~\Msol\ suffices to reproduce the observed metallicities 
and gas--to--luminosity fractions with $b \geq 0.8$; the resulting typical 
M/L ratio is $\sim$0.8.

In conclusion, the Kroupa and modified--Larson IMFs, adopting a steep 
Scalo slope of --1.7 for $M\gsim$1~\Msol, seem to meet the opposite problem 
than the IMFs previously discussed: they are too inefficient in terms 
of oxygen production to account easily for the observed metallicities 
in spirals, if a SFH
of the type of Sbc/Sc discs is required at the same time ($b > 0.8$).
See however Kroupa (2001) about the uncertainties on the slope of the IMF
at high masses: a value shallower than 1.7, possibly up to the Salpeter
slope $\sim$1.3, is by no means excluded from local observations. 

\subsection{An overview of the results}
\label{sect:overview}
For all the IMF considered, the full chemo--photometric models lie close 
to the 
predictions of the corresponding simple models with the same mass limits --- 
a complex result of combining realistic, infall--like and radially dependent 
SFHs with metallicity--dependent photometry. As a consequence, the conclusions
of \S\ref{sect:b-models} for the M/L ratios are confirmed.

The Salpeter IMF results in M/L ratios much higher than required to match
observations; moreover, a Salpeter slope over the whole stellar mass 
range is currently not supported by observations 
(see \S\ref{sect:introduction}).
As to the bottom--light IMFs, broadly speaking we see two types 
of behaviour. 

{\bf In the first case} the low locked--up fraction typical of
bottom--light IMFs leads to a high global yield (Kennicutt, Larson and Chabrier
models), too high to reproduce the typical gas fractions of
Sbc/Sc spirals at the same time as their metallicities.
These IMFs have yields comparable to what
is required to explain the observed metal enrichment of hot gas in galaxy 
clusters ({\mbox{$y \sim$2--3~\Zsol}}, Pagel 2002), and much higher than
what is typically inferred for the Solar Neighbourhood ($y$ \lsim \Zsol) ---
compare the oxygen yields $y_O$ in Table~\ref{tab:IMFyields} to the solar 
oxygen abundance $Z_{O,\odot} \sim 9 \times 10^{-3}$. A possible 
explanation is that disc galaxies, just like ellipticals in clusters, disperse
a large part of the metals they produce into the intergalactic medium, so that
we do not detect these metals in present--day discs.
Blow--out of material
from present--day discs is observed in the form of fountains, but this gas
does not seem to have the necessary energy to escape the galactic potential, 
so that it just falls back onto the disc on relatively short timescales
($\sim 10^8$~yrs, Bregman 1980; Fraternali \etal 2001; Heckman 2002). 
So, escape of large quantities of metals
from present--day discs does not seem very realistic. However, in the early 
phases of galaxy formation it is possible that dispersal of metals took place,
when the potential well of the galaxy was weaker; large outflows are in 
fact observed in high--redshift Lyman--break galaxies, though these are usually
identified with the progenitors of ellipticals (Pettini \etal 2000). Besides,
there are arguments advocating large outflows of baryons from galaxies 
in general (Silk 2003). 

Alternatively, a large part of the metals produced may be locked--up
in dust, which may correspondingly induce an observational underestimate 
of the true
metallicity in spiral galaxies. Theoretical models and observational
evidence suggest that in the interstellar medium of our Galaxy, 40\% of
the metals is locked--up in dust grains (Dwek 1998 and references therein).
As a consequence, the oxygen abundance probed by atomic emission lines might be
underestimated by almost a factor of 2. However, the oxygen abundance is 
typically measured in HII regions, where dust depletion is probably much lower
than the average for the diffuse interstellar medium; besides, oxygen 
abundances and abundance gradients derived from the spectra of OB stars,
which should be totally unaffected by dust depletion, agree very well with 
those from HII regions (Smartt \& Rolleston 1997; Gummersbach \etal 1998).
So we consider oxygen abundances measured in HII regions as a reliable estimate
of the real metallicity of a galaxy. Dust depletion does not seem to provide
a viable solution to the apparent overproduction of metals.

As an alternative solution to metal outflows, some fine--tuning of the upper 
mass limit of the IMF can be 
applied, corresponding to a tuning of the global yield, so that
the chemical properties of discs can be matched without requiring considerable
dispersal of metals.\footnote{The same reduction of the net yield is
obtained, if above a certain stellar mass there is fallback 
of metals onto a central black hole after the supernova explosion
(Colpi, Shapiro \& Wasserman 1996; Zampieri, Shapiro \& Colpi 1998; 
MacFadyen, Woosley \& Heger 2001;
Podsiadlowski \etal 2002). In the presence of fallback, 
the actual amount of metals enriching the interstellar medium is less 
than what is formally predicted by supernova models. See also Tsujimoto \etal
(1997).}
However the resulting I--band M/L ratio is 
quite robust to such fine--tuning (see \S\ref{sect:driving} 
and Tsujimoto \etal 1997), 
and for Sbc/Sc discs we consistently find 
for these IMFs
$M/L_I<1$, with values between 0.7 (Kennicutt IMF) and 0.9 (Chabrier IMF).

{\bf In the second category}, we find IMFs with a steep slope at the 
high--mass end (--1.7, Kroupa and modified--Larson IMFs) implying a low
oxygen yield. In this case, to reproduce the observed metallicities and
gas fractions combined with an Sbc/Sc--like SFH ($b \geq$0.8) the global yield
must be increased, via reducing the locked--up fraction with a higher
minimum stellar mass $M_i$. Correspondingly, since less mass is locked in very 
low--mass, low--luminosity stars, the M/L ratio also decreases, and for the
Kroupa and modified--Larson models of Sbc/Sc galaxies we find again
{{\mbox{$M/L_I \sim$0.7--0.8.}}

Our conclusions derived from the ``chemical properties'' of discs, i.e.\ the 
typical metallicities and gas fractions, are still subject to some 
uncertainties.
The constraint of the gas--to--luminosity fraction is rather uncertain 
for it has been derived by assembling a number of independent studies and by 
applying various corrections (see \S\ref{sect:chem-constraint}); besides,
it relies on the B luminosity, which is affected by uncertain
dust extinction. Furthermore, the global yield of a given IMF,
and the corresponding fine--tuning of the mass limits necessary
in chemical models, depends on the adopted stellar yields; here also
some uncertainties exist,
although the present study is mainly based on oxygen, which is 
the element with best estimated theoretical yields.
However, the M/L ratio is generally robust with respect to the 
fine--tuning of the IMF mass limits we applied to reproduce metallicities
and gas fractions. For a metallicity distribution resembling the observed one
(oxygen abundance of about 9.1~dex at R=$h_B$, gradient around --0.2~dex/kpc)
we get as a consistent result that $M/L_I = 0.7-0.9$ for 
Sbc/Sc disc models, with all the bottom--light IMFs we considered.
Many other models were computed than those presented here, with shallower
or steeper gradients (say, between --0.1 and --0.3 dex/kpc), or typical 
metallicities slightly different than the reference value of 9.1 
(say between 9.0 and 9.2), and the conclusions on the M/L ratio in the I band
for late--type discs are quite robust to such variations in the model 
constraints.

\subsection{The driving parameters for the $M/L$ ratio}
\label{sect:driving}
We have demonstrated that various ``bottom--light'' IMFs suggested in
recent literature result in low M/L ratios $M/L_I < 1$ for Sbc/Sc discs.
It is interesting to discuss what is the common feature, among these
different IMFs, that determines this result.

\begin{figure*}
\leavevmode
\centerline{ \psfig{file=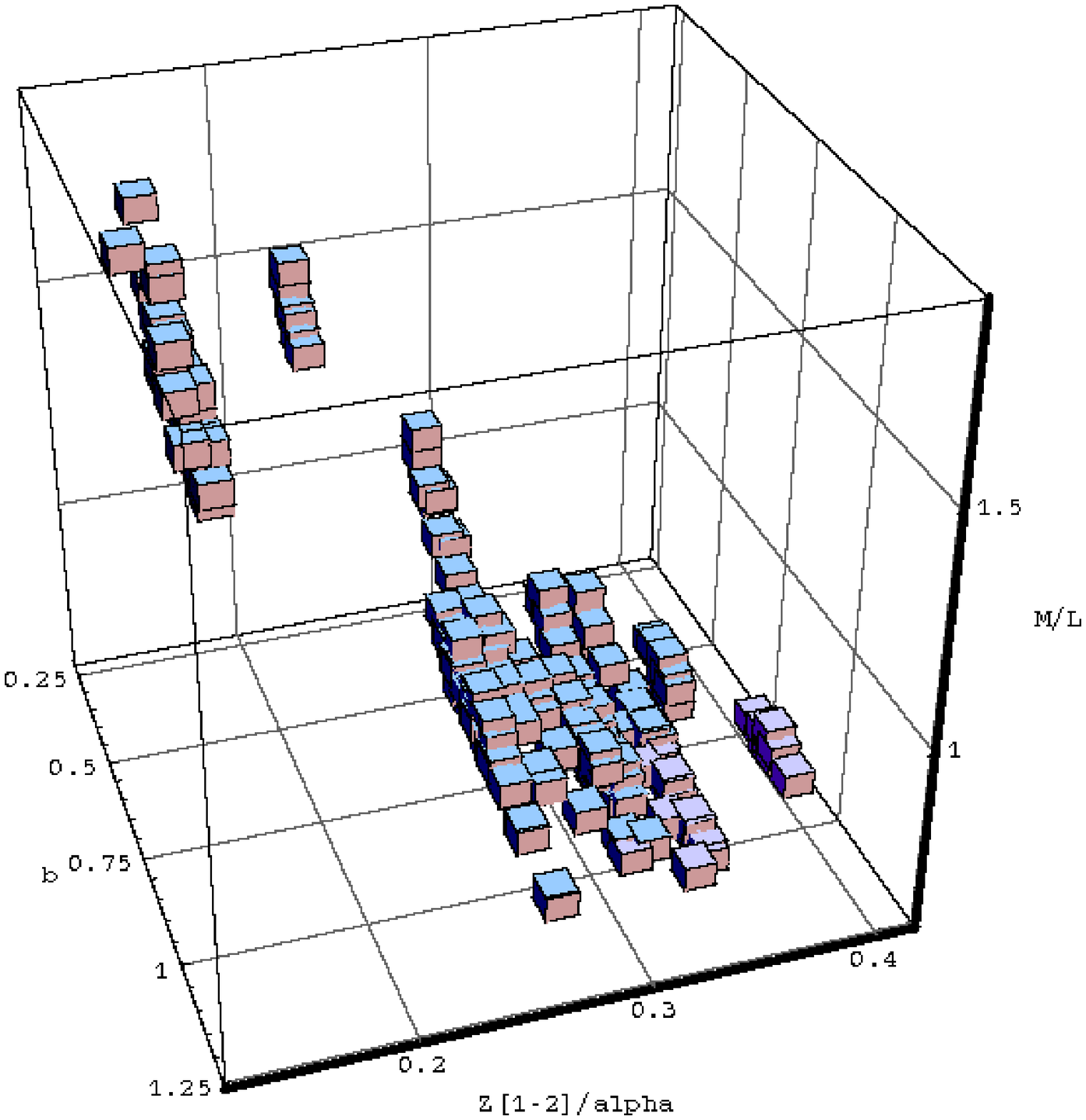,width=8truecm} 
\psfig{file=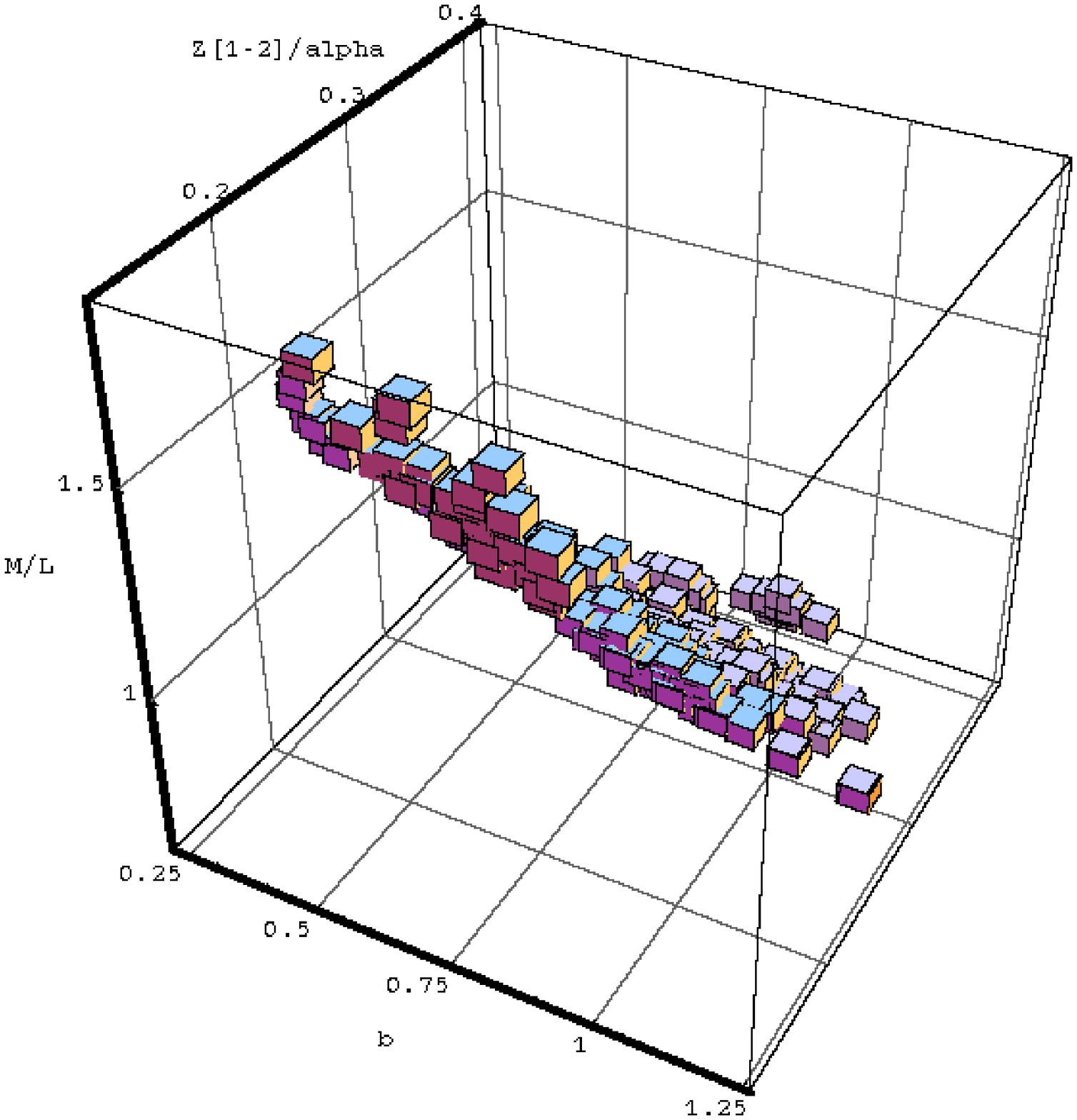,width=8truecm} }
\caption{3-D plot relating the $M/L_I$ ratio to the $b$--parameter and to the
IMF ``driving quantity'' $\zeta_{[1-2]}/\alpha$ (see text).}
\label{fig:3Dplots}
\end{figure*}

\begin{figure}
\centerline{\psfig{file=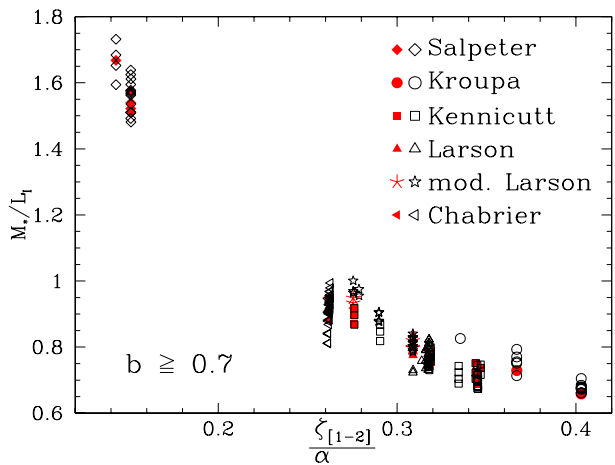,angle=270,width=8.5truecm}}
\caption{Relation between $M/L_I$ and the ``driving quantity'' 
$\zeta[1-2]/\alpha$ (see text), for models with $b>=0.7$.
Solid symbols for the
``calibrated'' models in Table~\protect{\ref{tab:models}}, open symbols for
other models. A given IMF-symbol can correspond to different values 
of the abscissa because the models explore different mass limits 
(cf.\ Table~\protect{\ref{tab:models}}).}
\label{fig:2Dslice}
\end{figure}

We analyzed in our models the correlations between the predicted M/L ratio 
and various
quantities characterizing the IMF, e.g.\ the mass fraction in stars
around 1~\Msol, or around 10~\Msol, or the mass fraction in ``ever--living''
stars below 1~\Msol, etc. The best correlation is found
with the quantity $\zeta_{[1-2]}/\alpha$, where $\zeta_{[1-2]}$ is the
mass fraction that a given IMF ``assigns'' to stars between
1--2~\Msol\ and $\alpha$ is the mass fraction that remains locked in 
ever--living stars and remnants. Fig.~\ref{fig:3Dplots} shows a 3-D plot
of the resulting $M/L_I$ as a function of the $b$--parameter and of the 
above mentioned
quantity $\zeta_{[1-2]}/\alpha$ characterizing the IMF. The plot shows  
the ``calibrated'' models in Table~\ref{tab:models}, as well as many other
models with a variety of IMF mass limits, metallicities, 
gas--to--luminosity fractions etc., run during the calibration process.
Besides the expected dependence on the $b$--parameter 
amply discussed in this paper, the typical $M/L_I$ evidently decreases at 
increasing $\zeta_{[1-2]}/\alpha$. Fig.~\ref{fig:2Dslice} underlines the
latter dependence by showing
a ``slice'' of the 3-D plot of Fig.~\ref{fig:3Dplots}, corresponding to large 
$b$--parameters $b \geq 0.7$ and projected on the 
($\frac{\zeta_{[1-2]}}{\alpha}$,$\frac{M}{L_I}$) plane.
These correlations indicate that, in general, the $M/L_I$ ratio is
lower when the IMF favours stars in the [1--2]~\Msol\ range 
versus stars of lower masses and remnants from more massive stars.

This ``driving quantity'' $\zeta_{[1-2]}/\alpha$ is relevant for the 
$M/L$ ratio in red optical bands like the I band. At different 
wavelengths, the luminosity is dominated by stars of different 
masses and the corresponding IMF---M/L ratio relation is driven by
a different quantity. For instance, a similar analysis of the M/L ratio
in the B band shows that this is influenced mostly by the mass fraction 
in the range [1--9]~\Msol,
as expected from the fact that B--band light is sensitive to 
the contribution of more massive stars. 

For the IMFs and the corresponding mass limits in the calibrated models 
of Table~\ref{tab:models}}, we list the characteristic quantities
$\zeta_{[1-2]}$, $\alpha$ etc. in Table~\ref{tab:IMFyields}. We list also
the global oxygen yield $y_O$ (see Eq.~\ref{eq:yield}) and the related IMF 
quantity $\zeta_9/\alpha$, where $\zeta_9$ is the mass fraction
in stars more massive than 9~\Msol --- since it is these stars
that contribute the bulk of the oxygen production.
Actually, stellar yields depend on metallicity and this effect is included 
in our chemical models (Portinari \etal 1998); the global yields 
and locked--up fractions listed 
for illustration in Table~\ref{tab:IMFyields} are computed for solar 
metallicity. 

Traditionally, in chemical models the stellar IMF has been characterized 
by the quantity 
$\zeta_1$, that is the mass fraction in stars more massive than 1~\Msol.
This corresponds to dividing the IMF into two mass ranges, the ``ever-living''
low-mass stars, locking up mass out of further chemical processing forever,
and stars with a lifetime shorter than a Hubble time 
which recycle gas and metals; this description is useful
and meaningful within the simplified Instantaneous Recycling Approximation
for chemical evolution. A similar approach is applied also when deriving the
IMF from the present--day mass function (PDMF), grossly separating ever-living
stars, whose PDMF is the integrated result of the IMF over the history of the
galaxy, from stars with finite stellar lifetimes assumed then to evolve 
``instantly'', whose PDMF is a direct signature of the IMF (Tinsley 1980;
Scalo 1986).

\begin{table}
\begin{tabular}{ |cc|cccc| }
\hline
 IMF & $[M_i-M_s]$ & $\alpha$ & $\frac{\zeta_{[1-2]}}{\alpha}$
 & $\frac{\zeta_9}{\alpha}$ & $y_O$ \\
\hline
Salpeter     &  [0.1--100] & 0.70 & 0.15 & 0.19 & 1.23E-02 \\ 
Salpeter     &  [0.1--70]  & 0.71 & 0.15 & 0.17 & 1.16E-02 \\
Salpeter     & [0.09--100] & 0.71 & 0.14 & 0.17 & 1.16E-02 \\
\hline
Kroupa       &  [0.1--100] & 0.71 & 0.23 & 0.10 & 7.05E-03 \\
Kroupa       &  [0.2--100] & 0.66 & 0.28 & 0.13 & 8.57E-03 \\
Kroupa       &  [0.3--100] & 0.62 & 0.34 & 0.15 & 1.03E-02 \\
Kroupa       & [0.35--100] & 0.60 & 0.37 & 0.17 & 1.12E-02 \\
Kroupa       &  [0.4--100] & 0.58 & 0.40 & 0.18 & 1.23E-02 \\
\hline
Kennicutt    &  [0.1--100] & 0.56 & 0.34 & 0.27 & 1.82E-02 \\
Kennicutt    &    [0--100] & 0.61 & 0.28 & 0.22 & 1.47E-02 \\
Kennicutt    &  [0.1--35]  & 0.58 & 0.35 & 0.19 & 1.51E-02 \\
Kennicutt    &  [0.1--30]  & 0.59 & 0.35 & 0.18 & 1.39E-02 \\
Kennicutt    & [0.05--35]  & 0.60 & 0.32 & 0.18 & 1.39E-02 \\
\hline
Larson       & [0.01--100] & 0.49 & 0.31 & 0.48 & 3.16E-02 \\
Larson       & [0.01--23]  & 0.55 & 0.32 & 0.24 & 1.75E-02 \\
Larson       & [0.01--22]  & 0.55 & 0.32 & 0.23 & 1.64E-02 \\
\hline
mod.\ Larson & [0.01--100] & 0.63 & 0.28 & 0.17 & 1.13E-02 \\
mod.\ Larson &  [0.2--100] & 0.60 & 0.31 & 0.19 & 1.27E-02 \\
\hline
Chabrier     & [0.01--100] & 0.59 & 0.26 & 0.24 & 1.62E-02 \\
Chabrier     & [0.01--33]  & 0.61 & 0.26 & 0.18 & 1.36E-02 \\
Chabrier     & [0.01--32]  & 0.61 & 0.26 & 0.17 & 1.34E-02 \\
\hline
\end{tabular}
\caption{Characteristic quantities for the IMFs adopted in the models of 
Table~\protect{\ref{tab:models}} (see text).
$\alpha$ is the locked--up fraction, $\zeta_{[1-2]}$ is the mass fraction 
in the [1-2]~\Msol\ range, $\zeta_9$ is the mass fraction in stars more massive
than 9~\Msol. $y_O$ is the global oxygen yield of the IMF.}
\label{tab:IMFyields}
\end{table}

In the light of our results for the M/L ratio, we suggest a gross description
of the IMF based on three mass ranges. (1) Ever-living stars 
($M \lsim$0.8--1~\Msol) contribute to the locked--up fraction; their 
contribution to luminosity becomes important only for very old systems, where
they are not much smaller than the typical ``turn-off mass''.
(2) Intermediate living stars ($M \sim$1--9~\Msol) contribute an important 
amount of recycled gas (for a Salpeter IMF, $\sim$2/3 of the total returned 
gas fraction) and dominate the luminosity; depending on the observational band
considered, the ``important'' sub--range for the luminosity may change:
e.g.\ [1--2]~\Msol\ for the I--band, [1--9]~\Msol\ for the B--band. 
(3) Short--lived stars ($M>$9~\Msol) contribute part of the recycled gas,
part of the locked--up fraction (through the remnants) and
most of the metals (oxygen in particular); their luminosity contribution
is dominant only for blue and UV bands, and/or for very young systems 
or systems dominated by a recent burst. We prefer here the terminology 
``ever, intermediate and short living'' stars rather than 
``low, intermediate and high mass'' stars, to distinguish this classification 
from the mass ranges classified on the base of stellar evolution properties
(Iben 1991; Chiosi, Bertelli \& Bressan 1992).

Broadly speaking, it is the relative
proportion of the mass fractions in these three ranges that determines 
the expected chemo-photometric properties of the system. Since for the
I--band, the mass range dominating the luminosity differs from the
mass range responsible for the oxygen production, it is easy to understand
how we could to tune the yields through the IMF upper mass limits 
and reproduce the observed gas--to--luminosity fraction, with negligible 
effects on the $M/L_I$ ratio (\S\ref{sect:overview}; see also Tsujimoto 
\etal 1997).
However, this dichotomy may not hold 
when one consider other bands (for instance blue or UV bands more sensitive 
to the contribution of massive stars) or other chemical elements (for instance 
helium, carbon, nitrogen and iron, produced in non--negligible amounts by 
``intermediate living'' stars --- including SN~Ia). On the contrary, one can
envisage a connection between the M/L ratio in certain bands and the 
production of certain elements. So for a full account 
of the chemo-photometric evolution of galaxies, it is important to maintain 
consistency between the computed luminosity and chemical outputs 
(Marigo \& Girardi 2001).

\subsection{The total baryonic M/L ratio}
\label{sect:baryonML}
In this section we discuss the global baryonic M/L ratio,
including the mass in gas, for those models of Sbc/Sc discs (large $b$ values)
that also comply with the constraint on the 
gas--to--luminosity ratio (\S\ref{sect:chem-constraint}) --- and hence have 
the gas amount indicated by observations.

With the Salpeter IMF, the typical global baryonic M/L ratio
in Sbc/Sc discs is $\gsim$2~\Msol/\Lsol\ (Table~\ref{tab:baryonML}, column 5), 
and the mass in gas is about 25\% of the total baryonic mass (column 6). 
With the bottom 
light IMFs, the gas contribution is more important since less mass is stored
in the stellar component, for the same luminosity. With the adopted
gas--to--luminosity ratio, the gas mass typically accounts
for 40\% of the total baryonic mass in Sbc/Sc discs, and the global
baryonic M/L ratio is 1.2--1.5~\Msol/\Lsol.

A gas fraction of 40\% may seem exceedingly large, but here we are dealing 
with late--type spirals which are notoriously gas--rich (Roberts \& Haynes 
1994; Sommer--Larsen 1996). Besides, the gas fraction is a strong function
of radius: in the inner parts, say within $R <1 h_B$, our 
Sbc/Sc models typically have a gas fraction less than 20\%, increasing
at higher radius. So, while the gas mass is relevant in computing
the global baryonic mass in the disc, in the inner regions the baryonic content
is definitely dominated by the stellar component.

Besides, the face-value gas fractions we find depend on the adopted 
gas--to--luminosity ratio, suggested from observations yet rather uncertain
(\S\ref{sect:chem-constraint}). This section is just meant to draw attention
on the fact that, with bottom--light IMFs, the gas mass is likely to contribute
significantly to the total baryonic mass in late--type discs, and it should 
not be neglected when estimating the global baryonic content. This may be 
crucial when discussing the physical nature of the Tully--Fisher relation,
since over a wide enough dynamical range the rotation velocity appears to 
correlate better with the global baryonic (stars+gas) mass, rather than with 
the stellar mass only (McGaugh \etal 2000). When discussing the baryonic
TF relation, the M/L ratio ascribed to the stars and, correspondingly, 
the relative contribution of the gas mass have important consequences for 
the derived slope (McGaugh 2001, 2003). Lower--than--Salpeter M/L ratios,
as favoured in the present paper, imply a somewhat shallower TF relation 
(Bell \& de Jong 2001; de Jong \& Bell 2001).

\begin{table}
\begin{tabular}{ |llc|ccc| }
\hline
\multicolumn{1}{c}{$^{(1)}$} & \multicolumn{1}{c}{$^{(2)}$~~~~~} & $^{(3)}$ 
 & $^{(4)}$ & $^{(5)}$ & $^{(6)}$ \\
\multicolumn{1}{c}{Model} & \multicolumn{1}{c}{IMF~~~~~} & $b$ 
 & $\frac{M_*}{L_I}$ & $\frac{M_{bar}}{L_I}$ & $\frac{M_{gas}}{M_{bar}}$ \\ 
\hline
{\sf SalpD} & Salpeter      [0.1--70]  & 0.78 & 1.58 & 2.09 & 0.24 \\
{\sf SalpE} & Salpeter     [0.09--100] & 0.79 & 1.67 & 2.23 & 0.25 \\
\hline						      	
{\sf KrouD} & Kroupa       [0.35--100] & 0.73 & 0.73 & 1.12 & 0.35 \\
{\sf KrouE} & Kroupa        [0.4--100] & 0.84 & 0.66 & 1.06 & 0.38 \\
\hline						      	
{\sf KennG} & Kennicutt     [0.1--30]  & 0.78 & 0.74 & 1.22 & 0.40 \\
{\sf KennH} & Kennicutt     [0.1--35]  & 0.97 & 0.69 & 1.21 & 0.43 \\
{\sf KennI} & Kennicutt    [0.05--35]  & 0.81 & 0.79 & 1.35 & 0.42 \\
{\sf KennJ} & Kennicutt    [0.05--35]  & 0.90 & 0.76 & 1.23 & 0.39 \\
\hline						      	       
{\sf LarsC} & Larson       [0.01--22]  & 0.81 & 0.78 & 1.37 & 0.43 \\
{\sf LarsD} & Larson       [0.01--22]  & 0.83 & 0.79 & 1.28 & 0.39 \\
{\sf LarsE} & Larson       [0.01--23]  & 0.95 & 0.74 & 1.30 & 0.43 \\
\hline						      	
{\sf LmodD} & mod.\ Larson  [0.2--100] & 0.86 & 0.82 & 1.36 & 0.40 \\
{\sf LmodE} & mod.\ Larson  [0.2--100] & 0.90 & 0.80 & 1.30 & 0.39 \\
{\sf LmodF} & mod.\ Larson  [0.2--100] & 0.93 & 0.80 & 1.26 & 0.36 \\
\hline						      	       			  
{\sf ChabD} & Chabrier     [0.01--32]  & 0.83 & 0.94 & 1.53 & 0.38 \\
{\sf ChabE} & Chabrier     [0.01--32]  & 0.87 & 0.91 & 1.45 & 0.37 \\
{\sf ChabF} & Chabrier     [0.01--33]  & 0.92 & 0.90 & 1.42 & 0.36 \\
\hline
\end{tabular}
\caption{Total baryonic M/L ratio of the model Sbc/Sc discs from
Table~\protect{\ref{tab:models}}, including the mass in gas.
(1) Model name. 
(2) Adopted IMF with corresponding mass limits.
(3) $b$--parameter of the SFH (Eq.~\protect{\ref{eq:b-psi}}).
(4) M/L ratio in the I--band of the stellar component (comprehensive of 
    living stars and stellar remnants).
(5) Total baryonic M/L ratio including the gas mass: $M_{bar}=M_*+M_{gas}$.
(6) Gas mass fraction in the baryonic disc.}
\label{tab:baryonML}
\end{table}

\subsection{An indirect comparison to the Solar Neighbourhood}
\label{sect:tsujimoto}
Tsujimoto \etal (1997, hereinafter T97) performed an analogous study 
to our present one, combining chemical evolution and M/L ratio constraints 
for the Solar Neighbourhood to determine a three--parameters IMF.
Although their study
was similar in spirit to ours, it is not trivial to compare
results, both because of substantial differences in the models (in  
stellar yields, SN~Ia rate formalism, SF law and photometric computations)
and because of the different approach (detailed chemical patterns and star 
counts in the Solar Neighbourhood, versus global average properties of external
Sbc/Sc disc galaxies). Still, 
a comparison can be interesting and we shall broadly outline it.

T97 considered single--slope IMFs with varying mass limits
so that their set of IMF parameters is ($M_i$, $M_s$, $x$) where $x$ is the
power--law slope.
A direct comparison with our results is possible
for the ``power--law IMFs'' considered in this paper (Salpeter, Kroupa and
Kennicutt): although the Kroupa and Kennicutt IMFs are not single--slope,
the slope changes only below 1~\Msol where what matters is effectively the 
global mass in low--mass stars, rather than the detailed mass distribution
(T97 and \S\ref{sect:Kroupa_models}). For our ``best models'' of Sbc/Sc discs 
listed in Table~\ref{tab:baryonML}, with the right gas--to--luminosity ratio
(i.e.\ level of metal  enrichment), we can compute the ``equivalent'' value
of $M_i$ that, for a single--slope IMF, would result in the same mass fraction 
below 1~\Msol\ ($M_{i,eq}$ in Table~\ref{tab:tsujimoto}).
We also compare the
mass fractions within the three significative mass ranges discussed
in \S\ref{sect:driving}: ``ever--living'' ($M<1$~\Msol), 
intermediate--living ($M=1-9$~\Msol) and short--lived ($M>1$~\Msol) stars. 
In Table~\ref{tab:tsujimoto} we list the various IMF parameters
as constrained by T97 and in this paper. For brevity, we consider the 
highest values of $M_s$ and of $M_i$ allowed in Fig.~3 of T97; considering
their lower values somewhat increases the discrepancy with our results 
(see below), but does not change the outcome of this qualitative comparison. 

For each of the IMFs in 
Table~\ref{tab:tsujimoto}, our results favour smaller low--mass fractions 
$\zeta_{<1}$, and higher fractions both in the intermediate [1--9]~\Msol\
and in the high $>9$~\Msol\ mass range. This implies both a higher
metal enrichment (roughly driven by $\zeta_{>9}/\alpha$ or 
$\zeta_{>9}/\zeta_{<1}$) and a lower M/L ratio (roughly driven by  
$\zeta_{[1-9]}/\zeta_{<1}$) in our models.
The higher level of metal enrichment might be related to the fact that
the gas--to--luminosity ratio of external spirals is certainly
a more uncertain and loose constraint on chemical evolution than the
detailed abundance patterns in the Solar Neighbourhood considered by T97.
For instance, a very recent compilation (Bettoni, Galletta \& 
Garc{\'\i}a--Burillo 2003) seems to favour 
lower gas fractions ($M_{HI}/L_B \sim 0.2$ and $M_{H_2}/L_B \sim 0.1$)
than we adopted in \S\ref{sect:chem-constraint}, which would correspond
to a lower efficiency of metal enrichment for external spirals. In the context
of our models, this could be obtained via a different tuning of the upper
mass limit $M_s$ of the IMF to reduce the characteristic yield, but as 
discussed in \S\ref{sect:overview} and \S\ref{sect:driving} the resulting 
M/L ratios would hardly change.

As to the different M/L ratio between T97 and this paper, it is difficult 
to understand whether the discrepancy is due to model differences or to the 
different approach (Solar Neighbourhood vs.\ external discs). The difference 
is not simply accounted for by different SFHs, since both the Solar 
Neighbourhood in T97 and the Sbc/Sc discs here have been modelled with 
the respective suitable SFHs. Possibly, while chemical evolution is much 
better constrained in the Solar Neighbourhood, the estimate of the local
luminosity from star counts (T97) is more cumbersome than the study of the 
global luminosity in external galaxies. Besides, dark matter may have a minor, 
but non--zero contribution to the local dynamical M/L ratio,
easing the M/L constraints adopted by T97. Alternatively, the discrepancy
might be resolved when, for the M/L ratio of external galaxies, allowance 
is made for the dispersion and the uncertainty in the zero--point of the 
TF relation (Appendix~A).

The issue deserves further investigation, but this is beyond the purpose 
of this paper. Reasonably, one should first assess whether there is an
effective discrepancy in the observationally deduced M/L ratio and IMF,
free from modelling issues. The two environments, the Solar Neighbourhood
and outer disc galaxies, should be analyzed with the same models so that
a significant comparison can be made.
We postpone to future work the study of the M/L ratio
and IMF in the Solar Neighbourhood with our chemo--photometric models.

\begin{table*}
\begin{tabular}{ |cc|ccccc|ccccc| }
\hline
 IMF & $x$ & $M_i^{\rm T97}$ & $M_s^{\rm T97}$ & $\zeta_{<1}^{\rm T97}$ & 
$\zeta_{[1-9]}^{\rm T97}$ & $\zeta_{>9}^{\rm T97}$ & $M_{i, eq}^{\rm PST}$ & 
$M_s^{\rm PST}$ & $\zeta_{<1}^{\rm PST}$ & $\zeta_{[1-9]}^{\rm PST}$ & 
$\zeta_{>9}^{\rm PST}$\\
\hline
Salpeter  & 1.35 & 0.05 & 50 & 0.71 & 0.21 & 0.08 & 0.09--0.1  & 70--100 & 0.62 & 0.26 & 0.12 \\
Kennicutt & 1.5  & 0.14 & 55 & 0.66 & 0.26 & 0.08 & 0.35--0.38 & 30--35  & 0.44 & 0.45 & 0.11 \\
  ---     & 1.6  & 0.17 & 58 & 0.68 & 0.26 & 0.06 &     ---    &   ---   &  --- & ---  & ---  \\
Kroupa(*) & 1.7  & 0.23 & 64 & 0.65 & 0.29 & 0.06 & 0.44--0.47 &   100   & 0.44 & 0.46 & 0.10 \\
\hline
\end{tabular}
\caption{Comparison between IMFs as constrained by T97 and in the present 
paper (PST). $M_i$ and $M_s$ are the mass limits of the IMF of slope $x$;
$\zeta_{<1}$, $\zeta_{[1-9]}$ and $\zeta_{>9}$ are the corresponding mass 
fractions in the three representative mass ranges. $M_i^{eq}$(PST) is the 
``single--slope equivalent'' lower mass limit for our IMFs 
{\mbox{(see text).~~~~~~~~~~~~~~~~~~~~~~~~~~~~~~~~~~~~~~~~~~~~~~~~~~~~~~~~~~~}}
{\mbox{(*)~~~Slopes $x>1.6$ are formally ruled out by T97,}}
because for steep slopes they find that no value of $M_i$
can fulfill both the chemical and the M/L ratio constraints.
Still, for qualitative comparison with our Kroupa models we consider 
their case $x=1.7$ with $M_i=0.23$, an intermediate value between those
allowed by chemical and M/L ratio arguments (Fig.~3 in T97).}
\label{tab:tsujimoto}
\end{table*}

\section{The bulge contribution}
\label{sect:bulges}
In this section we attempt to give a rough estimate of the contribution
of the bulge to the global M/L ratio of disc galaxies. Bulges are tendentially
older and dimmer than discs, although in some cases
a continuity of properties (colours etc.) is seen between the bulge
and the inner regions of the disc, especially for late--type spirals 
(Balcells 2003 and references therein).
A lower limit to the global M/L ratio is hence given by considering
the central regions as the extrapolation of the disc properties to the center,
which in included in our models. As described in \S\ref{sect:calibration},
from the analysis of our disc models we had excluded the inner regions 
($R\leq 0.5~h_B$); including them yields the M/L ratios in parenthesis 
in column~4 
of Table~\ref{tab:bulges}. We limit the list to models with large
$b$ parameters, as representative of Sbc/Sc spirals.
The inclusion of the central disc regions induces a minor increase 
(at most 10\%) in the global M/L ratio. In particular, for all the 
``bottom--light'' IMFs considered, the M/L ratio remains below 1 
for the Sbc/Sc models.

An upper limit to the bulge M/L ratio is given by the characteristic
M/L ratio of an old SSP, neglecting any possible tails of later star 
formation and younger stellar components --- expected among others 
because of gas recycling from the first population, and detected for instance 
in the bulge of the Milky Way (Ng \etal 1995; van Loon \etal 2003). 
We consider SSPs of age 
10~Gyr, as characteristic of the bulges of early--type spirals (Peletier 
\etal 1999); this is likely to overestimate the M/L ratio of 
the bulges of Sbc/Sc galaxies, which tend to be younger (Peletier \etal
1999; Carollo, Ferguson \& Wyse 1999). We consider SSPs of solar metallicity 
for simplicity. 
In the Bulge of our galaxy, the bulk of stars is between half--solar and solar 
metallicity (Mc William \& Rich 1994; Zoccali \etal 2003) with a super--metal 
rich tail (Bertelli \etal 1995);
varying the metallicity 
by a factor of 2 would change the bulge M/L ratio in the I--band by 15\% or so,
negligible for the sake of the rough estimates we are giving here.
Column~5 of Table~\ref{tab:bulges} lists the I--band M/L ratios of the stellar 
component (i.e.\ stars+remnants, excluding the re-ejected gas) of 10~Gyr old,
solar SSPs with the same IMF as in the disc models.

\begin{figure}
\centerline{\psfig{file=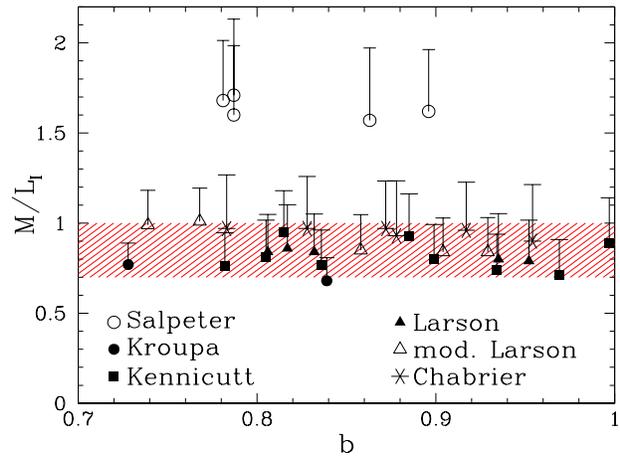,angle=270,width=8.5truecm}}
\caption{Estimated global M/L ratio of disc galaxies, including the 
contribution of the bulge. Dots represent the lower limit estimate (treating 
the bulge as an extension of the disc properties to the centre), bars indicate
the upper limit estimate (treating the bulge as a 10~Gyr old SSP contributing 
25\% of the total I--band luminosity). The shaded area marks the range 
M/L$_I$=0.7--1 favoured by observations of Sbc--Sc spirals.}
\label{fig:bulges}
\end{figure}

\begin{table*}
\begin{center}
\begin{tabular}{ |llccc|ccc|ccc|ccc| }
\hline
\multicolumn{1}{c}{$^{(1)}$} & \multicolumn{1}{c}{$^{(2)}$~~~~~} & $^{(3)}$ 
 & \multicolumn{2}{c}{$^{(4)}$} & $^{(5)}$ & $^{(6)}$ & $^{(7)}$ & $^{(8)}$ 
 & $^{(9)}$ & $^{(10)}$ \\
\multicolumn{1}{c}{Model} & \multicolumn{1}{c}{IMF~~~~~} & $b$ 
 & \multicolumn{2}{c}{$\frac{M_*}{L_I} (\cal D)$}
 & $\frac{M_*}{L_I} (\cal B)$ & $\frac{M_* (\cal B)}{M_* (\cal D)}$ 
 & $\frac{M_*}{L_I}(tot)$ & $\frac{M_*}{L_K} (\cal D)$ 
 & $\frac{M_*}{L_K} (\cal B)$
 & $\left( \cal \frac{B}{D} \right)_K$ \\
\hline
{\sf SalpA} & Salpeter      [0.1--100] & 0.79 & 1.54 & (1.60) & 3.77 & 0.61 & 1.99 & 0.67 & 1.31 & 0.31 \\ 
{\sf SalpB} & Salpeter      [0.1--100] & 0.86 & 1.52 & (1.57) & 3.77 & 0.62 & 1.97 & 0.65 & 1.31 & 0.31 \\ 
{\sf SalpC} & Salpeter      [0.1--100] & 0.90 & 1.51 & (1.62) & 3.77 & 0.62 & 1.96 & 0.64 & 1.31 & 0.30 \\ 
{\sf SalpD} & Salpeter      [0.1--70]  & 0.78 & 1.58 & (1.68) & 3.76 & 0.59 & 2.02 & 0.66 & 1.31 & 0.30 \\ 
{\sf SalpE} & Salpeter     [0.09--100] & 0.79 & 1.67 & (1.71) & 3.99 & 0.60 & 2.13 & 0.70 & 1.38 & 0.30 \\ 
\hline						      	
{\sf KrouD} & Kroupa       [0.35--100] & 0.73 & 0.73 & (0.77) & 1.53 & 0.52 & 0.89 & 0.31 & 0.53 & 0.31 \\ 
{\sf KrouE} & Kroupa        [0.4--100] & 0.84 & 0.66 & (0.68) & 1.41 & 0.53 & 0.81 & 0.28 & 0.49 & 0.31 \\ 
\hline						      	
{\sf KennA} & Kennicutt     [0.1--100] & 0.84 & 0.75 & (0.77) & 1.80 & 0.60 & 0.96 & 0.33 & 0.61 & 0.32 \\ 
{\sf KennB} & Kennicutt     [0.1--100] & 0.93 & 0.73 & (0.74) & 1.80 & 0.62 & 0.94 & 0.32 & 0.61 & 0.32 \\ 
{\sf KennC} & Kennicutt     [0.1--100] & 1.02 & 0.71 & (0.72) & 1.80 & 0.63 & 0.93 & 0.31 & 0.61 & 0.32 \\ 
{\sf KennD} & Kennicutt       [0--100] & 0.82 & 0.92 & (0.95) & 2.23 & 0.61 & 1.18 & 0.35 & 0.76 & 0.28 \\ 
{\sf KennE} & Kennicutt       [0--100] & 0.89 & 0.90 & (0.93) & 2.23 & 0.62 & 1.17 & 0.34 & 0.76 & 0.28 \\ 
{\sf KennF} & Kennicutt       [0--100] & 1.00 & 0.87 & (0.89) & 2.23 & 0.64 & 1.14 & 0.33 & 0.76 & 0.29 \\ 
{\sf KennG} & Kennicutt     [0.1--30]  & 0.78 & 0.74 & (0.76) & 1.78 & 0.60 & 0.95 & 0.31 & 0.61 & 0.31 \\ 
{\sf KennH} & Kennicutt     [0.1--35]  & 0.97 & 0.69 & (0.71) & 1.79 & 0.65 & 0.91 & 0.30 & 0.61 & 0.32 \\ 
{\sf KennI} & Kennicutt    [0.05--35]  & 0.81 & 0.79 & (0.81) & 1.94 & 0.61 & 1.02 & 0.34 & 0.66 & 0.31 \\ 
{\sf KennJ} & Kennicutt    [0.05--35]  & 0.90 & 0.76 & (0.80) & 1.94 & 0.64 & 1.00 & 0.32 & 0.66 & 0.31 \\ 
\hline						      	       
{\sf LarsA} & Larson       [0.01--100] & 0.82 & 0.84 & (0.86) & 2.16 & 0.64 & 1.10 & 0.37 & 0.74 & 0.32 \\ 
{\sf LarsB} & Larson       [0.01--100] & 0.94 & 0.78 & (0.80) & 2.16 & 0.69 & 1.06 & 0.34 & 0.74 & 0.32 \\ 
{\sf LarsC} & Larson       [0.01--22]  & 0.81 & 0.78 & (0.84) & 2.11 & 0.68 & 1.05 & 0.33 & 0.72 & 0.31 \\ 
{\sf LarsD} & Larson       [0.01--22]  & 0.83 & 0.79 & (0.84) & 2.11 & 0.67 & 1.05 & 0.33 & 0.72 & 0.31 \\ 
{\sf LarsE} & Larson       [0.01--23]  & 0.95 & 0.74 & (0.79) & 2.11 & 0.71 & 1.01 & 0.31 & 0.72 & 0.31 \\ 
\hline						      	
{\sf LmodB} & mod.\ Larson [0.01--100] & 0.74 & 0.93 & (0.99) & 2.18 & 0.59 & 1.18 & 0.40 & 0.75 & 0.31 \\ 
{\sf LmodC} & mod.\ Larson [0.01--100] & 0.77 & 0.95 & (1.01) & 2.18 & 0.57 & 1.20 & 0.40 & 0.75 & 0.30 \\ 
{\sf LmodD} & mod.\ Larson  [0.2--100] & 0.86 & 0.82 & (0.85) & 1.96 & 0.60 & 1.05 & 0.35 & 0.75 & 0.28 \\ 
{\sf LmodE} & mod.\ Larson  [0.2--100] & 0.90 & 0.80 & (0.84) & 1.96 & 0.61 & 1.03 & 0.34 & 0.68 & 0.31 \\ 
{\sf LmodF} & mod.\ Larson  [0.2--100] & 0.93 & 0.80 & (0.84) & 1.96 & 0.61 & 1.03 & 0.34 & 0.68 & 0.31 \\ 
\hline						      	       			  
{\sf ChabA} & Chabrier     [0.01--100] & 0.78 & 0.95 & (0.97) & 2.54 & 0.67 & 1.27 & 0.41 & 0.87 & 0.32 \\ 
{\sf ChabB} & Chabrier     [0.01--100] & 0.88 & 0.91 & (0.93) & 2.54 & 0.70 & 1.24 & 0.39 & 0.87 & 0.32 \\ 
{\sf ChabC} & Chabrier     [0.01--100] & 0.95 & 0.88 & (0.90) & 2.54 & 0.72 & 1.21 & 0.38 & 0.87 & 0.31 \\ 
{\sf ChabD} & Chabrier     [0.01--32]  & 0.83 & 0.94 & (0.97) & 2.53 & 0.67 & 1.26 & 0.40 & 0.87 & 0.31 \\ 
{\sf ChabE} & Chabrier     [0.01--32]  & 0.87 & 0.91 & (0.97) & 2.53 & 0.70 & 1.23 & 0.38 & 0.87 & 0.31 \\ 
{\sf ChabF} & Chabrier     [0.01--33]  & 0.92 & 0.90 & (0.96) & 2.53 & 0.70 & 1.23 & 0.38 & 0.87 & 0.31 \\ 
\hline
\end{tabular}
\end{center}
\caption{Estimated bulge contribution, total (M/L)$_I$ ratios and 
K--band $\cal B/D$ ratios for the Sbc/Sc models of 
Table~\protect{\ref{tab:models}}.
(1) Model name. 
(2) Adopted IMF with corresponding mass limits.
(3) $b$--parameter of the SFH (Eq.~\protect{\ref{eq:b-psi}}).
(4) M/L ratio in the I--band of the stellar component in the disc; 
    in parenthesis the M/L ratio when considering the disc extended 
    to the galactic center.
(5) Stellar M/L$_I$ ratio of a SSP of solar metallicity and age 10~Gyr, 
    representing the bulge, with the IMF indicated in column~2.
(6) Actual mass ratio between the disc and a bulge with I--band luminosity 25\%
    of that of the disc.
(7) Global galactic M/L$_I$ ratio assuming a 10~Gyr old bulge with 
    ${\cal(B/D)}_I$=25\%.
(8) M/L ratio in the K--band of the stellar component in our disc models.
(9) Stellar M/L$_K$ ratio of a SSP of solar metallicity and age 10~Gyr, 
    representing the bulge, with the IMF indicated in column~2.
(10) $\cal B/D$ luminosity ratio in the K--band corresponding to the mass 
     ratio in column~6, or equivalently to the adopted ${\cal (B/D)}_I$=25\%}.
\label{tab:bulges}
\end{table*}

Typically, the bulge--to--disc luminosity ratio in the I--band,
${\cal (B/D)}_I$, is around 25\% for Sbc/Sc spirals,
although with a large scatter (Sommer--Larsen \etal 2003). 
Assuming ${\cal (B/D)}_I$=0.25 and the respective M/L ratios 
for disc and bulge,
results in mass ratios between 50 and 70\% (column~6).
This illustrates that the luminosity ratio, even in the red bands, is not 
a direct indicator of the mass ratio of the two components: allowance must be 
made for the different M/L ratio between an old bulge and the disc.

With the adopted ${\cal (B/D)}_I$=25\% and the respective M/L ratios, 
we estimate the global galactic M/L ratios in column~7 of 
Table~\ref{tab:bulges}. For the bottom--light IMFs, $M/L_I(tot) = 1-1.2$ 
(see also Fig.~\ref{fig:bulges}). These values are strict upper limits,
for two reasons. Firstly, as mentioned above, bulges in Sbc/Sc spirals seem
to be younger and/or host a younger component than the assumed 10~Gyr, so
their characteristic M/L ratio is expected to be lower than in column~5.
Secondly, we are likely overestimating here the luminosity contribution from
the bulge: the typical ${\cal (B/D)}_I \sim$0.25 in Sommer-Larsen \etal (2003) 
are in fact derived from 2D light decompositions assuming a $r^\frac{1}{4}$ 
profile for the bulge (Byun 1992). Nowadays, exponential light profiles or 
shallower Sersic profiles are favoured for bulges, especially in late type 
spirals (de Jong 1996a; Carollo \etal 1999, 2001; Balcells 2003); 
with an exponential
profile, the decomposition yields significantly lower $\cal B/D$ ratios 
(de Jong 1996a) and the bulge contribution to the total M/L ratio 
is correspondingly less.
Notice also that the global (B-V)$_0$ colours in 
Table~\ref{tab:models}, obtained for the global disc (extended to the center)
are in good agreement with the observed {\mbox{(B-V)$\sim$0.55}} of Sbc/Sc 
spirals
(Roberts \& Haynes 1994); hence the contribution of an old, red bulge cannot be
very large. 

In conclusion, since bulges in Sbc/Sc spirals seem to display, at odds with 
earlier types, a continuity of properties with the inner regions of discs 
(Carollo \etal 1999);
the true global M/L ratio of an Sbc/Sc spiral should lie 
somewhere between the alternative values given in Table~\ref{tab:bulges} and 
in Fig~\ref{fig:bulges}.

\begin{table*}
\begin{tabular}{c c l c c c c c c c|c c c}
\hline
$^{(1)}$ & $^{(2)}$ & \multicolumn{1}{c}{$^{(3)}$} & $^{(4)}$ & $^{(5)}$ 
 & $^{(6)}$ & $^{(7)}$ & $^{(8)}$ & $^{(9)}$ & $^{(10)}$ & $^{(11)}$ 
 & $^{(12)}$ & $^{(13)}$ \\
Hubble type & $b$ & \multicolumn{1}{c}{IMF} & $\frac{M_*}{L_I} (\cal D)$ & 
$\frac{M_*}{L_I} (\cal B)$ & $\left( \cal \frac{B}{D} \right)_I$ & 
$\frac{M_* (\cal B)}{M_* (\cal D)}$ & 
$\frac{M_*}{L_I} (tot)$ & $\Delta (\frac{M_*}{L_I})$ & $\Delta (m_I)$ & 
$\frac{M_*}{L_K} (\cal D)$ & $\frac{M_*}{L_K} (\cal B)$ & 
$\left( \cal \frac{B}{D} \right)_K$ \\
\hline
Sa/Sab & 0.1  & Salpeter  & 2.70 & 3.77 & 0.45 & 0.63 & 3.03 & 1.54 & 0.47 & 0.99 & 1.31 & 0.48 \\
Sb     & 0.35 & Salpeter  & 2.09 & 3.77 & 0.35 & 0.63 & 2.53 & 1.28 & 0.27 & 0.80 & 1.31 & 0.38 \\
Sbc/Sc & 0.9  & Salpeter  & 1.52 & 3.77 & 0.25 & 0.62 & 1.97 & 1    & 0    & 0.61 & 1.31 & 0.29 \\
\hline
Sa/Sab & 0.1  & Kroupa    & 1.76 & 2.35 & 0.45 & 0.60 & 1.94 & 1.43 & 0.39 & 0.64 & 0.82 & 0.47 \\
Sb     & 0.35 & Kroupa    & 1.44 & 2.35 & 0.35 & 0.57 & 1.68 & 1.23 & 0.23 & 0.55 & 0.82 & 0.38 \\
Sbc/Sc & 0.9  & Kroupa    & 1.11 & 2.35 & 0.25 & 0.53 & 1.36 & 1    & 0    & 0.45 & 0.82 & 0.29 \\
\hline
Sa/Sab & 0.1  & Kennicutt & 1.29 & 1.80 & 0.45 & 0.63 & 1.45 & 1.41 & 0.37 & 0.47 & 0.61 & 0.48 \\
Sb     & 0.35 & Kennicutt & 1.02 & 1.80 & 0.35 & 0.62 & 1.22 & 1.26 & 0.25 & 0.39 & 0.61 & 0.39 \\
Sbc/Sc & 0.9  & Kennicutt & 0.76 & 1.80 & 0.25 & 0.59 & 0.97 & 1    & 0    & 0.31 & 0.61 & 0.30 \\
\hline
Sa/Sab & 0.1  & Larson(1) & 1.52 & 2.16 & 0.45 & 0.64 & 1.72 & 1.62 & 0.52 & 0.55 & 0.74 & 0.48 \\
Sb     & 0.35 & Larson(1) & 1.13 & 2.16 & 0.35 & 0.67 & 1.40 & 1.31 & 0.29 & 0.43 & 0.74 & 0.39 \\
Sbc/Sc & 0.9  & Larson(1) & 0.79 & 2.16 & 0.25 & 0.69 & 1.06 & 1    & 0    & 0.32 & 0.74 & 0.29 \\
\hline
Sa/Sab & 0.1  & Larson(2) & 1.56 & 2.18 & 0.45 & 0.63 & 1.75 & 1.50 & 0.44 & 0.57 & 0.75 & 0.47 \\
Sb     & 0.35 & Larson(2) & 1.22 & 2.18 & 0.35 & 0.62 & 1.47 & 1.26 & 0.25 & 0.47 & 0.75 & 0.38 \\
Sbc/Sc & 0.9  & Larson(2) & 0.91 & 2.18 & 0.25 & 0.60 & 1.17 & 1    & 0    & 0.37 & 0.75 & 0.29 \\
\hline
Sa/Sab & 0.1  & Chabrier  & 1.72 & 2.54 & 0.45 & 0.67 & 1.97 & 1.59 & 0.51 & 0.62 & 0.87 & 0.48 \\
Sb     & 0.35 & Chabrier  & 1.29 & 2.54 & 0.35 & 0.69 & 1.61 & 1.30 & 0.29 & 0.49 & 0.87 & 0.39 \\
Sbc/Sc & 0.9  & Chabrier  & 0.91 & 2.54 & 0.25 & 0.70 & 1.24 & 1    & 0    & 0.36 & 0.87 & 0.29 \\
\hline
\end{tabular}
\caption{Total M/L ratios, M/L and magnitude offsets as a function of Hubble
type, including the contribution of the bulge.
(1) Hubble type.
(2) Typical $b$--parameter for the disc.
(3) Adopted IMF (mass limits as in \S\protect{\ref{sect:IMFlist}}).
(4) M/L ratio in the I--band for the disc, from the simple models in 
    \S\protect{\ref{sect:b-models}}.
(5) M/L ratio in the I--band for the bulge, considered as a SSP of solar 
    metallicity and age 10~Gyr.
(6) $\cal B/D$ luminosity ratio in the I--band.
(7) Actual $\cal B/D$ mass ratio.
(8) Global M/L ratio in the I--band for the galaxy.
(9) Offset in M/L ratio with respect to Sbc/Sc galaxies.
(10) Offset in magnitude with respect to Sbc/Sc galaxies.
(11) M/L ratio in the K--band for the disc (from the simple models in 
     \S\protect{\ref{sect:b-models}}).
(12) M/L ratio in the K--band for the bulge (considered as a SSP of solar 
     metallicity and age 10~Gyr).
(13) $\cal B/D$ luminosity ratio in the K--band corresponding to the mass 
     ratio in column~7, or equivalently to the adopted ${\cal (B/D)}_I$
     in column~6.}
\label{tab:bulge-Hubble}
\end{table*}

\subsection{M/L offsets with Hubble type: the bulge contribution}
\label{sect:bulge-Hubble}
In this section we recompute the offsets in stellar M/L ratios -- 
and correspondingly in the TF relation -- as a function of Hubble type, 
discussed in \S\ref{sect:Hubbledependence}, including the effect of 
bulges. 

The $b$--parameter we used as Hubble type indicator, is an indicator 
of the SFH in the disc (see \S\ref{sect:Hubble}), hence the offsets
in stellar M/L ratios discussed in \S\ref{sect:Hubbledependence} are 
essentially due to the differences in disc properties. However, the bulge
also contributes to the global M/L ratio, possibly in a different fashion 
in different Hubble types. It is not clear beforehand how the bulge 
contribution might change the disc--based offsets in
\S\ref{sect:Hubbledependence}. On one hand, the $\cal B/D$ {\it luminosity} 
ratio increases toward earlier Hubble types; hence one might expect that
bulges in earlier Hubble types are more prominent and contribute more to 
increasing the global M/L ratio of galaxies; this would enhance the
systematic offset in M/L of earlier types with respect to Sbc/Sc.
On the other hand, in earlier Hubble types the stellar populations in the disc
{\it per se} are on average older and redder, so that the difference in M/L 
ratio between disc and bulge is less significant;
hence one might expect that including the bulge has a lesser impact on the
global M/L ratio for early types than for late types, reducing the net offsets.

In Table~\ref{tab:bulge-Hubble} we estimate the offsets in global M/L ratios
as a function of Hubble type, adding the effect of the bulge to the simple
disc models computed in \S\ref{sect:b-models}. The M/L ratios for
the discs (column 4) are taken from \S\ref{sect:b-models}, those of the bulges
(column 5) are derived from 10~Gyr old SSPs of solar metallicity. The
$\cal B/D$ 
ratios in the I--band as a function of Hubble type (column 6) are from
Sommer--Larsen \etal (2003). Computing the actual mass ratios (column 7)
shows that the $\cal B/D$ ratio in mass does not depend strongly
on Hubble type, hence the trends in $\cal B/D$  luminosity ratios are mostly
due to different stellar populations and SFHs in the discs --- as suggested by 
Kennicutt \etal (1994). For all Hubble types (and all IMFs) the typical
$\cal B/D$  
mass ratio is around 60\%. However, we remind that this estimates are derived
considering the bulges of all Hubble types to be similar: 10~Gyr old red 
objects. As discussed above, this is a good estimate for Sa to Sb galaxies,
less so for Sbc/Sc galaxies; if the latter have younger bulges, the mass
$\cal B/D$ 
ratio may still have a real trend with Hubble type. Besides, what we are 
discussing here are gross average trends, but the scatter in $\cal B/D$
ratios is known to be large (de Jong 1996a; Sommer--Larsen \etal 2003 
and references therein). In particular, bulge--less disc galaxies are also
observed, in particular among late Hubble types.

Including the effect of the bulge, the typical stellar M/L ratio is $\sim$1.5
times larger in Sa/Sab galaxies than in Sbc/Sc galaxies (column 9). The
corresponding systematic offsets in magnitude with respect to the Sbc/Sc
zero point are about 0.25-0.3~mag for Sb galaxies and 0.4--0.5~mag for Sa/Sab
(column 10). In summary, the effect of the bulge is to {\it reduce} the
offsets expected from the sole differences in disc SFH (\S\ref{sect:b-models}).
The offsets however remain significant.

The present estimate is a lower limit to the expected offsets, since the
effect of bulges is maximized. As discussed in \S\ref{sect:bulges}, the
I--band $\cal B/D$ ratios from Sommer--Larsen \etal (2003) are likely to be 
overestimated, because derived from a decomposition performed with a 
$r^\frac{1}{4}$ profile --- hence in Table~\ref{tab:bulge-Hubble} we are 
overestimating the effects of bulges.
In addition, if bulges in Sbc/Sc galaxies are younger, the global M/L ratios
of these galaxies are lower than listed in Table~\ref{tab:bulge-Hubble}, and
the difference to earlier types becomes more significant.
All in all, the typical magnitude offsets should lie between the values in
Table~\ref{tab:bulge-Hubble} and the estimates given in \S\ref{sect:b-models}
from the pure disc contributions.

\subsection{The bulge--to--disc ratio in the K--band}
It is very common nowadays to assess issues like the stellar mass distribution 
in galaxies and the $\cal B/D$ ratio from observations in NIR bands, 
supposedly better mass tracers than blue or optical bands, since an important
contribution to the light in these bands comes from the underlying old
stellar component. We stressed in \S\ref{sect:AGB} the uncertainties
in estimating {\it absolute} M/L ratios in NIR bands, especially due
to the role of AGB stars, a complex phase with uncertain modelling in 
population synthesis. However, as an estimate of {\it relative} mass content,
NIR bands are still useful. For instance, in Fig.~\ref{fig:marigoML}
different treatments of the AGB phase (G-SSPs vs.\ M-SSPs) seem
to result in overall offsets of the K--band M/L ratios, while the trend
as a function of the $b$--parameter is preserved. Therefore comparing
the NIR luminosity of two objects, we still get an indication of
their stellar mass content in relative terms, while deriving their
absolute stellar mass from the M/L ratio would be more model--dependent.

Since a lot of focus is set nowadays on NIR observations and surveys
(e.g.\ 2MASS, DENIS), it is useful to present here the $\cal B/D$ 
ratios we predict in the K--band, and discuss to what extent the K--band
luminosity can be considered a direct tracer of stellar mass.

In Table~\ref{tab:bulges}, we list the K--band M/L ratios
for our model Sbc/Sc discs (column~8) and for the bulges, modelled 
as 10~Gyr old SSPs of solar metallicity (column~9). We verified that changing 
the typical metallicity of the bulge by a factor of two with respect to solar, 
would change the bulge M/L$_K$ ratio by less than 10\%. In column~10 we list
the resulting $\cal B/D$ ratios in the K--band
when the mass ratios from column~6 are assumed; these are the ${\cal (B/D)}_K$
ratios corresponding to the adopted ${\cal (B/D)}_I$=25\% in the I--band, 
for Sbc/Sc galaxies. In all cases, the typical ${\cal (B/D)}_K$=30\%, which
is in good agreement, within the large scatter, with the findings of 
de Jong (1996a) for Sbc/Sc galaxies in his $r^{\frac{1}{4}}$ bulge--disc 
decompositions in the K--band. Still, the K--band $\cal B/D$ ratio is not 
a direct indicator of the corresponding mass ratio, due to the 
typical differences in stellar populations and M/L ratio between bulge and
disc (cf.\ column~10 vs.\ column~6).

In Table~\ref{tab:bulge-Hubble}, columns~11 to~13, we list the K--band
M/L ratios for discs of different Hubble type (from the simple models in 
\S\protect{\ref{sect:b-models}}), the M/L ratios of bulges, and the
K--band $\cal B/D$ ratios corresponding to the mass ratios in column~7,
i.e.\ corresponding to the ${\cal (B/D)}_I$ ratios 
adopted in column~6. The typical $\cal B/D$ ratio in the K--band is 30\%
for late--type spirals, 40\% for Sb spirals, and 50\% for early type, 
Sa/Sab spirals. These values and trends with Hubble type are in excellent
agreement with de Jong (1996a) for his $r^{\frac{1}{4}}$ bulge--disc 
decompositions; however, we are speaking of average
trends but the observational scatter is huge (de Jong 1996a). For no
Hubble type, the K--band $\cal B/D$ ratio can be considered a direct indicator
of the corresponding mass ratio (cf.\ column~13 vs.\ column~7), so that 
to derive the $\cal B/D$ mass ratio one should always account for the different
stellar populations and M/L ratios between bulge and disc.

\section{Summary and conclusions}
We reviewed in the introduction some theoretical and observational evidence 
that the M/L ratio of the stellar component in late--type spirals should be
low: $M/L_I <$1 in the I--band. This paper is devoted to discussing whether
M/L ratios so low are compatible with current understanding of the stellar
populations in galaxies. We explained in \S 2 why we should concentrate on 
late type spirals like Sbc/Sc, characterized by a SFH with ``birthrate 
parameter'' $b \geq$0.8, and we argued the the I--band is the optimal band
for estimates of M/L ratios. In this study we considered the ``classic''
Salpeter IMF, mostly for the sake of comparison, and other more
``bottom--light'' IMFs favoured by independent studies: the Kroupa IMF,
the Kennicutt IMF, the Larson IMF (also in a modified version) and the Chabrier
IMF (\S 3). We calculated up-to-date sets of SSPs for a wide range of 
metallicities with these different IMFs, to compute the corresponding galactic
photometry (\S 4). A simple approach, based on one--zone models with 
exponential SFHs and fixed metallicity, was presented in \S 5
showing that, aside from the Salpeter IMF
which has a high M/L ratio, all the other IMFs imply $M/L_I \leq$1 for Sbc/Sc
galaxies. Our models also predict a significant difference in typical M/L
ratios --- hence in the zero--point of the TF--relation --- between spirals
of different Hubble type. We used our simple models to show that the predicted
M/L ratio in the I--band is quite robust with respect to uncertainties
in the current SSPs (mostly the contribution of the AGB phase), in contrast
to NIR bands. Furthermore, we assessed uncertainties related to spectral 
libraries.
We estimated our M/L ratios to be robust within a 10\% or better
for the typical metallicities of spiral galaxies, and to be conservatively 
high.

For a more realistic analysis, we developed chemo--photometric models with
gas infall and radial gradients in SFH and metallicity, to discuss M/L ratios
in connection with the chemical properties of spiral discs of 
Sbc/Sc type. 
In the overall the M/L ratios from full 
chemo--photometric models agree quite well with the predictions of simple 
exponential models, but
some further insight is obtained when considering the corresponding chemical
evolution. 

Due to the low locked--up fractions, some of the bottom--light IMFs 
(Kennicutt, Larson and Chabrier) result in high metal yields,
which are hardly compatible with the observed 
metallicities and gas fractions --- unless metallicities are heavily 
underestimated due to dust depletion, but this seems unlikely 
(\S\ref{sect:overview}). Either we reduce the typical metal production
(by means of suitable tuning of the upper mass 
limits of the IMF, or by allowing for fallback of metals onto a black hole 
after the SN explosion), or we need to invoke dispersal of metals from disc 
galaxies into the intergalactic medium,
a behaviour reminiscent of cluster galaxies enriching the intra--cluster 
medium. 
In any case, the resulting M/L ratio for Sbc/Sc discs is 
$M/L_I \sim$0.7--0.9. 

In other cases (Kroupa and modified--Larson IMFs), the slope 
of the IMF at high masses is quite steep and the metal production turns out 
to bee low; then the net yield must be increased to obtain Sbc/Sc discs 
with the observed metallicities. This is accomplished by tuning the lower
mass end to larger values, thereby decreasing the locked--up fraction but also,
correspondingly, the M/L ratio. Also in this case, we find $M/L_I \sim$0.8 for
Sbc--Sc discs.

Although uncertainties in the gas fractions (a basic constraint for
chemical evolution models) and in the theoretical stellar yields may hamper the
quantitative details of the models, e.g.\ the exact tuning
on the IMF mass limits, the predicted M/L ratios appear to be rather robust
with respect to these details. We consistently find that low M/L ratios, 
$M/L_I \sim$0.8 for the typical SFHs of Sbc/Sc discs, naturally follow from
currently popular, ``bottom--light'' models of the stellar IMF.
Even after including the contribution of bulges to the overall M/L ratio 
of disc galaxies, we globally find a typical M/L$_I$ \lsim 1
with the ``bottom--light'' IMFs.

\section*{Acknowledgments}
We are grateful to L\'eo Girardi for many clarifications on stellar isochrones
and photometry.
We also benefited from suggestions and comments by 
P.\ Salucci, R.\ Kennicutt, A.\ Bressan, S.\ Yi, 
S.\ Courteau, J.\ Mould, A.\ Weiss, L.\ Piovan, B.\ Gibson.
We thank our anonymous referee for very useful remarks.
LP acknowledges kind hospitality from the Astronomy Department of Padua 
University and from the Observatory of Helsinki on various visits.

\noindent
This study has been financed by 
the Danmarks Grundforskningsfond, through the funding of TAC, and
by the Italian MIUR.
%

\section*{Appendix A: The zero point of the I--band TF relation}
One of the motivations for this study was the need for a low M/L 
ratio in galactic discs to match the observed TF relation
(\S\ref{sect:introduction}). However, there are differences in the zero point 
of the TF relation among different samples.

In Fig.~\ref{fig:zeroTF_kin} we compare different {\mbox{I--band}} TF relations
based on large samples of cluster and/or field galaxies, as rescaled by 
their authors to the respective kinematic zero--points. 
The TF relation by Giovanelli \etal
is based on a sample of 24 clusters, and the zero--point is set
by requiring that the galaxies in the 14 most distant ones move on average 
with the Hubble flow (negligible peculiar velocities). A better kinematic
zero--point, based on a sample of 52 distant clusters, has been determined by
Dale \etal (1999), making the TF relation 0.1 mag dimmer 
than the one by Giovanelli \etal Still, even with this correction 
there is a mismatch with other samples --- see Fig.~\ref{fig:zeroTF_kin},
top panel.

\begin{figure}
\psfig{file=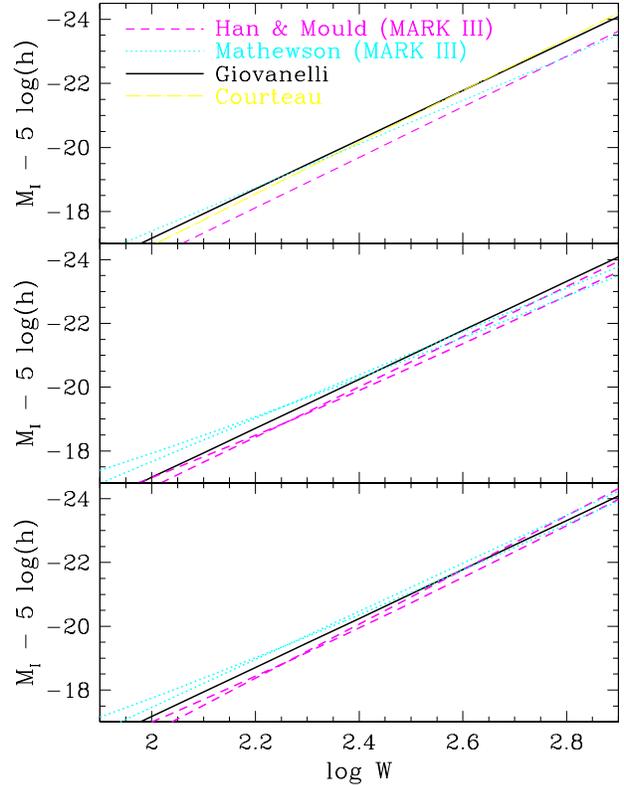,width=8.5truecm}
\caption{Comparison of different I--band TF relations with kinematic 
zero--points, from Giovanelli \etal (1997b), from the MARK~III catalogue
(Willick \etal 1995, 1996) and from Courteau \etal (2000). 
{\it Middle panel}: relations rescaled to a common $W_{50}$ (see text).
{\it Bottom panel}: \underline{inverse} MARK III TF relations, rescaled to 
$W_{50}$ (see text).}
\label{fig:zeroTF_kin}
\end{figure}

The sample by Han \& Mould (1992 and companion papers)
as reanalyzed in the MARK III catalogue by Willick \etal (1995), is also 
a large deep sample 
of 24 clusters, with the zero point set by requiring that the global
sample is on average moving as a pure Hubble flow. The sample by Mathewson,
Ford \& Buchhorn (1992), also reanalyzed in the MARK III catalogue,
is a large sample of relatively nearby, mostly field spirals, normalized
to the Han \& Mould sample in line widths and kinematic zero point
(Willick \etal 1996); compared to the other 
two TF relations, this is more influenced by peculiar motions and bulk flows 
due to
the vicinity of the galaxies. The TF relation by  Giovanelli et~al., which we
used as a reference in Fig.~\ref{fig:TF}, is the brightest. 
The large offset between the sample by Giovanelli \etal and that 
by Han \& Mould can be explained partly by the different definitions of 
line width used: at 50\% and 20\% of the maximum
HI flux, respectively ($W_{50}$ and $W_{20}$; Willick 1990; Han \& Mould 1992).
The two estimates can differ 
by up to a 10\% (Aaronson \etal 1986; Macri \etal 2000; Sakai \etal 2000),
or 0.04 in logarithm.
Since the TF relation is so steep, minor systematic offsets in the estimate 
of the circular velocity become crucial for the zero point in luminosity.

These offsets in velocity widths are unimportant  
when the TF relation is used as a relative distance indicator 
or to map
the peculiar velocity field (provided the definition is consistent within 
the sample), but they correspond to large effects in luminosity when the
TF relation is used to test galactic models.
The question obviously arises, which TF relation should galactic models 
be compared to. In Sommer--Larsen \& Dolgov (2001), Sommer--Larsen \etal
(2003) the circular velocity of the simulated disc galaxies is defined
as the circular velocity at $R=2.2 \, R_d$.
For Milky Way sized galaxies this corresponds quite well to the maximum
circular velocity, as obtained from optical rotation curves
(Sackett 1997; Courteau 1997).
As to radio line widths, $W_{50}$ is the one that best matches twice the 
optical maximum velocity (Buchhorn 1992; Mathewson \etal 1992;
Courteau 1997), and it is used in the Giovanelli \etal sample ---
with some further corrections to recover the true maximum rotational velocity
(Giovanelli \etal 1997a). In the middle
panel of Fig.~\ref{fig:zeroTF_kin} we rescale the two TF relations
of the MARK III samples from $W_{20}$ to $W_{50}$, dividing by 0.91
(Macri \etal 2000) or applying the corrections by Willick (1991), as in
Han \& Mould (1992).
These samples are now closer in zero point to the Giovanelli sample, 
yet with different conversions the difference in zero--point can reach
0.4~mag for Milky Way sized galaxies (log W $\sim$2.6), with the Giovanelli
\etal TF being the brightest for $V_c \gsim$130~km/sec..

Other effects may contribute to these magnitude offsets, most notably 
different 
internal dust corrections (Giovanelli \etal 1995 vs.\ Willick \etal 1996) 
or a different mixture of Hubble types.
The TF relation of Giovanelli \etal has been corrected to apply to
late--type spirals, while
the other relations are averaged over Hubble type. All samples
consist mostly of galaxies between Sb and Sc, but the systematic offset in 
magnitude of Sb galaxies may be larger than estimated by Giovanelli \etal
(see our model predictions in \S\ref{sect:Hubbledependence} and 
\S\ref{sect:bulge-Hubble}).

On the other hand, the Giovanelli \etal TF relation agrees well with the one by
Courteau \etal (2000), based on $\sim$300 galaxies at a distance of about 
6000~km~sec$^{-1}$ (shown only in the top panel 
of Fig.~\ref{fig:zeroTF_kin} for the sake of clarity).

\begin{figure}
\psfig{file=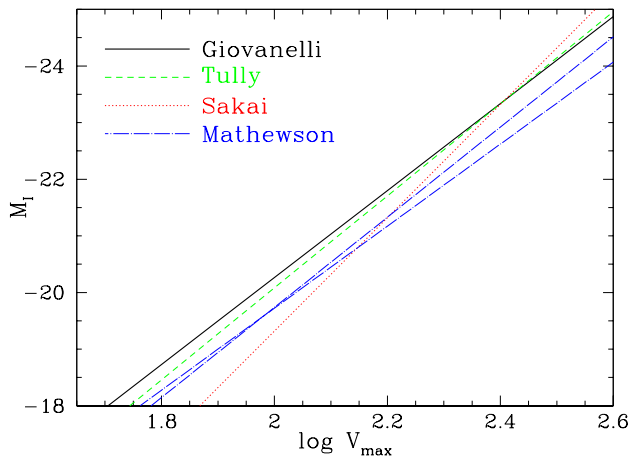,angle=-90,width=8.5truecm}
\caption{Comparison of different I--band TF relations with locally calibrated
zero--points; from Giovanelli \etal 1997b; Tully \etal 1998; Sakai \etal 2000;
Mathewson et~al.\ (1992); Buchhorn (1992).}
\label{fig:zeroTF_abs}
\end{figure}

Fig.~\ref{fig:zeroTF_abs} compares available I--band TF relations in  
absolute magnitude, calibrated using galaxies with Cepheid--based,
absolute distances. All relations are expressed as a function of the maximum
rotational velocity $V_{max}$. 

Here the Giovanelli \etal TF relation is plotted according
to their calibration using a local sample with known Cepheid distances,
corresponding to $h=0.69$ (Giovanelli \etal 1997b), hence very close to
the relation plotted in Fig.~\ref{fig:TF} where $h=0.65$ was used. Their 
linewidths $W_{50}$ are corrected so as to give directly a good estimate 
of twice $V_{max}$ (Giovanelli \etal 1997a).

The Tully \etal (1998) relation is based on two clusters
(Ursa Major and Pisces) and calibrated with local galaxies of known
Cepheid distances. To convert their linewidth $W_R$ to $V_{max}$
we adopted the corrections by Tully \& Fouqu\'e (1985).

The Sakai \etal (2000) TF relation is based on a 
combination of a local sample (with Cepheid distances obtained from 
the HST $H_0$ Key Project) and the sample by Giovanelli et~al. We assume here
that the linewidth they use, $W_{50}$, well corresponds to twice $V_{max}$ 
--- see above.

The Mathewson TF relations are those defined by Mathewson \etal (1992)
for Fornax, 
and by Buchhorn (1992) combining clusters and groups in the Mathewson sample
(hence these TF relations differ from the global 
Mathewson/MARK~III TF in Fig.~\ref{fig:zeroTF_kin}). 
We rescaled both relations to the recent,
Cepheid--based estimates of the Fornax distance modulus, 31.5 on average
(Mould \etal 2000). Their TF relations are expressed directly in terms of 
$V_{max}$, estimated mostly from optical rotation curves. 

In Fig.~\ref{fig:zeroTF_abs} also, the Giovanelli \etal TF relation 
is the brightest 
on average. There is very good agreement with the Tully \etal relation,
but large differences in slope and/or zero point ($\sim$0.6 mag) exist 
in comparison to the other samples.

Part of the differences in zero--point are related to the adopted 
technique for data analysis. One can derive ``direct'' TF relations
(least square fits minimizing the scatter in magnitude, with logarithmic 
velocity as independent variable), ``inverse'' TF relations (least square
fits with magnitude as independent variable), or bivariate TF relations
(as in Giovanelli \etal 1997a).
For the same sample, the direct TF relation tends to have a shallower slope,
hence be dimmer at the bright end, than the inverse TF relation
(e.g.\ Willick \etal 1995, 1996).
Fig.~\ref{fig:zeroTF_kin} and Fig.~\ref{fig:zeroTF_abs} seem to reveal
good agreement between bivariate (Giovanelli) and inverse (Courteau, Tully) 
TF relations drawn from different samples.
The Sakai TF relation is also bivariate; it is steeper than the Giovanelli 
\etal but the luminosities are comparable at the high velocity end. 
Direct TF relations
tend instead to be dimmer (MARK III in Fig.~\ref{fig:zeroTF_kin}, top and 
middle panels; Mathewson in Fig.~\ref{fig:zeroTF_abs}).
For the MARK III samples, inverse TF relations are also available
(Willick \etal 1995, 1996). In the bottom panel of Fig.~\ref{fig:zeroTF_abs}
we plot the {\it inverse} MARK III TF relations, also rescaled to a common
$W_{50}$ with the Giovanelli TF relation. There is now better agreement 
between the luminosities at the bright end. 

Various arguments favour an inverse or bivariate fit 
over a direct fit (Schechter 1980; Tully 1988; Giovanelli 1997a). In this 
sense we may disregard the dimmer direct TF relations and consider the 
Giovanelli \etal as in reasonable agreement, within uncertainties, with other 
(bivariate or inverse) TF relations.

\section*{Appendix B : Colour corrections to the stellar M/L ratio}
In this appendix we discuss the relation between the M/L ratio of the 
stellar (+remnants) component of a galaxy, and its colours --- hereinafter
MLC relation. MLC relations are useful, for instance, to colour--correct 
the TF relation to a
common zero point for all Hubble types, basing on the colours of the
individual galaxies (Kannappan \etal 2002), or to estimate 
the radial variation of the stellar M/L ratio 
in a disc, when deriving the mass profile deconvolving the
baryonic and the dark component (Kranz \etal 2003).
The luminosities and colours computed here do not include dust effects, hence
observational colours and magnitudes should be dust-corrected before 
our MLC relations can be applied. However, the reddening+dimming effects
of dust result in a vector that runs almost parallel to the MLC relation, 
so that to a first approximation dust has a minor influence on stellar mass 
estimates from the MLC relation (Bell \& de Jong 2001, hereinafter BdJ). 

For simplicity, in this appendix we consider only IMFs with the ``standard'' 
mass limits given in \S\ref{sect:IMFlist}: [0.1-100]~\Msol\ 
for the Salpeter, Kroupa and Kennicutt IMFs and [0.01-100]~\Msol\ for the 
Larson, modified Larson and Chabrier IMFs (in these cases whether the lower 
limit is 0.01 or 0.1~\Msol\ has negligible impact on the M/L ratios).
Other mass ranges, considered for example in Table~\ref{tab:models} 
when tuning the level of metal enrichment, would correspond to simple
changes in the zero--point of the MLC relation --- as will become apparent
from what follows.

Fig.~\ref{fig:colourML} shows some examples of MLC relations
for three representative IMFs among those considered in this paper. The plot
shows results from our chemo-photometric models of disc galaxies, both
the calibrated models in Table~\ref{tab:models} and several other models 
computed during the calibration process.
Open symbols represent results for the individual one-zone annuli that
constitute our disc models (see \S\ref{sect:chemo-photo}); solid symbols
represent results for multi--zone regions, namely they are the integrated 
luminosities and colours of the sum of several annuli (e.g.\
the ``bulge region'' within $R<0.5~h_B$, or the ``disc region'' $R>0.5~h_B$,
or the global galaxy, etc.; see \S\ref{sect:calibration}). Asterisks represent
results for the ``optical disc region'' ($0.5 \, h_B < R < 3 \, h_B$) discussed
extensively in the paper (\S\ref{sect:calibration}). Henceforth, points falling
in the same colour--M/L ratio region in the plots result from a variety
of star formation and chemical enrichment histories, both one--zone and
composite. This is important to see if colours can be good tracers of 
the underlying M/L ratio variations, with no need of independent 
knowledge of the star formation and metal enrichment history.

A well defined linear relation between colour and logarithm of the M/L ratio 
is apparent in the plots (see also BdJ), at least down 
to a certain limit in colour (e.g.\ $\sim$0.5 in 
$B-V$); blueward of this, models deviate toward M/L ratios lower
(higher in the K band) than the linear fit. In the ``linear range'', the 
MLC relation is rather tight in spite of the variety of SFH and chemical
evolution history that can result in a given colour. The worst correlations
are obtained with {\mbox{$V-K$}}, and in general with colours involving 
the K band
($B-K, R-K$ etc.) or in MLC relations involving the K luminosity.
The reason is that in optical colours, the age--metallicity
degeneracy favours convergence into a tight MLC relation, while in IR bands
and colours, age and metallicity effects are less degenerate  
(see BdJ, their Fig.~2 and related comments).
Hence the $K$ band (and NIR bands in general) offers the advantage of 
a smaller variation of the M/L ratio with colour, Hubble type, SFH and
metal enrichment history, but 
colour corrections and MLC relations based on the K band are less tight 
and compelling.

Fig.~\ref{fig:colourML} shows results for the Salpeter, Kroupa and Kennicutt 
IMFs as representative cases. 
To a first approximation, the slope of the MLC relation
does not depend on the IMF, which just sets the zero--point; here we
confirm the findings of BdJ. For all the 6 IMFs considered in this paper
we did verify that, {\it when the linear fit is performed over the same colour
range}, the slope of the MLC relation is independent of the IMF within 
the uncertainty given by the scatter. We provide in 
Table~\ref{tab:MLCslopes}
the typical, IMF-independent slopes of the MLC relations, as well as the 
zero--points corresponding to the different IMFs. 
We express the zero--points as the logarithmic M/L ratios corresponding 
to certain values of the colour; specifically:
\[ (B-V)~:~~~ \log \left( \frac{M}{L_X} \right) = s_X [(B-V)-0.6] + q_X \]
\[ (B-R)~:~~~ \log \left( \frac{M}{L_X} \right) = s_X [(B-R)-1.0] + q_X \]
\[ (V-I)~:~~~ \log \left( \frac{M}{L_X} \right) = s_X [(V-I)-0.9] + q_X \]
\[ (V-K)~:~~~ \log \left( \frac{M}{L_X} \right) = s_X [(V-K)-2.5] + q_X \]
Table~\ref{tab:MLCslopes} lists the slope $s$ and zero--point $q$ 
for the MLC relations in the $B,V,R,I,K$ bands; also given are the
color ranges over which the linear fits are valid. MLC relations with 
a different colour base (e.g.\ $V-R, B-I$, etc.) can be obtained as a linear 
combination of these.

\begin{figure*}
\centerline{\psfig{file=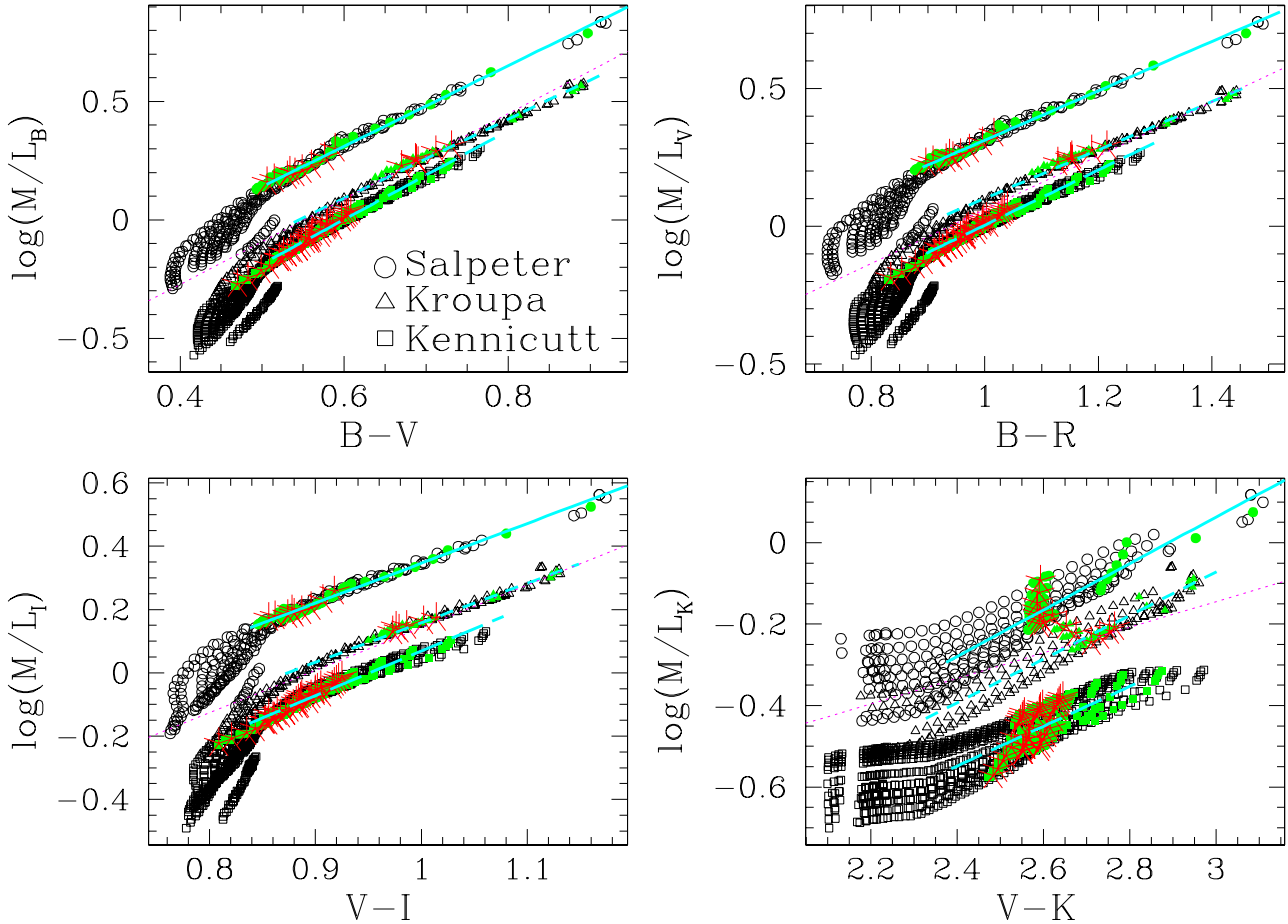,angle=-90,width=18truecm}}
\caption{MLC relations from chemo-photometric models of disc galaxies
with three representative IMFs. {\it Open symbols}: one--zone models for 
individual disc annuli; {\it solid symbols}: multi-zone models (integrate
photometry of bulge region, disc region, or global galaxy); {\it asterisks}: 
``optical disc'' region (see text). The solid, short dashed and long dashed
lines represent the linear fits for the different IMFs, over the respective 
relevant colour range. The thin dotted line is the linear fit from Bell \& de
Jong (2001).}
\label{fig:colourML}
\end{figure*}

\begin{table}
\begin{center}
\begin{tabular}{|c|ccccc|c|}
\hline
colour &    range   & $s_B$ & $s_V$ & $s_R$ & $s_I$ & $s_K$ \\ 
\hline
 $B-V$ & 0.55--0.90 & 1.69  & 1.29  & 1.11  & 0.97  & 0.73 \\
 $B-R$ & 0.95--1.45 & 1.17  & 0.89  & 0.77  & 0.68  & 0.51 \\
 $V-I$ & 0.88--1.14 & 2.17  & 1.66  & 1.43  & 1.26  & 0.96 \\
 $V-K$ &  2.4--3.0  & 1.23  & 0.97  & 0.84  & 0.76  & 0.57 \\
\hline
\end{tabular}

\begin{tabular}{|l|rrrrr|}
\hline
colour & $q_B$ & $q_V$ & $q_R$ & $q_I$ & $q_K$ \\
\hline
\multicolumn{6}{c}{Salpeter IMF [0.1--100]~\Msol} \\
 $B-V$ &   0.311 &   0.341 &   0.317 &   0.260 & --0.115 \\
 $B-R$ &   0.272 &   0.312 &   0.292 &   0.236 & --0.134 \\
 $V-I$ &   0.246 &   0.292 &   0.275 &   0.220 & --0.147 \\
 $V-K$ &   0.093 &   0.169 &   0.168 &   0.123 & --0.222 \\
\hline
\multicolumn{6}{c}{Kroupa IMF [0.1--100]~\Msol} \\
 $B-V$ &   0.089 &   0.120 &   0.099 &   0.045 & --0.298 \\
 $B-R$ &   0.061 &   0.099 &   0.080 &   0.027 & --0.311 \\
 $V-I$ &   0.068 &   0.104 &   0.086 &   0.032 & --0.309 \\
 $V-K$ &   0.002 &   0.048 &   0.036 & --0.015 & --0.343 \\
\hline
\multicolumn{6}{c}{Kennicutt IMF [0.1--100]~\Msol} \\
 $B-V$ & --0.002 &   0.028 &   0.007 & --0.050 & --0.421 \\
 $B-R$ & --0.034 &   0.004 & --0.014 & --0.068 & --0.435 \\
 $V-I$ & --0.041 & --0.002 & --0.020 & --0.074 & --0.438 \\
 $V-K \leq 2.8$ & --0.182 & --0.114 & --0.117 & --0.162 & --0.505 \\ 
\hline
\multicolumn{6}{c}{Larson IMF [0.01--100]~\Msol} \\
 $B-V$ &   0.081 &   0.111 &   0.088 &   0.029 & --0.359 \\
 $B-R$ &   0.045 &   0.085 &   0.065 &   0.010 & --0.373 \\
 $V-I$ &   0.023 &   0.068 &   0.050 & --0.004 & --0.383 \\
 $V-K \leq 2.8$ & --0.155 & --0.075 & --0.075 & --0.118 & --0.466 \\ 
\hline
\multicolumn{6}{c}{modified Larson IMF [0.01--100]~\Msol} \\
 $B-V$ &   0.064 &   0.095 &   0.073 &   0.016 & --0.356 \\
 $B-R$ &   0.031 &   0.069 &   0.051 & --0.004 & --0.371 \\
 $V-I$ &   0.025 &   0.065 &   0.047 & --0.006 & --0.374 \\
 $V-K$ & --0.123 & --0.055 & --0.059 & --0.104 & --0.446 \\
\hline
\multicolumn{6}{c}{Chabrier IMF [0.01--100]~\Msol} \\
 $B-V$ &   0.131 &   0.161 &   0.138 &   0.080 & --0.303 \\
 $B-R$ &   0.095 &   0.134 &   0.115 &   0.060 & --0.318 \\
 $V-I$ &   0.078 &   0.121 &   0.103 &   0.049 & --0.325 \\
 $V-K \leq 2.8$ & --0.107 & --0.028 & --0.029 & --0.072 & --0.419 \\ 
\hline
\end{tabular}

\end{center}
\caption{Coefficients for the MLC relations. The slopes $s_X$ are independent
of the IMF, while the zero--points $q_X$ depend on the IMF. ``Range''
indicates the colour range where the linear fit is a good representation
of the MLC relation; the lower limit is set by the ``deviations'' from
a linear relation (see text and Fig.~\ref{fig:colourML}), the upper limit 
by the reddest colour probed by the models. For the Kennicutt, Larson and
Chabrier IMFs, more strict upper limits apply in V-K, as indicated (see text).}
\label{tab:MLCslopes}
\end{table}

The models with the Kennicutt IMF in Fig.~\ref{fig:colourML} do not reach 
so red colours as the other two sets, because a significant SFR at late times
is maintained by the large returned gas fraction from the stellar populations 
($1-\alpha = 40-50$\% for the Kennicutt, Larson and Chabrier IMFs, 
Table~\ref{tab:IMFyields}). As a consequence, low
$b$-parameters (corresponding to red colours) are hardly achieved
with these IMFs, unless gas outflows occur, not included in our disc models. 
The slopes $s_X$ given in Table~\ref{tab:MLCslopes} have been derived from
models with the Salpeter, Kroupa and modified-Larson IMFs, for these models
extend to redder colours so the slopes could rely on a wider colour range.

For the Kennicutt (and Larson and Chabrier) case, the linear
relation seems to break down also toward the red in $V-K$
(Fig.~\ref{fig:colourML}, lower right panel), as well as in other
colours involving the K--band ($B-K$, $R-K$ etc.). It is not clear if
this is due to a selection effect --- all models with these IMFs
are blue in optical colours --- or whether it reflects the fact
that the evolution of $V-K$ is more sensitive to the IMF
than bluer colours (Vazdekis \etal 1996, 1997). Anyways, in such cases
we give both lower and upper limits to the ``linear range'' where the
linear fit can be applied.

A thorough discussion of MLC relations and their implications on the TF 
relation is given by BdJ. The dotted lines in 
Fig.~\ref{fig:colourML} are the linear MLC relations derived by BdJ
for their favourite set of galactic models (their Table 1). Overall, 
their slopes
are in good agreement with ours, albeit slightly steeper (with the exception 
of the MLC relations in $V-K$). Our slopes
are somewhat intermediate between those of their full galactic models
(their Table~1 and Fig.~\ref{fig:colourML}) and those of their simple 
exponential models at fixed metallicity (their Table~4).
We impute this slight difference to a tighter calibration 
in metallicity for our galactic models (\S\ref{sect:calibration}),
while BdJ select their models mostly on the base of their
K--band magnitude and central surface brightness.

Their normalization of the MLC relation, based on maximal disc arguments,
corresponds roughly to a Kroupa IMF. Other IMFs with lower M/L ratios, 
such as the Kennicutt IMF 
(Fig.~\ref{fig:colourML}), would then imply slightly submaximal discs.
The Salpeter IMF seems to be excluded from maximal disc arguments 
(BdJ). 

The deviation from the MLC relation at blue colours seen
in Fig.~\ref{fig:colourML} is not apparent in the results by BdJ.
However, in their Fig.~10 their galactic models with a more regular
SFH (their closed box, infall, and outflow case), which are probably 
more similar to our chemo--photometric models, do show a slight bend-over
in the MLC relation and do not stretch to very blue colours.
In their favourite set of models, small bursts of star formation are
included and the bend-over
is wiped out by the larger scatter. The different behaviour toward
blue colours is probably
due to the fact that, while their models are global galactic models selected
so as to have K band magnitudes and central surface brightnesses 
representative of spiral galaxies, our blue models (say, bluer than 
B-R$\sim$0.9 for instance) corresponds to individual external annuli of our 
galactic disc models, characterized by overall young ages and low 
metallicities.

Our steeper slope in the MLC relations in V-K is probably due to the fact
that it is defined over a narrower colour range, out of which our MLC relation
tends to flatten (Fig.~\ref{fig:colourML}, bottom right panel).

Considering that both our galactic chemical evolution models and our 
population synthesis code
differ from BdJ, we regard our results as in good agreement with theirs,
which supports the use of MLC relations as a robust theoretical prediction.

\end{document}